\documentclass[10pt]{article}

\usepackage[utf8]{inputenc}
\usepackage{amsmath}
\numberwithin{equation}{section}
\usepackage{amssymb,amsfonts,amsthm}
\usepackage{mathrsfs}
\usepackage{bbold}
\newcommand{\bbGreek}[1]{\mathbb{#1}}
\IfFileExists{stmaryrd.sty}{\usepackage{stmaryrd}}{\newcommand{\llbracket}{[\mkern-3.5mu[}\newcommand{\rrbracket}{]\mkern-3.5mu]}}
\usepackage{cite}
\usepackage{verbatim}
\usepackage{float}
\usepackage{color}
\usepackage[a4paper, total={6in, 8in}]{geometry}
\usepackage{appendix}
\usepackage{hyperref}
\hypersetup{
  colorlinks=true,
  linkcolor=blue,
  citecolor=blue,
  urlcolor=blue,
  pdfauthor={Matteo Inglese, Dario Martelli and Antonio Pittelli},
  pdftitle={Supersymmetry and localization on three-dimensional orbifolds}
}
\usepackage{authblk}
\usepackage{upgreek}

\newcommand{\samf}{{\mathfrak s_\omega}}
\newcommand{\kcs}{{ \rm k }}
\newcommand{\linbun}{{ \rm L }}
\newcommand{\UN}{{ \mathcal U_+ }}
\newcommand{\US}{{ \mathcal U_- }}
\newcommand{\eqp}{{ \epsilon }}
\newcommand{\blens}{{\bbGreek{\Lambda}}}
\newcommand{\flux}{{\mathfrak f}}
\newcommand{\rfug}{{\gamma_R}}
\newcommand{\gfug}{{\gamma_G}}
\newcommand{\ag}{ { \mathcal A  } }
\newcommand{\acf}{ { A^{(1)} } }
\newcommand{\subm}{ { \mathrm c } }
\newcommand{\knorm}{k_0}

\newcommand{\st}{{\upsigma}}
 \newcommand{\amf}{{\omega }}
\newcommand{\ff}[1]{{\left\lfloor{#1}\right\rfloor}}
\newcommand{\cf}[1]{{\left\lceil{#1}\right\rceil}}
\newcommand{\rem}[2]{{\left\llbracket{#1}\right\rrbracket_{#2}}}
\newcommand{\f}[2]{{\frac{#1}{#2}}}
\newcommand{\lc}{\varepsilon}
\newcommand{\p}[1]{{\left({#1}\right)}}
\newcommand{\comm}[1]{{\left[{#1}\right]}}
\newcommand{\acomm}[1]{{\left\{{#1}\right\}}}

\newcommand{\nn}{\nonumber}
\newcommand{\spindle}{{\bbGreek{\Sigma}}}

\newcommand{\arXivLink}[1]{\href{https://arxiv.org/abs/#1}{\texttt{arXiv:#1}}}
\newcommand{\doiLink}[1]{\href{https://doi.org/#1}{\texttt{doi:#1}}}

\newcommand{\ee}{ {  \rm e } }

\newcommand{\aN}{{\mathfrak a_+}}
\newcommand{\aS}{{\mathfrak a_-}}
\newcommand{\kN}{k_+}
\newcommand{\kS}{k_-}
\newcommand{\mN}{{\mathfrak m_+}}
\newcommand{\mS}{{\mathfrak m_-}}
\newcommand{\nN}{{n_+}}
\newcommand{\nS}{{n_-}}
\newcommand{\nbN}{p_+}
\newcommand{\nbS}{p_-}
\newcommand{\pN}{\mathfrak p_+}
\newcommand{\pS}{\mathfrak p_-}
\newcommand{\qN}{ q_+}
\newcommand{\qS}{q_-}
\newcommand{\qe}{{\rm q} }
\newcommand{\tN}{{t_+}}
\newcommand{\tS}{{t_-}}
\newcommand{\wN}{w_+}
\newcommand{\wS}{w_-}
\newcommand{\zN}{z_+}
\newcommand{\zS}{z_-}
\newcommand{\mm}{{\mathfrak m}}
\newcommand{\hh}{{\mathfrak h}}

\newcommand{\rhoN}{{\rho_+}}
\newcommand{\rhoS}{{\rho_-}}

\newcommand{\Dr}{{\rm D}}

\newcommand{\dd}{{\rm d}}
\newcommand{\im}{{\rm i}}

\newcommand{\Bt}{{\widetilde B}}
\newcommand{\Ct}{{\widetilde C}}
\newcommand{\Ft}{{\widetilde F}}
\newcommand{\Pt}{{\widetilde {\rm P} }}

\newcommand{\Ac}{{\mathcal A}}
\newcommand{\Cc}{{\mathcal C}}
\newcommand{\Dc}{{\mathcal D}}
\newcommand{\Fc}{{\mathcal F}}
\newcommand{\Gc}{{\mathcal G}}
\newcommand{\Lc}{{\mathcal L}}
\newcommand{\Rc}{{\mathcal R}}

\newcommand{\g}{\gamma}
\newcommand{\phit}{{\widetilde\phi}}
\newcommand{\la}{\lambda}
\newcommand{\lat}{{\widetilde\lambda}}
\newcommand{\psit}{{\widetilde\psi}}
\newcommand{\z}{\zeta}
\newcommand{\zt}{{\widetilde\zeta}}
\newcommand{\zzt}{{v}}
\newcommand{\Tht}{{\widetilde\Theta}}

\newcommand{\sigmavar}{\varphi}

\newcommand{\rf}{{ \mathfrak l }}

\title{Supersymmetry and localization on three-dimensional orbifolds}

\author{Matteo Inglese, Dario Martelli and Antonio  Pittelli}
\affil{\emph{Department of Mathematics ``Giuseppe Peano'', University of Turin,}\\
\emph{Turin, 10123, Italy}}
\date{}

\begin{document}

\maketitle

\begin{abstract}
\noindent  
We consider three-dimensional ${\mathcal N}=2$ supersymmetric field theories defined on general complex-valued backgrounds of Euclidean new minimal supergravity admitting two Killing spinors of opposite $R$-charges.  We  compute  partition functions for   theories defined on general circle bundles over spindles $\spindle$, including  $\spindle \times S^1$ as well as branched and squashed lens spaces, thus obtaining novel  observables  characterizing   three-dimensional supersymmetric gauge theories. We discuss both twisted and anti-twisted theories compactified on $\spindle \times S^1$ and demonstrate that their partition functions are encoded by  a single formula that we  refer to as the  \emph{spindle index}, unifying    and generalizing      superconformal and topologically twisted indices in the limit where   orbifold singularities are absent. Furthermore, we test our new  index using non-perturbative dualities and obtain one-loop determinants of  two-dimensional supersymmetric gauge theories  compactified on the spindle.

\end{abstract}

\newpage

\setcounter{tocdepth}{3}
\tableofcontents 


\section{Introduction}

  The investigation of supersymmetric quantum field theories (SQFTs) stands as a longstanding pursuit in theoretical physics as it represents  an invaluable avenue for exploring non-trivial phenomena,  
particularly in the regime of strong interactions, where perturbation theory fails. From a physics standpoint the exact computation of SQFTs' partition functions on curved backgrounds, made possible through localization techniques  \cite{Pestun:2007rz}, has  produced profound insights on   non-perturbative 
aspects of    gauge theories, including  the discovery of  dualities among seemingly completely different models. 
 From a mathematical point of view, such exact partition functions may serve as generating functions for a number of interesting  invariants 
 characterizing the manifolds where the theory is defined,  see \emph{e.g.} \cite{Manschot:2023rdh} and references therein for a recent review.

 Partition functions of SQFTs defined on the direct product of a circle  and  a compact manifold play a special role as they correspond to supersymmetric indices, which may be interpreted in terms of ``counting'' of various quantities. The study of indices 
has revealed  unforeseen correlations across different areas of physics and mathematics,  establishing connections between SQFTs, the  geometry of the moduli space in the context of string theory, as well as   in purely mathematical fields such as symplectic and algebraic geometry and representation theory, including    Gromov-Witten invariants and characters of infinite-dimensional algebras. Recent advancements in localization techniques have yielded significant breakthroughs, facilitating explicit computations of supersymmetric indices as well as other partition functions corresponding to manifolds with diverse  topology and dimension.

Supersymmetric partition functions  play also an important role in the AdS/CFT correspondence, where they provide  precision checks of dualities between gravitational backgrounds and holographically dual field theories. The most striking instance of this aspect is 
the use of supersymmetric indices to perform a precise counting of the microstates of black holes, thereby offering a glimpse into the fundamental tenets of quantum gravity. In particular, the  large-$N$ limit of  supersymmetric indices has  been extensively  used to obtain a  microscopic derivation of the entropy of supersymmetric black holes \cite{Benini:2015eyy,Cabo-Bizet:2018ehj,Choi:2018hmj,Benini:2018ywd}.  Several other works have   studied different asymptotic limits and determined subleading corrections, matching these to higher-derivative or quantum corrections in supergravity
\cite{Liu:2017vbl,Bobev:2020egg,Bobev:2021oku,Bobev:2023dwx}.  Another set of  precision tests in holography involves matching the large-$N$ limit of    partition functions of 
  three-dimensional ${\cal N}=2$ gauge theories on manifolds with $S^3$ topology \cite{Imamura:2011wg,Hama:2011ea,Alday:2013lba} to various supergravity solutions \cite{Martelli:2011fu,Martelli:2011fw,Martelli:2012sz,Martelli:2013aqa,Farquet:2014kma}.

In the past two years, starting with \cite{Ferrero:2020laf}, a substantial body of literature has demonstrated that supergravity solutions displaying orbifold singularities should not be discarded.
On the contrary, such solutions are expected to play a crucial role as non-trivial saddle points in the quantum-gravity path integral or in its semi-classical limit. In this extended landscape of solutions  supersymmetry can be realized in novel fashions,   raising the intriguing question of how typical these supergravity backgrounds are.  One of the simplest and perhaps most interesting  scenarios  is that of four-dimensional 
supersymmetric\footnote{See also \cite{Klemm:2013eca} for a discussion of supersymmetry of the solutions of \cite{Plebanski:1976gy,Podolsky:2006px}.} and accelerating black holes \cite{Plebanski:1976gy,Podolsky:2006px}, wherein  the horizon is a spindle $\spindle$ \cite{Ferrero:2020twa}, with  $\spindle=\mathbb{WCP}^1_{\comm{\nN, \nS}}$ being  the two-dimensional weighted complex projective  line whose topology is that of a two-sphere with conical singularities $\mathbb C/\mathbb Z_{n_\pm}$ at the poles. The conformal boundary of these geometries comprises a metric on  $\spindle \times \mathbb{R}_t$ together with a gauge field with ``anti-twisted'' flux on $\spindle$; therefore, via holography, their microstates are expected to be captured by the supersymmetric partition function of the dual ${\cal N}=2$ three-dimensional field theories defined on a rigid supersymmetric background on $\spindle \times S^1$ \cite{Cassani:2021dwa}. 
In this work we will discuss in detail the calculation of this new observable, that we named   \emph{spindle index} in \cite{Inglese:2023wky}.  

More broadly, the recent supergravity constructions involving orbifolds prompted  a number of related questions concerning  superconformal field theories (SCFTs). For example, it would be interesting to systematically understand  the data defining  
a SCFT at an orbifold point and to identify  the additional properties that one can hope to detect on such backgrounds. For SCFTs in even space-time dimension  some progress has been made by studying the 't Hooft anomalies of   theories compactified on orbifolds, extending anomaly-polynomial techniques to spaces with conical singularities \cite{Ferrero:2020laf,Hosseini:2021fge,Martelli:2023oqk}. This approach is somewhat insensitive to the detailed description of the microscopic fields at the orbifold singularities, suggesting that a similar point of view may be fruitful for studying field theories on orbifolds also in odd space-time dimensions. In particular, we expect that    \emph{orbifold}   indices \cite{Pittelli:2024ugf} and other  orbifold partition functions should unveil  novel features  of SQFTs in both odd and even space-time dimension, also beyond the context  of holography.

Although one of the  main motivations of  this work was the microstate counting of the supersymmetric and accelerating AdS$_4$ 
black holes found in \cite{Ferrero:2020twa,Ferrero:2021ovq}, in this paper we shall  discuss a much more general setting, for which the ``anti-twisted'' spindle index, relevant for the black holes of \cite{Ferrero:2020twa},  is a particular case. 
 Indeed, we will consider  ${\mathcal  N} = 2$ supersymmetric field theories placed on compact three-orbifolds and compute their full supersymmetric partition functions.  Technically, a key ingredient in our calculations is the  equivariant index theorem on orbifolds  \cite{vergne,MEINRENKEN1998240}. We expect that our methods can be readily extended to other setups, including theories in dimension higher than three and possibly incorporating boundaries. 

\paragraph{Outline.}   
In Section \ref{sec: susybackgrounds}  we characterize a very general class of {\it complex}, local and rigid  supersymmetric backgrounds. We will show that these comprise all the known supersymmetric geometries in three dimensions, as well as several generalizations that we discuss in detail.  For instance, we describe general backgrounds on $\spindle \times S^1$, with both twist and  anti-twist for the 
background $R$-symmetry connection. Moreover, the conformal boundary of the accelerating black hole of \cite{Ferrero:2020twa} as well as its complex deformation \cite{Cassani:2021dwa} are shown to fit into our general canonical form.
In Section \ref{sec: localizationspindle} we calculate  the full supersymmetric partition function of  three-dimensional  ${\cal N}=2$ Chern-Simons-matter theories on $\spindle  \times S^1$, whose  one-loop determinants are obtained by employing 
both the  unpaired-eigenvalues method  and the equivariant  index theorem on orbifolds. The ultimate outcome  encompasses     anti-twisted and twisted partition functions of gauge theories on $\spindle \times S^1$, demonstrating that the final expression  can be written in a unified form, corresponding to  the spindle index \cite{Inglese:2023wky}. We also include flavour fugacities as well as several classical terms, which  were not discussed in \cite{Inglese:2023wky}, and discuss  subtleties related to the breaking of gauge invariance and its restoration 
through the introduction of effective  terms. Our final result for the spindle index is tested via  some simple field-theory dualities that hold in a form non-trivially  affected  by the presence of the orbifold singularities of $\spindle$. As a further application of our formalism,   
 in Section \ref{sec: localizationbranchedlens} we compute the full supersymmetric partition function of  three-dimensional  ${\cal N}=2$ gauge  theories on a {\it branched and squashed  lens  space}, again 
 finding a result that unifies and generalises previous constructions regarding non-trivial circle (orbi-)bundles over two-dimensional (orbifold) Riemann surfaces. In particular, our partition functions encompass the path integrals of gauge theories  on the squashed lens space \cite{Imamura:2012rq} and on the branched three-sphere \cite{Nishioka:2013haa} as special cases. Features  of this observable and its possible applications will be discussed further in 
Section \ref{sec: conclusions}. 
In Section 
\ref{sec: spindlefromthreesphere} we show how to recover the one-loop determinants relevant for the spindle index by viewing the three-sphere as an orbi-bundle over the spindle and subsequently  
reducing the three-sphere one-loop determinant. This procedure extends  a method proposed in \cite{Benini:2012ui} to  quickly obtain the two-sphere partition function out of the three-sphere one. 
In Section \ref{sec: conclusions} we summarize and  discuss our findings, providing  a number of immediate applications and outlining a broader program for continuing the 
exploration of SQFTs on orbifolds that we initiated here. Further technical material is reported in the appendices.


\section{Supersymmetric complex backgrounds}\label{sec: susybackgrounds}

\subsection{General analysis}

 We will employ the well-established framework of the rigid limit of new minimal supergravity in three-dimensions  to study three-dimensonal supersymmetric ${\cal N}=2$ theories in curved backgrounds. The bosonic fields in the gravity multiplet are the metric and the auxiliary fields $A_\mu,V_\mu,H$, which in the rigid limit become background fields.
In Euclidean signature $A_\mu,V_\mu,H$ are allowed to take complex values, while traditionally 
the metric $g_{\mu\nu}$ has been assumed to be real-valued. In view of some recent developments, in particular in the context of BPS black holes this turn out to be too restrictive and in this paper we will therefore allow the metric to take \emph{complex values}.

The  Killing-spinor equations (KSEs) read
\begin{align}\label{eq: kses}
& \Dc_\mu \z = (\nabla_\mu - \im A_\mu)\z = - \f H2 \g_\mu \z - \im V_\mu \z - \f12 \lc_{\mu\nu\rho} V^\nu \g^\rho \z ~ , \nn \\
& \Dc_\mu \zt = (\nabla_\mu + \im A_\mu)\zt = - \f H2 \g_\mu \zt + \im V_\mu \zt + \f12 \lc_{\mu\nu\rho} V^\nu \g^\rho \zt ~ ,  
\end{align}
where in Euclidean the two Killing spinors $(\z , \zt)$ of $R$-charges $(+1, -1)$ are independent.
On the other hand, in Lorentzian signature, the second equation is simply as the charge conjugate of the first one, with $\zt=\z^c$. In previous
literature \cite{Closset:2012ru,Alday:2013lba} the attention has been restricted to such ``real'' backgrounds, while here we will relax this assumption. 

 In order to encompass general backgrounds in which the metric can be complex and the Killing spinors $(\z , \zt)$ are not related by charge conjugation, we will work with a ``holomorphic'' set of spinor bilinears, that are constructed without performing charge conjugation. In this way, we will never have to work with equations that contain the complex conjugate of the metric (or of the other background fields). In particular, 
we define the following complex-valued bilinears of Killing spinors:
\begin{align}\label{eq: ksbilinears}
 \zzt = \z  \zt ~ , \qquad K^\mu = \z \g^\mu \zt ~ , \qquad P^\mu =  \z \g^\mu \z / \zzt ~ , \qquad \widetilde P^\mu =  \zt \g^\mu \zt / \zzt ~ ,
\end{align}
where $\widetilde P^\mu \neq (P_\mu)^*$.
Using the Fierz identities we obtain
\begin{align}\label{eq: kpptorthogonality}
& \zzt^{-2}K_{(\mu}K_{\nu)} - \widetilde P_{(\mu}  P_{\nu )} = g_{\mu\nu} ~ , & \iota_K K^\flat = \zzt^2 ~ , \qquad  \iota_{\widetilde P} P^\flat = \iota_{ P} \widetilde P^\flat = -2 ~ ,
\end{align}
 where $X^\flat = \iota_X g$ is the 1-form dual to the vector $X$ with respect to the three-dimensional metric tensor $g$.
as well as  the identities
\begin{align}\label{eq: kpptalgebra}
K_{[\mu} P_{\nu]} = \f{\im \zzt}2 \lc_{\mu\nu\rho}P^\rho ~ , \qquad K_{[\mu} \widetilde P_{\nu]} = - \f{\im \zzt}2 \lc_{\mu\nu\rho}\widetilde P^\rho ~ , \qquad \widetilde P_{[\mu} P_{\nu]} = \f{\im }\zzt \lc_{\mu\nu\rho}K^\rho ~ ,
\end{align}
so that the complex one-forms $(K, \widetilde P , P)$ constitutes a three-dimensional ``complex frame''. 
It is then straightforward to obtain the covariant derivatives of the Killing-spinor bilinears, that read 
\begin{align}\label{eq: ksbd}
 \nabla_\mu \zzt & = \lc_{\mu\nu\rho}V^\nu K^\rho ~ ,  \nn \\
 \nabla_\mu K_\nu & = \lc_{\mu\nu\rho}(\im H K^\rho - \zzt V^\rho) ~ , \nn \\
\p{ \nabla_\mu - 2 \im  A_\mu } P_\nu & = + \f{2H}\zzt K_{[\mu} P_{\nu]} - 2 \im V_\mu P_\nu + \f{ \im }{\zzt^2} \comm{ (\iota_P V) K_{\mu}  - ( \iota_K V) P_{\mu} } K_\nu ~ , \nn \\
 \p{ \nabla_\mu + 2 \im  A_\mu } \widetilde P_\nu  & = -  \f{2H}\zzt K_{[\mu} \widetilde P_{\nu]} + 2 \im V_\mu \widetilde P_\nu - \f{\im}{\zzt^2} \comm{ ( \iota_{\widetilde P}V)  K_{\mu} - ( \iota_K V)\widetilde P_{\mu} } K_\nu ~ .
\end{align}
In particular,  $\mathcal L_K \zzt = 0$ and 
\begin{align}
\nabla_{(\mu} K_{\nu)} = 0 ~ , \qquad \comm{ \nabla_\mu - 2 \im \p{ A_\mu - \f12 V_\mu } }P^{\mu} = 0 ~ , \qquad  \comm{ \nabla_\mu + 2 \im \p{ A_\mu - \f12 V_\mu } } \widetilde P^{\mu} = 0 ~ ,
\end{align}
ensuring that $K^\mu$ is a \emph{complex} Killing vector. The equations (\ref{eq: ksbd}) can be inverted to obtain expressions of the background fields in term of the bilinears. Firstly, we have
\begin{align}
\lc^{\mu\nu\sigma}\nabla_\mu K_\nu & = 2 \p{ \im H K^\sigma - \zzt V^\sigma } ~ , \nn \\
\lc^{\mu\nu\sigma}\nabla_\mu \zzt & = 2 V^{[\nu}K^{\sigma]} ~ , \nn \\
\lc^{\mu\nu\sigma} \p{ \nabla_\mu - 2 \im A^C_\mu } P_\nu & = \f{(\iota_P V)}\zzt K^\sigma + \f2\zzt\comm{ \im \zzt H - ( \iota_K V) }P^\sigma ~ , \nn \\
\lc^{\mu\nu\sigma} \p{ \nabla_\mu + 2 \im A^C_\mu } \widetilde P_\nu & = \f{( \iota_{ \widetilde P} V )}\zzt K^\sigma + \f2\zzt\comm{ \im \zzt H - (\iota_K V) }\widetilde P^\sigma ~ ,
\end{align}
enjoying the shift symmetry
\begin{align}\label{eq: AHVshiftsym}
H \to H + \im \, h ~ , \qquad V_\mu \to V_\mu - (h/\zzt) K_\mu ~ , \qquad A_\mu \to A_\mu - \f{3  }{ 2  } (h/\zzt) K_\mu ~ ,
\end{align}
with $h$ being a complex function satisfying $\mathcal L_K h = 0$. The identities above immediately imply
\begin{align} 
V^\mu &  = \f{( \iota_K V)}{\zzt^2} K^\mu + \im \widetilde P^{[\mu} P^{\nu]} \nabla_\nu \log \zzt = \f\im \zzt H K^\mu - \f1{2 \zzt} \lc^{\mu \nu \rho}\nabla_\nu K_\rho  ~ ,
\end{align}
namely
\begin{align}\label{eq: Vbf}
V^\mu &  = \f{1}{\zzt^2} \p{  \im \zzt H  - \f1{2 \zzt} K_\lambda \lc^{\lambda \nu \rho}\nabla_\nu K_\rho } K^\mu + \im \widetilde P^{[\mu} P^{\nu]} \nabla_\nu \log \zzt  ~ .
\end{align}
Moreover, by using (\ref{eq: ksbd}), we can write down $A_\mu$ in terms of the derivatives of $K, P, \widetilde P$:
\begin{align}\label{eq: Abf}
A_\mu & = + \f\im4 \widetilde P^\nu \nabla_\mu P_\nu + \f{\im H}{2 \zzt}K_\mu + V_\mu  \nn \\
& = - \f\im4  P^\nu \nabla_\mu \widetilde P_\nu + \f{\im H}{2 \zzt}K_\mu + V_\mu    \nn\\
& = + \f\im8 \p{ \widetilde P^\nu \nabla_\mu P_\nu -  P^\nu \nabla_\mu \widetilde P_\nu } + \f{\im H}{2 \zzt}K_\mu + V_\mu     ~  .
\end{align}
In particular, the 1-form $A^C_\mu  = A_\mu - \f32 V_\mu$  appearing in the conformal Killing spinor equation is independent of the background scalar field $H$, indeed
\begin{align}\label{eq: Acbf}
A^C_\mu & = A_\mu - \f32 V_\mu  \nn\\
 & = \f\im4 \widetilde P^\nu \nabla_\mu P_\nu + \f{\im H}{2 \zzt}K_\mu - \f12 V_\mu      \nn\\
 & = \f\im4 \widetilde P^\nu \nabla_\mu P_\nu  +  \f{1}{2\zzt^2} \p{    \f1{2 \zzt} K_\lambda \lc^{\lambda \nu \rho}\nabla_\nu K_\rho } K^\mu + \f\im2  P^{[\mu} \widetilde P^{\nu]} \nabla_\nu \log \zzt     ~  .
\end{align}

The background scalar field $H$ appears in the cohomological complex and, in turn, in the partition function of the theory only through the combination 
\begin{align}
 h_R = \iota_K V - \im \zzt H    =  - \f1{2 \zzt} K_\mu \lc^{\mu\nu\rho} \nabla_\nu K_\rho ~ , 
\end{align}
which is completely fixed by the Killing spinors $\z, \zt$ via the 1-form $K$ and the scalar $\zzt$.  As an example, the covariant Lie derivative $L_K$ can be written as follows:
\begin{align}
L_K &   =   \Lc_K  - \im q_R\p{ \iota_K A - \f12 \iota_K V } - \im q_Z ( \iota_K \Cc )    + \zzt(  q_Z - q_R H  )  \nn\\  
& = \Lc_K  - \im q_R\p{ \iota_K A^C + h_R } - \im q_Z ( \iota_K \Cc + \im \zzt)   ~ ,
\end{align}
with $h_R$ representing the deviation from the background 1-form $A^C$ appearing in the CKSE. The field $h_R$ is invariant under the shift symmetry reported in (\ref{eq: AHVshiftsym}). The linear combination
\begin{align}\label{eq: rsymfug}
\Phi_R = \iota_K A^C + h_R = \iota_K (A^C + V) - \im \zzt H    ~ ,
\end{align}
   is analogous to the field $\Phi_G$, built out of vector multiplet fields $\mathcal A_\mu$ and $\sigma$, as it is apparent by performing the replacements  $(A^C + V )  \to \mathcal A $ and $H \to \sigma$. The quantity $\Phi_R$ can be seen as a fugacity related to the $R$-symmetry and appears e.g. in the Lie derivatives along $K$ of the Killing spinors:
   \begin{align}
   & \mathcal L_K \z  = + \im\, \Phi_R \, \z ~ , & \mathcal L_K \zt  = - \im \, \Phi_R \, \zt ~ .
   \end{align}
   
   Explicitly, we parametrize the Killing vector on the three-dimensional orbifold as
   \begin{align}\label{eq: masterkillingvector}
     K = K^\mu \partial_\mu  = \knorm \p{  \amf  \partial_{\varphi_1} +  \partial_{\varphi_2}      } =    \knorm \p{  \amf  \partial_{\sigmavar} +  \partial_{\psi}      }  ~ , 
   \end{align}
   with $\knorm$   being a normalization constant and $\amf$ a fugacity for the angular momentum, while $ \varphi_1 = \sigmavar$ and  $ \varphi_2 = \psi $    are  angular coordinates with period $2\pi$. In these coordinates, the Killing vector equation $\nabla_{(\mu} K_{\nu)} = 0$ reads
    \begin{align}
     \p{  \amf  \partial_{\sigmavar} +  \partial_{\psi}      } g_{\mu\nu} \, = \, 0 \, ,
      \end{align}
   which, if the metric is real, implies that its components are independent of the coordinates $\sigmavar$, $\psi$. More generally, we require that the metric components are periodic functions of $\sigmavar$, $\psi$. 
   Expanding in Fourier modes in one of the two coordinates, we have
        \begin{align}
   g_{\mu\nu} (\sigmavar , \psi)\, = \, \sum_n e^{\im n\psi} g^{(n)}_{\mu\nu} (\sigmavar) \, ,
      \end{align}
     so that
            \begin{align}
   \amf  \partial_{\sigmavar}  g^{(n)}_{\mu\nu} (\sigmavar)+  \im n  g^{(n)}_{\mu\nu} (\sigmavar)\, =\, 0\qquad \Rightarrow \qquad g^{(n)}_{\mu\nu} (\sigmavar) = e^{-\im \frac{n}{\amf} \sigmavar} \tilde g^{(n)}_{\mu\nu} \, . 
      \end{align} 
   Thus, $g_{\mu\nu} (\sigmavar , \psi)$ cannot be periodic  in $\sigmavar$ and the only admissible solution is that the metric does not depend on $\sigmavar$ and $\psi$.
   Denoting by $x$ the third coordinate, we then have that the metric takes the form 
    \begin{align}
      & \dd s^2 = f^2 \dd x^2 + \subm_{ij} (\dd \varphi_i +  c_i \dd x)(\dd\varphi_j + c_j \dd x )   ~ , & i,j=1, 2   ~ , 
   \end{align}
where the metric functions $f=f(x)$, $c_i(x)$, and $\subm_{ij} = \subm_{ij}(x)$  are allowed to take complex values and are independent of $\varphi_{1,2}$. After reabsorbing $c_i(x)$ by a (complex) diffeomorphism, we arrive at the form    
   \begin{align}
      & \dd s^2 = f^2 \dd x^2 + \subm_{ij} \dd \varphi_i \dd\varphi_j  ~ . 
   \end{align} 
   Introducing the dreibein $e^m$ 
   \begin{align}
    e^1 = - f  \dd x ~ , \qquad e^2 = \sqrt{\frac{\det \subm}{\subm_{22}}} \dd \sigmavar  ~ , \qquad e^3 =   \sqrt{\subm_{22}} \p{ \frac{\subm_{12}}{\subm_{22}} \dd\sigmavar +  \dd\psi } ~ ,
   \end{align}  
   the three-dimensional orbifold can be viewed as a circle fibration over a two-dimensional orbifold parametrized by $\p{x,\sigmavar}$ with line element $ \dd s^2_{(2)}$, namely
   \begin{align}
      & \dd s^2 = \dd s^2_{(2)} + \subm_{22}\p{ \acf + \dd \psi }^2 ~ ,   &  \dd s^2_{(2)} = f^2 \dd x^2 + \frac{\det \subm}{\subm_{22}} \dd \varphi^2  ~ , \qquad \acf = \frac{\subm_{12}}{\subm_{22}} \dd\sigmavar  ~ .
   \end{align}
   In this setting, the orthogonality relations (\ref{eq: kpptorthogonality}) and the algebra (\ref{eq: kpptalgebra}) imply  
   \begin{align}
   \zzt^2  & = \knorm^2 \p{   \amf^2 c_{11}   + 2 \amf c_{12}  + c_{22} } ~ , \nn\\
       K^\flat   & = \knorm\comm{ \p{\amf \subm_{11} + \subm_{12}}\dd\sigmavar + \p{\amf \subm_{12} + \subm_{22}}\dd\psi }   ~ , \nn\\
       P^\flat   & = \ee^{2 \im \theta}\comm{ \im f \dd x - \knorm v^{-1} \sqrt{\det \subm}\p{  \dd \sigmavar - \amf \dd \psi}  }  ~ , \nn\\
       \widetilde P^\flat   & =  \ee^{-2 \im \theta}\comm{ \im f \dd x + \knorm v^{-1} \sqrt{\det \subm}\p{  \dd \sigmavar - \amf \dd \psi}  }   ~ , 
   \end{align}
    where we parametrize the phase factor $\theta=\theta(x , \sigmavar, \psi)$ as\footnote{Notice that $\alpha_{2,3}|_{\rm here}=\alpha_{1,2}|_{\rm there}$ with respect to \cite{Inglese:2023wky}.}
    \begin{align}
    		\theta = \frac{1}{2}\comm{ \alpha_1\p{x} + \alpha_2 \sigmavar + \alpha_3 \psi  }  ~ .
    \end{align}
    The dual vectors  $K, P $ and $\widetilde P$ can be found by contraction with the metric tensor, for instance
   \begin{align}
       P  & = \ee^{2 \im \theta}\acomm{ \im f^{-1} \partial_x -    \frac{\knorm}{v \sqrt{\det \subm}}\comm{ \p{\amf \subm_{12} + \subm_{22}}\partial_\sigmavar   - \p{\amf \subm_{11} + \subm_{12}}\partial_\psi } }      ~ . 
   \end{align}
   In turn, the definitions (\ref{eq: ksbilinears}) determine the form of the  Killing spinors $\zeta, \widetilde \zeta$  in a purely algebraic way, with no need to solve any differential equations. Indeed, the spinors compactly written as
   \begin{align}\label{eq: masterkillingspinors}
   	 & \zeta_\alpha = \ee^{ \im \theta} \begin{pmatrix} u_1 \\ - u_2 \end{pmatrix}_\alpha ~ ,  \qquad  \widetilde \zeta_\alpha  = -  \ee^{ - \im \theta} \begin{pmatrix} u_2 \\ u_1 \end{pmatrix}_\alpha ~ ,  \nn\\
   	u_{1,2} & =\sqrt{ \frac{v}{2} \mp  \frac{ \knorm \, \amf }{2}    {e^2}_\sigmavar } =  \sqrt{ \frac{v}{2} \mp   \frac{\knorm \,  \amf}{2}   \sqrt{\frac{\det \subm }{\subm_{22} }} }  ~ ,  
   \end{align}
   satisfy the Killing spinor equations (\ref{eq: kses}) if the background fields given by  (\ref{eq: Vbf}), (\ref{eq: Abf}) and (\ref{eq: Acbf}) are turned on, namely
   \begin{align}\label{eq: masterbackgroundfields}
   	   	 A^C  & = \dd \theta - \frac{v^3}{4 \knorm f \sqrt{\det \subm} } \comm{ \p{\frac{\subm_{11}}{v^2} }'\dd \sigmavar -   \amf^{-1}  \p{\frac{\subm_{22}}{v^2} }'\dd \psi  } ~ , \nn\\
   	 V  & = \dd \theta - A^C -  \frac{ \knorm}{2 v f}  \p{ \sqrt{\det \subm} }' \p{\dd\sigmavar - \amf   \dd\psi }     + \frac{\im H}{v}  K^\flat  \nn\\
   	 & = \frac{1}{4 \knorm f \sqrt{\det \subm} }\acomm{ \comm{ v^3  \p{\frac{\subm_{11}}{v^2}}' -\frac{ \knorm^2}{v}  \p{ \det \subm}' }\dd\sigmavar -  \comm{  \frac{v^3}{\amf}  \p{\frac{\subm_{22}}{v^2}}' - \amf  \frac{ \knorm^2}{v}  \p{ \det \subm}' }\dd\psi     } + \frac{\im H}{v}  K^\flat ~ , \nn\\
   	 A  & = A^C + \frac{3}{2} V = - \frac12 A^C + \frac{3}{2} \comm{ \dd \theta   -  \frac{ \knorm}{2 v f}  \p{ \sqrt{\det \subm} }' \p{\dd\sigmavar - \amf   \dd\psi }     + \frac{\im H}{v}  K^\flat  }  ~ , 
   \end{align}
   where     $X'$ stands for $\partial_x X$.  Despite the appearance  of  $\amf^{-1}$ factors, by inspection the background fields above are finite in the limit $\amf\to0$.
  

\subsection{\texorpdfstring{$\spindle \times S^1$}{Sigma x S1} with twist and anti-twist}\label{eq: spindles1background}

On $\spindle \times S^1 = \mathbb{WCP}^1_{\comm{\nN, \nS }} \times S^1$ we adopt   the  line element 
\begin{align}\label{eq: lineelementspindles1}
\dd s^2 = f^2 \dd x^2 + \p{1-x^2}\p{\dd \varphi - \Omega \dd \psi}^2 + \beta^2 \dd \psi^2 ~ , 
\end{align}
with   $x\in \comm{-1,+1}$ and $\varphi \in [0, 2 \pi)$ being coordinates on  $\spindle$ while $\psi\in [0, 2 \pi)$ is the compactified  Euclidean time parametrizing $S^1$. The function $f=f(x)$ satisfies 
\begin{align}
\lim_{x\to \pm1} f \to \frac{n_\pm }{\sqrt{2\p{1 \mp x}}} ~ .
\end{align}
We require that $f$ is regular on the complementary domain $\p{-1,+1}$, but we leave arbitrary  the specific profile of $f$. The positive integers $n_\pm$ are coprime and encode the  conical singularities   $\mathbb C/\mathbb Z_{n_\pm}$ at the poles of the spindle. The dimensionless complex parameter $\Omega$ induces a refinement   of the partition function by a fugacity for the angular momentum on the spindle, as it happens on the round sphere \cite{Benini:2015noa}. The parameter $\beta$ is the ratio between the radius   of $S^1$  and the radius of the equatorial circle of $\spindle$. The integral of the Ricci scalar of the spindle is
\begin{align}
\frac{1}{4 \pi}\int_\spindle  \dd x \dd \varphi \sqrt{g_\spindle}R_\spindle = \frac{1}{\nN} + \frac{1}{\nS}  ~ ,
\end{align}
correctly  reproducing the orbifold Euler characteristic $\chi_{\rm orb}$ of $\spindle = \mathbb{WCP}^1_{\comm{\nN, \nS}}$.

We use an  orthonormal frame explicitly realizing $\spindle \times S^1$ as a $U(1)$ fibration over the spindle base:
\begin{align}
e^1 & = - f \dd x ~ , \nn\\
 e^2 & = \beta \sqrt{\frac{1-x^2}{\beta^2 + \Omega^2\p{1-x^2}}}\dd \varphi ~ ,  \nn\\
 e^3 & = \sqrt{ \beta^2 + \Omega^2\p{1-x^2}}\p{   \acf +  \dd \psi }   ~ , \qquad \acf = \frac{ -  \Omega  \p{1-x^2}}{\beta^2 + \Omega^2\p{1-x^2}}\dd \varphi ~ ,
\end{align}
where the flux of $\acf$ through $\spindle$ is vanishing, as it behooves a trivial circle fibration over a spindle. Near the north pole of $\spindle$ at $x=+1$ we have
\begin{align}
	& e^1_{\UN} = - \frac{\nN \dd x}{\sqrt{2\p{1-x}}} = \dd \rhoN ~ , \qquad e^2_{\UN}  =   \sqrt{2\p{1-x}} \dd \sigmavar = \frac{\rhoN}{\nN} \dd \sigmavar ~ , & \rhoN = \nN \sqrt{2\p{1-x}}  ~ .
\end{align}
Hence, in an open neighbourhood $\UN$  of $x=+1$ the spindle $\spindle$ can be parametrized by the complex coordinate $\zN $ satisfying
\begin{align}\label{eq: ccsnp}
	& \zN = \rhoN e^{\im \sigmavar/\nN} ~ , \qquad  \zN  \sim \wN \zN ~ , &  \wN = e^{2 \pi \im /\nN}  ~ ,
\end{align}
where the identification involving the root of unity $\wN$ makes manifest that the topology of  $\UN$ is that of $\mathbb C/\mathbb Z_{\nN}$. On the other hand, near the south pole of the spindle at $x=-1$ we have
\begin{align}
	& e^1_{\US} = - \frac{\nS \dd x}{\sqrt{2\p{1+x}}} = - \dd \rhoN ~ , \qquad e^2_{\US} =   \sqrt{2\p{1+x}} \dd \sigmavar = \frac{\rhoS}{\nS} \dd \sigmavar ~ , & \rhoS = \nS \sqrt{2\p{1+x}}  ~ .
\end{align}
Thus, a proper coordinate in an open neighbourhood $\US$  of $x=-1$ is 
\begin{align}\label{eq: ccssp}
	& \zS = \rhoS e^{-\im \sigmavar/\nS} ~ , \qquad  \zS  \sim w_-^{-1} \zS ~ , &  \wS = e^{2 \pi \im /\nS}  ~ ,
\end{align}
meaning that  $\US \cong \mathbb C/\mathbb Z_{\nS}$. The minus sign in the exponential of $\zS = \rhoS e^{-\im \sigmavar/\nS}$ takes into account the reversed  orientation of $\US$ with respect to that of $\UN$. A necessary condition for the orbifold $\spindle\times S^1$ to be supersymmetric is that the magnetic flux of the $U(1)$ $R$-symmetry   field $A^C$ through the spindle takes the form
\begin{align}
& \frac{1}{2\pi}\int_\spindle \dd A^C = \frac{1}{2}\p{\frac{1}{\nS} + \frac{\st}{\nN}} = \frac{\chi_\st}{2}    ~ , & \st =\pm 1  ~ .
\end{align}
 The values of the sign $\st=\pm1$ represent two inequivalent ways to preserve  supersymmetry on $\spindle \times S^1$. In the case of $\st=+1$, the magnetic flux of the $R$-symmetry background field $A^C$ equals half of the orbifold Euler characteristic of the spindle  $\chi_{\rm orb}=\nS^{-1} + \nN^{-1}$. This configuration is referred to as \emph{topological twist} as it is the orbifold generalization of the analogous setup found in spheres \cite{Benini:2015noa, Closset:2015rna} and Riemann surfaces \cite{Benini:2016hjo}. On the other hand, when $\st=-1$, the magnetic flux of the  $R$-symmetry background field equals half of $\chi_- = \p{ \nS^{-1} - \nN^{-1}}$. This specific configuration is dubbed \emph{anti-twist} to  highlight  its contrast with the topological twist, primarily due to the presence of a relative minus sign between $\nN^{-1}$ and $\nS^{-1}$, resulting in $\chi_-$ being radically  distinct from $ \chi_{\rm orb} = \chi_+$. In the special case where $\nN=\nS=1$  the spindle become a smooth sphere, and the anti-twisted  $\spindle\times S^1$ configuration reduces to  the generalized superconformal index background studied in \cite{Imamura:2011su,Kapustin:2011jm}, characterized by the absence of magnetic flux for the  $R$-symmetry connection. Eventually, after    shrinking to zero the radius of $S^1$ in $\spindle \times S^1$ and keeping $\nN=\nS=1$,  the anti-twisted  spindle becomes the    supersymmetric sphere with no  $R$-symmetry twist explored in \cite{Benini:2012ui}.
 

The Killing spinors corresponding to the  twisted $\spindle\times S^1$ geometry, occurring if  $\st = +1$, are
\begin{align} 
    & \zeta_\alpha   =  e^{ \im \theta }  \begin{pmatrix}   u_1 \\  - u_2  \end{pmatrix}_\alpha  ~ ,  \qquad  \widetilde \zeta_\alpha    = -   e^{- \im \theta}  \begin{pmatrix}   u_2 \\  u_1 \end{pmatrix}_\alpha  ~ , \nn\\
    & u_{1,2} = \sqrt{  - \frac{\knorm \beta }{2}\p{ 1  \pm     \Omega \sqrt{ \frac{1-x^2}{\beta^2 + \Omega^2 \p{1-x^2}} } } }  ~ , 
\end{align}
which satisfy the conformal Killing spinor equations provided that the background fields $A^C, V, A$ take  the form
\begin{align}
	A^C & = \dd\theta -  \frac{x}{2 f \sqrt{1-x^2}}  \p{ \dd \sigmavar - \Omega \dd \psi}    ~ , \nn\\
	V &= -   \im \beta H \dd \psi ~ , \nn\\
	A &= A^C + \frac{3}{2} V ~ .
\end{align}
Unlike the conventional topological twist, the  $R$-symmetry field $A^C$ above is unrelated to the spin connection. Nevertheless, the flux of $A^C$ through the spindle $\spindle$ precisely equals $\chi_{\rm orb}/2$, as desired. Moreover, we observe that the  fluxes of $A^C$ and $A$ coincide with each other. The 1-form fields $\p{A^C, V, A}$ and $\zeta, \widetilde \zeta$ are regular in the northern patch $\UN$ and the southern patch $ \US$ of the spindle if $\theta$ takes the form
    \begin{align}
	 & \alpha_2|_{\UN}  = \frac1\nN ~ , \qquad  \alpha_2|_{\US} = - \frac1\nS ~ , & \alpha_3\ & \in \mathbb Z ~ ,  \nn\\
	& \theta  = \frac12\comm{ \alpha_1(x) + \alpha_2 \sigmavar + \alpha_3 \psi }  ~ ,
\end{align}
where $\alpha_2|_{\UN, \US}$ is the value of $\alpha_2$ in the patch covering $x=\pm1$, respectively.

   Conversely, the Killing spinors associated with the anti-twisted $\spindle\times S^1$ geometry, which occurs when $\st = -1$, exhibit a formal similarity to the Killing spinors corresponding to  the twisted background. However, they differ in terms of their components $u_{1,2}$:
\begin{align} 
    &  u_{1,2}  =\sqrt{  \frac{\knorm \beta  }{2} \comm{ x \mp   \p{\Omega - \im \beta }   \sqrt{\frac{1-x^2}{\beta^2 + \Omega^2\p{1-x^2}}} \,  } }   ~ , 
\end{align}
fulfilling  the conformal Killing spinor equations if  the background fields $A^C, V, A$ are 
\begin{align}
	A^C &   = \dd\theta +   \frac{1}{2 f \sqrt{1-x^2}} \comm{   \dd \sigmavar - \p{\Omega + \im \beta} \dd \psi }   ~ , \nn\\
	V &=  \p{x^{-1} + x }H \dd \sigmavar + \im \acomm{\frac{\beta}{f\sqrt{1-x^2}} + x^{-1}\comm{\beta +  \im\, \Omega\p{1-x^2}} H } \dd\psi  ~ , \nn\\
	A &= A^C + \frac{3}{2} V ~ . 
\end{align} 
Consistently, the flux of the  $R$-symmetry connection $A^C$ through the spindle is   $\p{\chi_-/2}$. To ensure that the 1-form $V$ remains regular over the entire $\spindle\times S^1$, we require that $H\sim \mathcal O(x)$ in the vicinity of the spindle's equator at $x=0$. The background fields $\p{A^C, V, A}$ and $\zeta, \widetilde \zeta$ are non-singular in the patches $\UN$ and  $ \US$ of $\spindle$ if $\theta$ is
    \begin{align}
	 & \alpha_2|_{\UN}  = - \frac1\nN ~ , \qquad  \alpha_2|_{\US} = - \frac1\nS ~ , & \alpha_3 & \in \mathbb Z ~ ,  \nn\\
	& \theta  = \frac12\comm{ \alpha_1(x) + \alpha_2 \sigmavar + \alpha_3 \psi }  ~ ,
\end{align}
where again $\alpha_2|_{\UN, \US}$ are the values of $\alpha_2$ in the patches that respectively include the poles $x=\pm1$. 


\subsection{Conformal boundary of the accelerating black hole}\label{eq: abhbackground}

In \cite{Cassani:2021dwa} the   thermodynamic features of  asymptotically ${\rm AdS}_4$  black holes carrying electromagnetic charge, angular momentum and acceleration along the axis of rotation were explored in the context of  Einstein-Maxwell theory. In particular, it was shown that the conformal boundary of such a class of  black holes is   the  three-dimensional orbifold $\mathbb R \times \spindle$, where $\mathbb R$ encodes time while the spindle $\spindle$ characterizes  the spatial part of the boundary.  Specifically, the conformal boundary of this accelerating black holes is  described  by the Lorentzian line element
\begin{align}
	\dd \widehat s^2  = - \p{\widehat e^0}^2 + \p{\widehat e^1}^2 + \p{\widehat e^2}^2 ~ ,
\end{align}
with 
\begin{align}
	\widehat e^0  = \sqrt{\widetilde P}\p{\frac{\dd t}{\kappa} - a f \dd \phi } ~ , \qquad \widehat e^1 = - \sqrt F \, \dd x ~ , \qquad  \widehat e^2 = \sqrt G \, \dd \phi ~ ,
\end{align}
being the 1-forms $\widehat e^{1,2,3}$  comprising the corresponding    Lorentzian frame. The functions appearing in the line element $\dd \widehat s^2$ read
\begin{align}
	P & = 1 - 2 \alpha m x + \comm{\alpha^2 \p{a^2 + e^2 + g^2} - a^2}x^2 ~ , \nn\\
	\widetilde P & = 1 - \alpha^2\p{1-x^2}P ~ , \nn\\
	f & = \widetilde P^{-1} \p{1 - \alpha^2 P}\p{1-x^2} ~ , \nn\\
	F & = \frac{\p{1+a^2\alpha^2x^4}^2}{P\p{\widetilde P + a^2 \alpha^2 x^4}\p{1-x^2}} ~ , \nn\\
	G & = \widetilde P^{-1}  P\p{\widetilde P + a^2 \alpha^2 x^4}\p{1-x^2} ~ , 
\end{align}
while the   parameters $\p{a, e, g, m, \alpha}$  can be expressed in terms of  fewer constants $\p{b, s, \mu}$  according to the following relations \cite{Cassani:2021dwa}:
\begin{align}
	\mu & = \frac{\nS + \nN}{\nS - \nN} ~ ,  \qquad g  = \alpha m ~ , \qquad e  =  \frac{b s }{\alpha^2 c} ~ , \qquad g =  \frac{ s }{\alpha^2 c} ~ , \qquad a =  \frac{ s }{\alpha } ~ , \nn\\
	c & =  \frac{2\p{1+b^2}s \mu}{1 - 2 b s - s^2} ~ ,  \qquad \kappa  =  \frac{ \p{b+s}\p{1- b s}}{\alpha\p{ 1+b^2}\p{1+s^2}} ~ ,  \qquad \alpha^2  =  \frac{ 4 \mu^2\p{1 - b s}^2 - \p{1 - 2 b s - s^2}^2}{4 \mu^2 \p{ 1+b^2}\p{1+s^2}} ~ .  
\end{align}
The Euclidean version of the Lorentzian conformal boundary above is obtained upon performing the Wick rotation $t = - \im \tau $ and moving to the  dreibein $e^{1,2,3}$ given by
\begin{align}
	   e^1 = \widehat e^1 = - \sqrt F \, \dd x ~ , \qquad    e^2 = \widehat e^2 = \sqrt G \, \dd \phi ~ , \qquad e^3  =  \im \, \widehat e^0 = \sqrt{\widetilde P}\p{\frac{\dd \tau}{\kappa} - \im a f \dd \phi } ~ , 
\end{align}
which constitute  an orthonormal  frame yielding in turn the Euclidean line element
\begin{align}
	\dd  s^2  =    \p{ e^1}^2 + \p{  e^2}^2 + \p{  e^3}^2 ~ .
\end{align}
This background is endowed with the Killing vector 
\begin{align}
	 K = \im  \kappa\p{1 + a^2 \alpha^2}\partial_\tau + \alpha\p{\frac{e}{g} + a \alpha}\partial_\phi  ~ ,
\end{align}
which can be written in the notation of (\ref{eq: masterkillingvector}) if
\begin{align}
	 & \knorm = \im  \kappa\p{1 + a^2 \alpha^2}  ~ , & \amf = \frac{   \alpha\p{\frac{e}{g} + a \alpha}}{\im  \kappa\p{1 + a^2 \alpha^2}} ~ .
\end{align}
The  norm of $K$ is
\begin{align}
	  \zzt   = e^{\im \pi/2}\comm{  \frac{1 - 2 b s - s^2}{2 \mu}\p{1-x^2} + \p{1+s^2}x}    ~ ,
\end{align}
and plugging $v$ and ${e^2}_\phi$ into our general formulae for the Killing spinors written in (\ref{eq: masterkillingspinors}) indeed  yields objects of the form
\begin{align} 
    & \zeta_\alpha   =  e^{ \im \theta }  \begin{pmatrix}   u_1 \\  - u_2  \end{pmatrix}_\alpha  ~ ,  \qquad  \widetilde \zeta_\alpha    = -   e^{- \im \theta}  \begin{pmatrix}   u_2 \\  u_1 \end{pmatrix}_\alpha  ~ ,  
\end{align}
with arbitrary $\theta$ and 
\begin{align} 
    &  u_{1,2}  = e^{\im \pi/4} \sqrt{  \frac{1   }{2} \comm{ \frac{1 - 2 b s - s^2}{2 \mu}\p{1-x^2} + \p{1+s^2}x  \pm \im \alpha\p{b + s } \sqrt{G} } }   ~ . 
\end{align}
Especially, if we choose 
\begin{align}
	 \alpha_1(x) & =  -  \frac\pi2  ~ , \nn\\
	\alpha_2 & =  -  \frac{ 4 \mu^2 (1 + s^2) (1 - 2 b s - s^2)}{
4 \mu^2 (1 - b s)^2 - (1 - 2 b s - s^2)^2}  ~ , \nn\\
	\alpha_3 & = \frac{\im \p{1+s^2}}{1 - b s } ~ ,  \nn\\
	\theta & = \frac12\comm{ \alpha_1(x) + \alpha_2 \phi + \alpha_3 \tau }  ~ ,  
\end{align}
our   $\zeta_\alpha$ and $\widetilde \zeta_\alpha$ coincide with the boundary Killing spinors found in  \cite{Cassani:2021dwa}. Consistently, inserting the values above into our general formulae for the background fields reported in  (\ref{eq: masterbackgroundfields})  gives  
\begin{align}
   	   	 A^C  & =  -\frac{x\comm{g + a^2 g  \alpha^2 x^2 - a e \alpha \p{1-x^2} }}{ 1 + a^2 \alpha^2 x^4 } \dd \phi +  \frac{\im \, x \alpha \p{ e - a g \alpha x^2 }}{\kappa \p{1 + a^2 \alpha^2 x^4} } \dd \tau   ~ , \nn\\
   	 V  & = \dd \theta - A^C -  \frac{ \knorm}{2 v f}  \p{ \frac{ 1 + a^2 \alpha^2 x^4}{\kappa  \, f  }}' \p{\dd\sigmavar - \amf   \dd\psi }     + \frac{\im H}{v}  K^\flat ~ ,  \nn\\
   	 A  & = A^C + \frac{3}{2} V  ~ , 
   \end{align}
   where, in particular, $A^C$ matches the $U(1)_R$  gauge field  appearing in the (boundary) conformal Killing spinor equations  solved in \cite{Cassani:2021dwa}.


\subsection{The squashed lens space}\label{eq: sqlensbackground}

Given a squashed three-sphere $S^3_{b_1, b_2}$ with deformation parameters $\p{b_1,b_2}$ and a positive integer $n$,  the corresponding squashed lens space $L_{b_1,b_2}\p{n,1}$ can be obtained from the free quotient  $S^3_{b_1, b_2}/\mathbb Z_n$,  where the orbifold projection acts on the Hopf fiber of  $S^3_{b_1, b_2}$. Topologically, $L_{b_1,b_2}\p{n,1}$ is an $\mathcal O(-n)$ circle fibration with base a two-dimensional ellipsoid. The non-trivial features of  gauge theories on lens spaces were first investigated in \cite{Gang:2009qdj,Benini:2011nc}, the corresponding    large-$N$ limit was explored in  \cite{Alday:2012au} while their dualities were  studied in \cite{Imamura:2012rq}. Furthermore, lens spaces are  important  in the context of holomorphic blocks and factorization of supersymmetric partition functions, as shown in \cite{Imamura:2013qxa,Nieri:2015yia}.

On $L_{b_1,b_2}\p{n,1}$ we use the line element
\begin{align}\label{eq: lineelementsls}
    \dd s^2 = f^2 \dd x^2 +  b_1^2 \sin^2 x \p{ \dd \sigmavar + \frac{\dd \psi}{n} }^2 + \frac{b_2^2}{n^2} \cos^2 x \, \dd \psi^2  ~ ,
\end{align}
with $\sigmavar, \psi \in \comm{0, 2 \pi}$  being periodic coordinates and   $x\in\comm{0, \pi/2}$. The function
\begin{align}\label{eq: squashingfunction}
    f = - \sqrt{ b_1^2 \cos^2 x + b_2^2 \sin^2 x  }  ~ ,
\end{align}
produces a smooth squashing of the manifold tuned  by the complex numbers $b_{1,2}$. The presence of the  positive integer $n$ encodes the $\mathbb Z_n$ orbifold acting on the $S^1$ parametrized by the coordinate  $\psi$. If $n=1$,   the squashed lens  space reduces to the squashed three-sphere explored in \cite{Hama:2011ea}.

We employ the following orthonormal frame:
\begin{align}
    e^1 & = - f \, \dd x   ~ , \nn\\
    e^2 & =   \frac{b_1 b_2 \sin x \cos x}{\sqrt{b_1^2 \sin^2 x + b_2^2 \cos^2 x  }}  \dd\sigmavar  ~ , \nn\\
    e^3 & = n^{-1} \sqrt{b_1^2 \sin^2 x + b_2^2 \cos^2 x  } \p{ \acf +  \dd \psi }       ~ , \qquad \acf = \frac{ n \, b_1^2 \sin^2 x}{ b_1^2 \sin^2 x + b_2^2 \cos^2 x  } \dd\sigmavar  ~ ,
\end{align}
which realize the squashed lens space as a non-trivial circle fibration over the base $\spindle$, where the latter is an ellipsoid with coordinates  $\p{x,\sigmavar}$. Indeed,  the Euler characteristic of $\spindle$ coincides with that of a two-sphere, 
\begin{align} 
    \frac{1}{4 \pi} \int_\spindle \dd x \dd   \varphi \sqrt{g_\spindle} R_\spindle = 2 = \chi_{S^2} ~ ,
\end{align}
while  the  1-form   $\acf$  is  a representative of an  $\mathcal O(-n)$ bundle over $\spindle$:
\begin{align} 
    \frac{1}{2 \pi} \int_\spindle \dd \acf =  n ~ .
\end{align}
This geometry admits a Killing vector\footnote{Here  we have fixed the sign choice $\samf = -1$. The general case with $\samf = \pm1$ can be recovered as a special case of Section \ref{eq: blensbackground} by setting $n_\pm = 1$.} of the form (\ref{eq: masterkillingvector}):
\begin{align}
	 K = \knorm\p{ -  \frac{b_1 + b_2}{n \,  b_1}\partial_\sigmavar + \partial_\psi } ~ ,
\end{align}
 with norm
\begin{align}
	  \zzt   = \knorm b_2/n    ~ .
\end{align}
In this setting,    the Killing spinors (\ref{eq: masterkillingspinors}) have components
\begin{align} 
    u_{1,2} & =\sqrt{ \frac{\knorm b_2}{2 n}\p{   1 \pm         \frac{ \p{b_1 + b_2}\sin x \cos x }{\sqrt{b_1^2 \sin^2 x + b_2^2 \cos^2 x  }}  } }   ~ ,
\end{align}
where the phase $\theta$ can be fixed by requiring that the background fields (\ref{eq: masterbackgroundfields}), namely  
\begin{align}
   	   	 A^C  & = \dd\theta - \frac{b_1}{2 f} \dd \sigmavar - \frac{b_1 - b_2}{2 n f}\dd\psi     ~ , \nn\\
   	 V  & = \frac{ b_1\p{1 - \im f H}\sin^2x }{  f}\dd\sigmavar + \frac{  \comm{ b_1 - b_2 - \p{b_1+b_2}\cos\p{2 x} }\p{ 1 -  \im  f H} }{2 n f}\dd \psi   ~ ,  \nn\\
   	 A  & = A^C + \frac{3}{2} V  ~ , 
   \end{align}
    are non-singular  on the whole squashed lens space. By inspection, regularity of both $\p{A^C, V, A}$ and $\zeta, \widetilde \zeta$ is achieved if  
    \begin{align}
	\alpha_2 & = - 1     ~ , \nn\\
	\alpha_3 & =  0 ~ ,  \nn\\
	\theta & = \frac12\comm{ \alpha_1(x) + \alpha_2 \sigmavar + \alpha_3 \psi }  ~ .
\end{align}


\subsection{The branched-squashed three-sphere}

 Supersymmetric gauge theories on branched spheres, namely branched coverings of  $S^3$,  were explored in \cite{Nishioka:2013haa} because of  their     connection  to  R\'enyi entropy as well as entanglement entropy. Analogous results for free   theories on branched spheres were previously obtained in \cite{Klebanov:2011uf}. Here we consider supersymmetric gauge theories on a more general orbifold, namely a branched squashed three-sphere $S^3_{b_1, b_2, \nbN , \nbS }$, whose line element reads 
\begin{align}\label{eq: lineelementbss}
    \dd s^2 = f^2 \dd x^2 + \frac{b_1^2}{\nbN^2} \sin^2 x \, \dd \varphi_1^2 + \frac{b_2^2}{\nbS^2} \cos^2 x \, \dd \varphi_2^2  ~ ,
\end{align}
where $\varphi_1, \varphi_2 \in \comm{0, 2 \pi}$ parametrize the two circles   in $S^3_{b_1, b_2, \nbN , \nbS }$, while  $x\in\comm{0, \pi/2}$. The  (a priori complex) numbers $b_1,b_2$ are the squashing parameters also appearing in the function
\begin{align}
    f = - \sqrt{ b_1^2 \cos^2 x + b_2^2 \sin^2 x  }  ~ .
\end{align}
The positive integers $\p{\nbN, \nbS}$ parametrize the orbifold singularities at $x=0$ and $x = \pi/2$. If $\nbN = \nbS=1$, we obtain   the squashed sphere studied in \cite{Imamura:2011wg}. Instead,  the metric of the branched sphere considered in \cite{Nishioka:2013haa} is obtained from (\ref{eq: lineelementbss}) after choosing $b_1 = b_2 =1$ and identifying $\nbN=q^{-1}_{\rm there}$ as well as $\nbS=p^{-1}_{\rm there}$. Further setting $\nbN=1$ yields the line element of the  orbifold employed  in \cite{Klebanov:2011uf}.

We  work in  the diagonal frame
\begin{align}
    e^1 & = - f \, \dd x   ~ , \nn\\
    e^2 & = b_1 \sin x \, \dd \varphi_1  /\nbN    ~ , \nn\\
    e^3 & =  b_2 \cos x \, \dd \varphi_2  /\nbS   ~ .
\end{align}
The branched, squashed three-sphere is linked to the other orbifold backgrounds explored in this paper as $S^3_{b_1, b_2, \nbN , \nbS }$  is a non-trivial $S^1$ fibration with base a spindle  $\spindle$. Indeed, by applying a  $SL\p{2, \mathbb Z}$ map acting on $\p{\varphi_1, \varphi_2}$  as  
\begin{align}
     & \varphi_1 = t_{11}   \varphi + t_{12}   \psi      ~ ,  \qquad  \varphi_2 =  t_{21}  \varphi + t_{22}   \psi      ~ , & t_{11} t_{22} - t_{12} t_{21} = 1 ~ , 
\end{align}
with $t_{ij} \in \mathbb Z$, the line element (\ref{eq: lineelementbss}) can be rewritten as 
\begin{align}\label{eq: lineelementbsss1fibration}
    \dd s^2 = f^2 \dd x^2 + c_{22} \dd  \varphi^2 + h\p{ \dd \psi + \acf }^2  ~ ,
\end{align}
where  $\psi \in \comm{0, 2\pi} $ parametrizes the  $S^1$ fiber while $x \in \comm{0, \pi/2}$ and $\varphi \in \comm{0, 2\pi}$ are coordinates on the base $\spindle$. The objects appearing in the line element (\ref{eq: lineelementbsss1fibration})  read 
\begin{align} 
     c_{22} & =  \frac{g_{22} g_{33}}{  g_{22} t_{12}^2  + g_{33} t_{22}^2  } ~ , \nn\\
     h & =   g_{22} t_{12}^2 + g_{33} t_{22}^2 ~ , \nn\\
     \acf & =  \frac{ g_{22} t_{11} t_{12} + g_{33} t_{21} t_{22}  }{  g_{22} t_{12}^2  + g_{33} t_{22}^2  } \dd  \varphi ~ , \nn\\
     g_{22} & =  b_1^2 \sin^2 x /\nbN^2  ~ , \nn\\
     g_{33} & =  b_2^2 \cos^2 x /\nbS^2  ~ .
\end{align}
The Euler characteristic of $\spindle$ is
\begin{align} 
    \frac{1}{4 \pi} \int_\spindle \dd x \dd  \varphi \sqrt{g_\spindle} R_\spindle = \frac{1}{\nbS t_{12}} + \frac{1}{\nbN t_{22}} = \chi_{\rm orb} ~ ,
\end{align}
confirming  that the branched squashed three-sphere is a $S^1$ fibration over the base encoded  by the spindle  $\spindle = \mathbb{WCP}^1_{\comm{ \nbN t_{22} , \nbS t_{12} }}$. Especially, the flux of $\acf$ through $\spindle$ is non-vanishing and fractional:
\begin{align} 
    \frac{1}{2 \pi} \int_\spindle \dd \acf = \frac{t_{11}}{  t_{12}} - \frac{t_{21}}{t_{22}} = \frac{\nbN t_{11}}{ \nbN t_{12}} - \frac{\nbS t_{21}}{ \nbS t_{22}}  = \frac{\nbN \nbS }{\p{\nbN t_{22}}\p{\nbS t_{12}}} ~ ,
\end{align}
verifying  that  the fibration is non-trivial and that the  background field   $A^{(1)}$  is  a 1-form representing the orbibundle $\mathcal O(-\nbN \nbS)$   on $\spindle$. The line element (\ref{eq: lineelementbsss1fibration}) in a neighborhood of $x=0$ reads
\begin{align}
    \dd s^2|_{x=0} = b_1^2 \comm{ \dd x^2 + x^2 \p{  \frac{      \dd  \varphi  }{  \nbN t_{22}  } }^2  } +     \p{\frac{b_2 t_{22}}{\nbS}}^2 \p{ \dd \psi +   \frac{     t_{21}  }{      t_{22}  } \dd  \varphi }^2    ~ ,
\end{align}
which can be covered by the smooth manifold $\mathbb C \times \mathbb C^*$ by defining the complex coordinates
\begin{align}
	& z_0^{(1)} = b_1 x \exp\p{\frac{\im \varphi}{\nbN t_{22} }}  ~ , & z_0^{(2)} = \frac{b_2 t_{22}}{\nbS} \exp\comm{ \im \p{ \psi  + \frac{t_{21}}{t_{22}} \varphi } }  ~ ,
\end{align}
where $z_0^{(2)}\in \mathbb C^*$ embeds $S^1$ in $\mathbb C$ as  $|z_0^{(2)} |^2 = | {b_2 t_{22}}/{\nbS}|^2$ is constant. Moreover,  $z_0^{(2)} $ is invariant under  $\psi \sim \psi + 2 \pi $, while the rotational invariance of the base provides the identifications
\begin{align}
	& \varphi \sim \varphi + 2 \pi \implies \p{ z_0^{(1)} ,  z_0^{(2)} } \sim \p{ \wN z_0^{(1)} , \wN^{ \nbN t_{21} }  z_0^{(2)} } ~ ,  & \wN = \exp\p{\frac{2 \pi \im}{\nbN t_{22} }} ~ .  
\end{align}
In particular, going $t_{22}$ times around the point  $x=0$ gives 
\begin{align}
	 \varphi \sim \varphi + 2 \pi \, t_{22} \implies \p{ z_0^{(1)} ,  z_0^{(2)} } \sim \p{ \wN^{t_{22}} z_0^{(1)} ,  z_0^{(2)} } = \p{ e^{2 \pi \im / \nbN } z_0^{(1)} ,  z_0^{(2)} } ~ ,    
\end{align}
indicating  that $x=0$ is the fixed point of a non-free $\mathbb Z_{\nbN}$ action affecting  the base of the fibration but preserving  the fiber. This confirmes that the branched squashed sphere displays  a genuine  conical singularity at $x=0$ encoded by the quotient $\mathbb C/\mathbb Z_{\nbN}$. Analogously, near $x=\pi/2$ we find
\begin{align}
    \dd s^2|_{x=\pi/2} = b_2^2 \comm{ \dd x^2 +  \p{\frac\pi2-x}^2  \p{ \frac{  \dd  \varphi  }{ \nbS     t_{12}} }^2 } + \p{\frac{b_1 t_{12}}{\nbN}}^2\p{ \dd \psi +   \frac{   t_{11}  }{   t_{12}    } \dd  \varphi }^2  ~ ,
\end{align}
which can be lifted to the smooth manifold $\mathbb C\times \mathbb C^*$ after introducing the complex coordinates
\begin{align}
	& z_{\pi/2}^{(1)} = b_2  \p{\frac\pi2-x} \exp\p{ \frac{ -  \im  \varphi  }{ \nbS     t_{12}} }   ~ , & z_{\pi/2}^{(2)} = \frac{b_1 t_{12}}{\nbN} \exp\comm{ \im \p{  \psi +   \frac{   t_{11}  }{   t_{12}    }   \varphi } }  ~ ,
\end{align}
with $z_{\pi/2}^{(2)}\in\mathbb C^*$ embedding $S^1$ into $\mathbb C$. Alike $z_0^{(2)} $, the complex coordinate $z_{\pi/2}^{(2)} $ is not affected by  shifts $\psi \to \psi + 2 \pi $, whereas  
\begin{align}
	& \varphi \sim \varphi + 2 \pi \implies \p{ z_{\pi/2}^{(1)} ,  z_{\pi/2}^{(2)} } \sim \p{ \wS^{-1} z_{\pi/2}^{(1)} , \wS^{ - \nbS t_{11} }  z_{\pi/2}^{(2)} } ~ ,  & \wS = \exp\p{\frac{2 \pi \im}{\nbS t_{12} }}~ .
\end{align}
Winding $t_{12}$ times around the point $x=\pi/2$ yields  
\begin{align}
	 \varphi \sim \varphi + 2 \pi \, t_{12} \implies \p{ z_{\pi/2}^{(1)} ,  z_{\pi/2}^{(2)} } \sim \p{ \wS^{-t_{12}} z_{\pi/2}^{(1)} ,  z_{\pi/2}^{(2)} } = \p{ e^{-2 \pi \im / \nbS } z_{\pi/2}^{(1)} ,  z_{\pi/2}^{(2)} } ~ ,    
\end{align}
making manifest that $\mathbb Z_{\nbS}$ does not act  freely on the base at   $x=\pi/2$ and leaves the fiber invariant. Thus, the branched squashed sphere  exhibits  a $\mathbb C/\mathbb Z_{\nbS}$  conical singularity  at $x=\pi/2$.

The Killing spinors  
\begin{align}\label{eq: ksbss}
     \zeta_\alpha  &  = \sqrt{\frac{\knorm b_2}{2 \, \nbS }}\, e^{\frac{\im}{2}\p{\alpha_2 \varphi_1  + \alpha_3 \varphi_2  }} \sqrt{1- \samf  \sin x }\begin{pmatrix} 1 \\ \sec x  + \samf \tan x \end{pmatrix}_\alpha  ~ , \nn\\
     \widetilde \zeta_\alpha  &  = - \sqrt{\frac{\knorm b_2}{2 \, \nbS }}\, e^{-\frac{\im}{2}\p{\alpha_2 \varphi_1  + \alpha_3 \varphi_2  }} \sqrt{1- \samf \sin x }\begin{pmatrix} \sec x + \samf  \tan x \\ 1 \end{pmatrix}_\alpha  ~ , 
\end{align}
with $\samf=\pm1$ being a sign, solve the three-dimensional $\mathcal N=2$ Conformal Killing-spinor equation  if the  background-field 
\begin{align}
    A^C & =  \frac{1}{2}\p{ -\frac{b_1}{\nbN f} + \alpha_2 } \dd \varphi_1  + \frac{1}{2}\p{ -\frac{ \samf b_2 }{\nbS f} + \alpha_3 } \dd \varphi_2   ~ ,
\end{align}
is turned on. The 1-form  $A^C$ is smooth on the branched-squashed three-sphere  if
\begin{align}\label{eq: ACsmoothnessonbss}
     & \alpha_2 = - 1 / \nbN    ~ , & \alpha_3 =  - \samf / \nbS    ~ .
\end{align}
Furthermore, the spinors (\ref{eq: ksbss}) solve the three-dimensional $\mathcal N=2$ Killing-spinor equation with  background fields 
\begin{align}
    V & = \p{ \frac{1}{f} + \im \, \samf H  } \p{ \frac{b_1}{\nbN}\sin^2 x \, \dd \varphi_1  + \samf \frac{ b_2}{\nbS}\cos^2 x \, \dd \varphi_2   }  ~ , \nn\\
    A & = A^C + \frac{3}{2} V   ~ .
\end{align}
  The Killing spinors given in (\ref{eq: ksbss}) are distinguished by the sign  $ \samf = \pm 1 $, which is analogous to the sign $ \upsigma $ that classifies the twist and anti-twist in Section \ref{eq: spindles1background}. Indeed, in Section \ref{sec: spindlefromthreesphere}, we explicitly show that the gauge theory on the spindle base, obtained via dimensional reduction along the $ S^1 $ fiber, is ultimately defined on a twisted or anti-twisted $ \spindle $, depending on whether $ \samf = \pm 1 $, respectively.


\subsection{The branched-squashed lens space}\label{eq: blensbackground}

 The squashed lens space and the branched-squashed three-sphere are specific instances of  the branched-squashed lens space $\blens$, which  is a Seifert orbifold described  by a general $\mathcal O(-n)$ circle fibration over a spindle $\spindle = \mathbb{WCP}^1_{\comm{\nN, \nS}}$. A family of  line elements on the  three-dimensional orbifold  $\blens$  reads 
\begin{align}\label{eq: lineelementgsf}
	\dd s^2 = \p{e^1}^2 + \p{e^2}^2  + \p{e^3}^2   ~ ,
\end{align}
with $\p{e^1, e^2, e^3}$ being the  orthonormal frame
\begin{align}\label{eq: framegsf}
     e^1  & = - f \, \dd x   ~ , \nn\\
      e^2 & =  b_1 b_2  \,\widetilde f^{-1}  \sin x \cos x \,  \dd\sigmavar ~ , \nn\\
       e^3 & = \frac{\widetilde f}{n } \p{ \acf +  \dd \psi }       ~ ,  \nn\\
    \widetilde f & = \sqrt{\p{\nS b_1 \sin x}^2 + \p{\nN b_2  \cos x }^2 }   ~ , \nn\\
     \acf & =  n \, \widetilde f^{-2}\p{ \nS \tS b_1^2 \sin^2 x+ \nN \tN b_2^2 \cos^2 x  } \dd\sigmavar  ~ ,  
\end{align}
where the coordinates on the spindle base $\spindle$ are again  $\p{x,\sigmavar}$, the function $f$ is reported in (\ref{eq: squashingfunction})  and $\p{\tN, \tS}$ are integers that satisfy the constraint $\p{\nN \tS - \nS \tN } = 1$ thanks to Bezout's lemma. The Euler characteristic of the base and  the flux of  $\acf$ through $\spindle$ are 
\begin{align} 
    & \chi_{\rm orb}  = \frac{1}{4 \pi} \int_\spindle \dd x \dd   \varphi \sqrt{g_\spindle} R_\spindle = \frac1{\nN} + \frac1{\nS} ~ , & \frac{1}{2 \pi} \int_\spindle \dd \acf =  \frac{n \, \tS}{\nS} - \frac{n\, \tN}{\nN} =  \frac{n}{\nN \nS}  ~ ,
\end{align}
confirming that  $\blens$ is a $\mathcal O\p{ - n  }$  circle bundle  over   $\spindle$. If $n=1$, then $\blens$ coincides with a squashed three-sphere, which is indeed a $\mathcal O\p{ - 1  }$  circle fibration  over   a spindle. Furthermore, if  $n= \nN \nS$, then   $\blens$  becomes the branched-squashed three-sphere that we examined in a previous  section. Moreover,    if $n= n_0 \nN \nS $ for some positive integer $n_0$, then $\blens$  reduces to the free quotient $S^3_{b_1, b_2, \nN, \nS}/\mathbb Z_{n_0}$, with $\mathbb Z_{n_0}$  acting upon the $S^1$  fiber of the branched-squashed three-sphere. For example, a   line element on     $S^3_{b_1, b_2, \nN, \nS}/\mathbb Z_{n_0}$,   is
\begin{align} 
    \dd s^2 = f^2 \dd x^2 +  \p{\frac{b_1}{\nN}}^2 \sin^2 x \p{ \dd \sigmavar + \frac{\dd \psi}{n_0} }^2 + \p{\frac{b_2}{n_0 \,\nS}}^2 \cos^2 x \, \dd \psi^2  ~ ,
\end{align}
where, as usual, $\sigmavar, \psi \in \comm{0, 2 \pi}$  are   periodic     and    $x\in\comm{0, \pi/2}$.  

Let us now explore the geometric features of general $\blens$ labelled by  arbitrary integers $\p{n, \nN, \nS, \tN, \tS}$.  Expanding (\ref{eq: lineelementgsf}) in a neighborhood of $x=0$ gives 
\begin{align}
	\dd s^2|_{x=0} =  b_1^2\p{\dd x^2 + x^2 \frac{\dd\varphi^2}{\nN^2} } + \p{\frac{\nN b_2}{n} }^2 \p{ \dd\psi + \frac{n \, \tN }{\nN} \dd\varphi }^2 ~ ,
\end{align}
which can be rewritten via the complex coordinates
\begin{align}
	& z_0^{(1)} = b_1 \,  x \, e^{\im \varphi / \nN }  ~ , & z_0^{(2)} = \frac{\nN b_2 }{n} e^{ \im \p{  \psi + \frac{n  \tN }{\nN} \varphi  }  }  ~ ,
\end{align}
which are invariant under $\psi\sim\psi+2\pi$ and are subject to the identifications
\begin{align}
	& \varphi \sim \varphi + 2 \pi \implies \p{ z_0^{(1)} ,  z_0^{(2)} } \sim \p{ \wN z_0^{(1)} , \wN^{n \, \tN }    z_0^{(2)} } ~ ,  & \wN = e^{2 \pi \im/\nN } ~ .
\end{align}
If  $\rem{n \, \tN}{\nN}=0$, where $\rem{  \bullet }{ \diamond}$  is the  reminder of the integer division of $\bullet$ by $\diamond$, then $\wN^{n \, \tN }  =1$, implying that $\p{ z_0^{(1)} ,  z_0^{(2)} }$ are coordinates on the orbifold $\p{\mathbb C/\mathbb Z_{\nN}}\times \mathbb C^*$, where $\mathbb C/\mathbb Z_{\nN}$ encodes a genuine conical singularity with deficit angle $2\pi\p{1 - \nN^{-1}}$.  Otherwise,  if  $n \, \tN = \kN \tN' $ and $ \nN = \kN \nN' $ share a common factor $\kN$, then winding $\nN' $ times around the point $x=0$ shows that the north pole of the spindle base $\spindle$ is the fixed point of a $\mathbb Z_{\kN}$ action:
\begin{align}
	& \varphi \sim \varphi   + 2 \pi \, \nN'  \implies \p{ z_0^{(1)} ,  z_0^{(2)} } \sim \p{ e^{2 \pi \im /\kN }  z_0^{(1)} ,   z_0^{(2)} }  ~ .
\end{align} 
Therefore, the topology of this fibration in a neighborhood of  $x=0$ is that of $\p{\mathbb C/\mathbb Z_{\kN}}\times \mathbb C^*$, with the quotient $\mathbb C/\mathbb Z_{\kN}$ describing the corresponding conical singularity. Eventually, if $n \, \tN $ and $\nN$ turn out to be  mutually coprime, trivializing the $\mathbb Z_{\nN}$ action on the fiber also trivializes that on the base:
\begin{align}
	& \varphi \sim \varphi + 2 \pi \nN \implies \p{ z_0^{(1)} ,  z_0^{(2)} } \sim \p{ \wN^{\nN} z_0^{(1)} , \wN^{n  \nN \tN }    z_0^{(2)} } = \p{ z_0^{(1)} ,  z_0^{(2)} } ~ .
\end{align}
In this case, the $\mathbb Z_{\nN}$ action is free as it exhibits  no fixed points and the fibration is smooth near the point $x=0$. On the other hand,
\begin{align}
	\dd s^2|_{x=\pi/2} =  b_2^2\comm{\dd x^2 +\p{\frac\pi2-x}^2 \frac{\dd\varphi^2}{\nS^2} } + \p{\frac{\nS b_1}{n} }^2 \p{ \dd\psi + \frac{n \, \tS }{\nS} \dd\varphi }^2 ~ ,
\end{align}
suggesting the introduction of the complex coordinates
\begin{align}
& z_{\pi/2}^{(1)} = b_2 \p{\frac\pi2 - x} e^{-\im \varphi / \nS }  ~ , & z_{\pi/2}^{(2)} = \frac{\nS b_1 }{n} e^{ \im \p{  \psi + \frac{n  \tS }{\nS} \varphi  }  }  ~ ,
\end{align}
which are left unchanged by  $\psi\sim\psi+2\pi$, while they fulfil the equivalence relations
\begin{align}
	& \varphi \sim \varphi + 2 \pi \implies \p{ z_{\pi/2}^{(1)} ,  z_{\pi/2}^{(2)} } \sim \p{ \wS^{-1} z_{\pi/2}^{(1)} , \wS^{- n \, \tS }    z_{\pi/2}^{(2)} } ~ ,  & \wS = e^{2 \pi \im/\nS } ~ , 
\end{align}
hold. Again, if $\rem{n \, \tS}{\nS}=0$, the action on the fiber is trivial as $\wS^{n \, \tS }  =1$ and the complex coordinates $\p{ z_{\pi/2}^{(1)} ,  z_{\pi/2}^{(2)} }$ parametrize the orbifold $\p{\mathbb C/\mathbb Z_{\nS}}\times \mathbb C^*$.  Otherwise,  if  $n \, \tS = \kS \tS' $ and $ \nS = \kS \nS' $, then the action on the fiber is trivialized by going $\nS'$ times around the south pole of the spindle base, giving  
\begin{align}
	& \varphi \sim \varphi   + 2 \pi \, \nS'  \implies \p{ z_{\pi/2}^{(1)} ,  z_{\pi/2}^{(2)} } \sim \p{ e^{-2 \pi \im /\kS }  z_{\pi/2}^{(1)} ,   z_{\pi/2}^{(2)} }  ~ ,
\end{align} 
implying that the topology of the fibration near the point $x=\pi/2$ is that of $\p{\mathbb C/\mathbb Z_{\kS}}\times \mathbb C^*$. Finally, if $n \, \tS $ and $\nS$ are coprime, there is no way to trivialize the $\mathbb Z_{\nS}$ action  on the fiber without trivializing   its action on the base and  $\mathbb Z_{\nS}$ has no fixed points. Hence,  $\mathbb Z_{\nS}$ acts freely and the  fibration is  smooth in a neighborhood of $x=\pi/2$. If both $\mathbb Z_{\nN}$ and $\mathbb Z_{\nS}$ act freely everywhere, the fibration is a smooth manifold,  as in the case of the lens space.

The Killing vector of the form  (\ref{eq: masterkillingvector})  enjoyed by this fibration  reads
\begin{align}
	 & K = \knorm\p{ \amf \partial_\sigmavar + \partial_\psi } ~ , & \amf = -  \frac{ \nS b_1  - \samf  \nN b_2}{n \p{ \tS   b_1 - \samf  \tN b_2}}  ~ , \qquad \zzt = \frac{\knorm b_1 b_2}{n\p{\tS b_1 - \samf  \tN b_2}}   ~ ,
\end{align}
where $\samf=\pm1$ is the same sign introduced before,  $\zzt = \zeta \widetilde \zeta$ is the contraction of the Killing spinors (\ref{eq: masterkillingspinors}), with components
\begin{align}
    u_{1,2} 
    %
     %
     %
     & =\sqrt{ \frac{\knorm b_1 b_2}{2 n\p{\tS b_1 - \samf  \tN b_2}} \comm{ 1  \pm    \p{ \nS b_1  - \samf  \nN b_2}   \,\widetilde f^{-1}  \sin x \cos x   } }    ~ ,
\end{align}
    %
and fulfilling the Killing spinor equation for an arbitrary phase $\theta$ if one turns on the background fields
\begin{align}
   	   	 A^C  & = \dd\theta -    \frac{ \tS  b_1 + \samf \tN b_2 }{2 f} \dd\varphi -   \frac{\nS b_1 + \samf \nN b_2 }{2 n f}  \dd\psi  ~ ,   \qquad  \theta = \frac12\comm{ \alpha_1(x) + \alpha_2 \sigmavar + \alpha_3 \psi }   ~ , \nn\\
   	 V   & = \p{\frac{1}{f} + \im \, \samf  H   }\comm{ \p{ \tS b_1 \sin^2 x +  \samf \tN b_2 \cos^2 x}\dd\varphi + \p{ \nS b_1 \sin^2 x  + \samf \nN b_2 \cos^2x }\frac{\dd\psi}{ n} } ~ ,  \nn\\
   	 A  & = A^C + \frac{3}{2} V  ~ . 
   \end{align}
   To ensure regular R-symmetry background fields on the three-dimensional orbifold described by $ \blens $, it is convenient to work in two coordinate patches, \( \mathcal{U}_\pm \), which respectively include either the northern or the southern pole of the spindle base $\spindle $. These patches are parametrized by the following coordinates:   
\begin{align}
& \mathcal U_+ :   \p{\varphi, \psi_+} = \p{\varphi, \psi + \frac{n \, \tN}{\nN}} ~ , & \mathcal U_- :   \p{\varphi, \psi_-} = \p{\varphi, \psi + \frac{n \, \tS}{\nS}} ~ . 
\end{align}
In such coordinates the R-symmetry connection $A^C$ appearing in the CKSE reads
\begin{align}
	A^C_+ & =  \frac12\p{ - \frac{b_1}{\nN f} +  \alpha_2^+  - \frac{n \tN}{\nN}\alpha_3^+ } \dd\varphi + \frac12\p{ - \frac{ b_1 \nS + \samf \nN b_2}{n f} + \alpha_3^+ } \dd\psi_+ ~ , \nn\\ 
	A^C_- & =  \frac12\p{ \frac{\samf b_2}{\nS f} +  \alpha_2^-  - \frac{n \tS}{\nS}\alpha_3^- } \dd\varphi + \frac12\p{ - \frac{ b_1 \nS + \samf \nN b_2}{n f} + \alpha_3^- } \dd\psi_- ~ .
\end{align}
Then, the 1-forms $A^C_\pm$ are regular in the corresponding patches if
\begin{align}
	 & \alpha_2^+ = \frac{-1 + n \tN \alpha_3^+}{\nN} ~ ,   & \alpha_2^- = \frac{\samf + n \tS \alpha_3^-}{\nS}  ~ .
\end{align}
The simplest choice to have background fields that  are  regular on the fibration is
    \begin{align}
	\alpha_2^\pm = \alpha_2 = - \samf  \tN - \tS     ~ , \qquad  \alpha_3^\pm  = \alpha_3  =  - \frac{\samf \nN + \nS}{n  } ~ .
\end{align}
As anticipated, in Section \ref{sec: spindlefromthreesphere}, we demonstrate that the signs $\samf = \pm 1 $ correspond to the presence of an R-symmetry twist or anti-twist on the base $ \spindle $ of the circle fibration, respectively.


\section{Localization on \texorpdfstring{$\spindle \times S^1$}{Sigma x S1} with twist and anti-twist}\label{sec: localizationspindle}

\subsection{BPS locus}

We parametrize an Abelian gauge or flavour 1-form field   on  $\spindle \times S^1$ as 
\begin{align}\label{eq: bpsgaugeconnection}
& \ag = \ag_\sigmavar\p{x} \dd \sigmavar + \ag_\psi\p{x} \dd \psi  ~ , & \flux_G = \frac{1}{2\pi} \int_\spindle \dd \ag = \frac{\mm}{\nN \nS} ~ , \qquad \exp\p{\im \oint_{S^1}  \ag} = \hh ~ ,
\end{align}
with  $\mm\in \mathbb Z$ and   $\hh \in U(1)$, where we removed the component $\ag_x\p{x}$ along $\dd x$ via a gauge transformation. In the non-Abelian case, we consider a Lie group $G = G_G \times G_F$ that  is the direct product of the gauge group $G_G$ and the flavour group $G_F$. The 1-forms in the vector multiplet are then $\mathfrak g$-valued, where  $\mathfrak g = \mathfrak g_G \oplus \mathfrak g_G$ is the Lie algebra of $G$. In general, fields transform in a representation $\mathfrak R_G$ of $G$ with weight $\rho = \rho_G + \rho_G$. 

The 1-form in (\ref{eq: bpsgaugeconnection}) is the representative of an $\mathcal O\p{-\mm}$ orbibundle on the spindle $\spindle=\mathbb{WCP}^1_{\comm{\nN, \nS}}$ and of the $U(1)$ holonomy group of $S^1$. In particular, the   component $\ag_\sigmavar\p{x}$ evaluated at the north pole at $x=1$ and at the south  pole at $x=-1$ of the spindle reads
\begin{align}
  & \ag_\sigmavar\p{+1} = \frac{\mN}{\nN}  ~ , \qquad \ag_\sigmavar\p{-1} = \frac{\mS}{\nS}  ~ ,  & \nN \mS - \nS \mN = \mm ~ ,
\end{align}
where the integers $\p{\mN, \mS}$  can be expressed in terms of $\mm$ and  two other integers $\p{\aN, \aS}$ satisfying  
\begin{align}
  & \mN = \mm \, \aN  ~ , \qquad  \mS = \mm \, \aS  ~ ,   & \nN \aS - \nS \aN = 1 ~ .
\end{align}
Especially, given a pair $\p{  \aN,  \aS}$ satisfying the constraint above, any pair $\p{ \aN + \nN \delta \mathfrak a,  \aS + \nS \delta \mathfrak a}$ with $\mathfrak \delta \mathfrak a\in \mathbb Z$ fulfils the constraint too. Physical observables are supposed to be independent of $\delta \mathfrak a$.

The 1-form (\ref{eq: bpsgaugeconnection}) is consistent with the gauge fields used in   \cite{Ferrero:2021etw} provided that
\begin{align}
 \ag^N =  \ag^N_{\p{0}} + \frac{\mathfrak m_N}{\nN}  \dd \sigmavar = \ag^S_{\p{0}} + \p{ {\rm p} + \frac{\mathfrak m_S}{\nS} }  \dd \sigmavar =  \ag^S +  {\rm p} \,   \dd \sigmavar   ~ , 
\end{align}
with ${\rm p}\in \mathbb Z$ modelling  a gauge transformation in the overlap of the two patches $(N,S)$, where the connections $\p{\ag^N, \ag^S}$ are well defined. The dictionary between our notation and that of \cite{Ferrero:2021etw} is
\begin{align}
   \ag^N_{\p{0}}  & = \ag -  \frac{\mN}{\nN}  \dd \sigmavar   ~ , \qquad n_N  = \nN ~ ,  \qquad \mathfrak m_N  = \mN ~ ,  \nn\\
   \ag^S_{\p{0}}  & = \ag -  \frac{\mS}{\nS}  \dd \sigmavar   ~ ,  \qquad n_S  = \nS ~ ,  \qquad \mathfrak m_S  = \mS - \nS {\rm p} ~ . 
\end{align}
In this language, the flux of $\ag$ reads \cite{Ferrero:2021etw}
\begin{align}
	\frac{1}{2\pi}\int_\spindle  \dd \ag =  {\rm p} + \frac{\mathfrak m_S}{n_S}   - \frac{\mathfrak m_N}{n_N}  = \frac{\mm}{\nN \nS}  ~ ,
\end{align}
in agreement with (\ref{eq: bpsgaugeconnection}). If we add a flat connection $\p{\beta_2\dd\sigmavar}$ to $\ag$ in  (\ref{eq: bpsgaugeconnection})  as 
\begin{align}\label{eq: bpsgaugeconnectionshifted}
& \ag \to \ag' = \p{\ag_\sigmavar +\beta_2} \dd \sigmavar + \ag_\psi \dd \psi  ~ ,  
\end{align}
we can make $\ag$ vanish at the  poles of  $\spindle$ at $x=\pm1$ by setting $\beta_2|_{\UN}  = - \mN/\nN$ in the northern patch $\UN$ including $x=+1$ and $\beta_2|_{\US} = - \mS/\nS$ in the southern patch $\US$ covering $x=-1$.

 The vector-multiplet BPS equations are obtained by setting to zero  the gauginos $\p{\lambda, \widetilde \lambda}$ and their variations $\p{\delta \lambda , \delta \widetilde \lambda }$, giving  
 \begin{align}
 & {\Phi_G}  = \iota_K \Ac - \im \zzt \sigma = \knorm \varphi_G =   {\rm constant} ~ ,  \nn \\
   & D + \f\im 2 P^\mu \widetilde P^\nu \Fc_{\mu\nu} +\f\im\zzt \sigma \comm{   \iota_K V - \im \zzt H }  = 0  ~ . 
\end{align}
We do not impose any reality conditions on fields, for the moment. Inserting the gauge field (\ref{eq: bpsgaugeconnection}) into the BPS equations above yields the BPS values of the scalar  $\sigma$ and the auxiliary scalar   $D$. Specifically, if $\st=+1$, we have topologically twisted $\spindle\times S^1$ and the BPS locus 
\begin{align}
	& \sigma|_{\rm BPS} =   \im \beta^{-1} \comm{- \varphi_G + \amf \ag_\sigmavar\p{x} + \ag_\psi\p{x} } ~ ,  & D|_{\rm BPS} = - \frac{\ag_\sigmavar'\p{x}}{f \sqrt{1-x^2}} ~ ,  
\end{align}
whereas, if $\st=-1$, we have anti-twisted $\spindle\times S^1$ and
\begin{align}
	& \sigma|_{\rm BPS} =  - \im \comm{- \varphi_G + \amf \ag_\sigmavar\p{x} + \ag_\psi\p{x} }/\p{\beta x} ~ , \nn\\
	 & D|_{\rm BPS} =  \frac{\beta \, \sigma|_{\rm BPS} + \comm{\beta + \im \Omega \p{1-x^2} } \ag_\sigmavar'\p{x} + \im \p{1-x^2}\ag_\psi'\p{x} }{\beta x f \sqrt{1-x^2} } ~ .
\end{align}
Such supersymmetric values of $\p{\sigma, D }$  provide a non-trivial classical contribution to the partition function of the corresponding gauge theory thanks to the presence of (possibly mixed) Chern-Simons and Fayet–Iliopoulos terms, for example. Furthermore, the profiles $\p{\sigma|_{\rm BPS}, D|_{\rm BPS} }$ implicitly affect the one-loop determinant of supersymmetric fluctuations via the gauge fugacity $\varphi_G$ and the flux $\flux_G$.

Analogously, the   BPS locus for the chiral multiplet is found by setting to zero  the   spinor fields $\p{\psi, \widetilde \psi}$ together with their supersymmetric variations $\p{\delta \psi , \delta \widetilde \psi }$.  The outcome of this procedure is the following set of BPS equations:
\begin{align}
	& \p{\mathcal L_K - \im r \Phi_R - \im q_G \Phi_G}\phi = 0 ~ , & \p{\mathcal L_K + \im r \Phi_R + \im q_G \Phi_G}\widetilde\phi = 0 ~ , \nn\\
	& F + \im L_{\widetilde P}\phi  = 0 ~ ,   & \widetilde F + \im L_{ P}\widetilde\phi = 0 ~.
\end{align}
For arbitrary values of $\p{\Phi_R, \Phi_G}$ the unique solution to the chiral-multiplet BPS equations is  
\begin{align}
	\phi|_{\rm BPS} = \widetilde \phi|_{\rm BPS} = F|_{\rm BPS} = \widetilde F|_{\rm BPS} = 0 ~ .
\end{align}
As a consequence, we do not expect classical contributions to the partition function to come from F-terms or  superpotentials, for instance. In this setup,     matter fields affect the theory at the quantum level only, via one-loop determinants.


\subsection{Chiral-multiplet one-loop determinants}

\subsubsection{Unpaired eigenvalues}

The chiral-multiplet one-loop determinant entering the partition function can be obtained by exploiting the cancellations between bosonic and fermionic degrees of freedom due to supersymmetry. This procedure is implemented by the formula
\begin{align}
	 Z^{\rm CM}_{\text{1-L}} = \frac{\det_{\text{Ker} L_P} \delta^2}{\det_{\text{Ker} L_{\widetilde P}} \delta^2} = \frac{\det_{\text{Ker} L_P} \p{L_K + \mathcal G_{\Phi_G}}}{\det_{\text{Ker} L_{\widetilde P}} \p{L_K + \mathcal G_{\Phi_G}}} ~ .
\end{align}
Indeed, the operators $L_{P}$ and $L_{\widetilde P}$  pair bosonic and fermionic fields in the supersymmetry transformations written in cohomological form. The formula above follows from the fact that the  squared supersymmetry variation  $\delta^2 \propto \p{L_K + \mathcal G_{\Phi_G}} $ commutes with such pairing operators. Thus,  the unpaired eigenvalues  surviving the cancellations and contributing to $Z^{\rm CM}_{\text{1-L}}$ are those corresponding to the eigenfunctions   spanning the kernels of $L_{P}$ and $L_{\widetilde P}$. In particular, the eigenfunctions of $\delta^2$ that are annihilated by the differential operator $L_{P}$ have the form 
\begin{align}\label{eq: kerLPeigenfunctions}
	& \mathcal B_{ m_\sigmavar ,  m_\psi} = e^{\im m_\sigmavar \sigmavar + \im m_\psi \psi } B_{ m_\sigmavar ,  m_\psi}\p{x} ~ , & m_\sigmavar ,  m_\psi \in \mathbb Z  ~ ,
\end{align}
while the eigenfunctions of $\delta^2$ that are  annihilated by the operator $L_{\widetilde P}$ have the form 
\begin{align}\label{eq: kerLPteigenfunctions}
	& \Phi_{ n_\sigmavar ,  n_\psi} = e^{\im n_\sigmavar \sigmavar + \im n_\psi \psi } \phi_{ n_\sigmavar ,  n_\psi}\p{x} ~ , & n_\sigmavar ,  n_\psi \in \mathbb Z  ~ .
\end{align}
Here, we have set $\alpha_2=0$ for simplicity, in analogy with \cite{Benini:2015noa}. The  identical final result  follows for    $\alpha_2\neq0$ after consistently restoring $\alpha_2$ everywhere: for instance, if  $\alpha_2\neq0$, then (\ref{eq: kerLPteigenfunctions})  becomes
\begin{align}\label{eq: kerLPteigenfunctionsgeneralalpha2}
	 \widehat \Phi_{ n_\sigmavar ,  n_\psi} = e^{\im \p{r/2}\alpha_2  + \im n_\sigmavar \sigmavar + \im n_\psi \psi } \phi_{ n_\sigmavar ,  n_\psi}\p{x} ~ ,  
\end{align}
where $r$ is the $R$-charge of the chiral multiplet. The form of (\ref{eq: kerLPteigenfunctionsgeneralalpha2}) ensures that $\widehat \Phi_{ n_\sigmavar ,  n_\psi} $ has the correct transition function on $\spindle$, as it behooves a section of the $R$-symmetry bundle on the spindle.

The eigenfunctions contributing to $Z^{\rm CM}_{\text{1-L}}$ are those that are non-singular on the whole three-dimensional orbifold. For this reason, we now study the  regularity of the eigenfunctions  $\p{\mathcal B_{ m_\sigmavar ,  m_\psi} , \Phi_{ n_\sigmavar ,  n_\psi} }$ on $\spindle\times S^1$ in presence of both twist and anti-twist for the  $R$-symmetry. As in \cite{Inglese:2023wky} we keep the notation compact by defining the following quantities: 
\begin{align}\label{eq: bcpfug}
	 & \pN  = q_G \mN - \st \frac{r}{2} ~ ,  &  \pS  =q_G \mS +  \frac{r}{2} ~ , \nn\\
	 & \mathfrak b  = 1 + \st \ff{ \st \frac{\pN}{\nN} } +  \ff{ - \frac{\pS}{\nS} } ~ ,   & \mathfrak c  =  \frac{ \rem{ - \pS}{ \nS} }{\nS} - \st \frac{ \rem{ \st \,   \pN}{ \nN} }{\nN} ~ , \nn\\
	 & \rfug  = - \frac{\alpha_3}{2} +  \frac{\amf}{4} \chi_{-\st} ~ ,  &  \gfug  =    - \varphi_G + \frac{\amf}{2}\p{\frac{\mS}{\nS}  + \frac{\mN}{\nN}  }  ~ ,   \nn\\
	 & q = e^{2 \pi \im \amf } ~ ,  & y = q^{\mathfrak c/2} e^{2 \pi \im \p{ r \,  \gamma_R + q_G \gamma_G } } ~ ,
\end{align}
 In particular, $\p{\mathfrak b-1}$ is the degree of the line bundle $\mathcal O( - \mathfrak p  )  = \mathcal O\p{ - \p{r/2}\p{\nN + \nS} - q_G \mathfrak m }$, with $\mathfrak p = \p{ \nN  \pS  -  \nS \pN } $. Also,  $\amf$, $\gamma_G$ and $\gamma_R$ are  effective fugacities  for  angular momentum,    gauge  and    $R$-symmetry, respectively. The objects written in (\ref{eq: bcpfug}) are valid for a theory invariant under a  $U(1)$ gauge or flavour group. For a general, possibly non-Abelian, Lie group $G$, the substitutions 
 \begin{align}
 & \mathfrak m_\pm \to \rho\p{\mathfrak m_\pm } ~ , & \gfug \to \rho\p{\gfug } ~ ,
 \end{align}
 are in order, with $\rho  $ being the weight of the representation $\mathfrak R_G$ that we defined previously.


\paragraph{Topological twist.}

In the case of the topologically twisted  $\spindle \times S^1$ the behaviour of $B_{ m_\sigmavar ,  m_\psi}\p{x}$ near the north pole of the spindle is 
\begin{align}
	\lim_{x \to +1} B_{ m_\sigmavar ,  m_\psi}\p{x} \to \p{1-x}^{\comm{2 m_\sigmavar \nN + \p{r-2} - 2  q_G \mN }/4}  ~ ,
\end{align}
which is non-singular if 
\begin{align}
	 m_\sigmavar  \geq \cf{  \frac{ 1 - \p{r/2} + q_G \mN}{\nN} } =   \cf{  \frac{ 1 + \pN}{\nN} } = 1 +  \ff{  \frac{  \pN}{\nN} } ~ ,
\end{align}
where $\cf{\bullet}$ is the ceiling   of $\bullet$, namely the smallest integer greater than or equal to $\bullet$, while $\ff{\bullet}$ is  the floor of $\bullet$, which is  the greates integer lesser  than or equal to $\bullet$. Instead, at the south pole of the spindle we have  
\begin{align}
	\lim_{x \to -1} B_{ m_\sigmavar ,  m_\psi}\p{x} \to \p{1+x}^{\comm{ - 2 m_\sigmavar \nS + \p{r-2} + 2  q_G \mS }/4}  ~ ,
\end{align}
which is non-singular if 
\begin{align}
	 m_\sigmavar  \leq  \ff{  \frac{  \p{r/2} - 1 + q_G \mS}{\nS} }  =  \ff{  \frac{   - 1 +  \pS}{\nS} } =  - 1 +  \cf{  \frac{     \pS}{\nS} } =  - 1 -  \ff{ -  \frac{     \pS}{\nS} } ~ .
\end{align}
Analogously, the modes  $\phi_{ n_\sigmavar ,  n_\psi}\p{x}$ behave near the north pole of the spindle as
\begin{align}
	\lim_{x \to +1} \phi_{ n_\sigmavar ,  n_\psi}\p{x} \to \p{1-x}^{\p{-2 n_\sigmavar \nN - r + 2  q_G \mN }/4}  ~ ,
\end{align}
which is non-singular if 
\begin{align}
	 n_\sigmavar  \leq \ff{  \frac{ - \p{r/2} + q_G \mN}{\nN} } =  \ff{  \frac{ \pN}{\nN} } ~ ,
\end{align}
while at the south pole of the spindle we find  
\begin{align}
	\lim_{x \to -1} \phi_{ n_\sigmavar ,  n_\psi}\p{x} \to \p{1+x}^{\p{ 2 n_\sigmavar \nS - r - 2  q_G \mS }/4}  ~ ,
\end{align}
which is non-singular if 
\begin{align}
	 n_\sigmavar  \geq  \cf{  \frac{  \p{r/2}  + q_G \mS}{\nS} }  = \cf{    \frac{ \pS}{\nS} }   = - \ff{  -  \frac{ \pS}{\nS} }  ~ .
\end{align}
The inequalities above imply that there are no regular functions  in either  ${\rm Ker } L_{\widetilde P}$ or    ${\rm Ker } L_{ P}$ if   $\mathfrak b = 0$; thus, $Z^{\rm  CM}_{\text{1-L}}|_{\mathfrak b=0}=1$.  If $\mathfrak b \leq -1$, there are no regular fields in the kernel of $L_{\widetilde P}$ and the eigenfunctions  $\Phi_{ n_\sigmavar ,  n_\psi}$ do not contribute to the one-loop determinant of the chiral-multiplet partition function. Nonetheless,  the eigenfunctions  $\mathcal B_{ m_\sigmavar ,  m_\psi}$ are non-singular if $\mathfrak b \leq -1$  and do    contribute to $Z^{\rm  CM}_{\text{1-L}}$ via  
\begin{align}\label{eq: zBtwist}
	 Z^{\mathcal B}_{\text{1-L}} & = \prod_{m_\psi \in \mathbb Z} \prod_{m_\sigmavar =   1 +  \ff{ \pN/ \nN } }^{  - 1 -  \ff{ -  \pS/ \nS } } \comm{m_\psi + \amf m_\sigmavar + \p{2 - r }\frac{\alpha_3}{2} -  q_G \varphi_G  }  \nn\\
	 & = \prod_{m_\psi \in \mathbb Z} \prod_{ j =   0 } ^{  - 1 -  \mathfrak b } \comm{m_\psi + \amf \p{j + \frac{1+\mathfrak b + \mathfrak c}{2} }   + r \gamma_R +  q_G \gamma_G  } ~ .	 
\end{align}
We can regularize in a single step the infinite product above by writing
\begin{align}
	 Z^{\mathcal B}_{\text{1-L}} &  = \prod_{m_\psi \in \mathbb Z} \prod_{ j \in \mathbb N }  \frac{m_\psi + \amf \p{j + \frac{1+\mathfrak b + \mathfrak c}{2} }   + r \gamma_R +  q_G \gamma_G  }{m_\psi + \amf \p{j + \frac{1-\mathfrak b + \mathfrak c}{2} }   +r \gamma_R +  q_G \gamma_G  }   \nn\\	 
	 &  =  \p{-y}^{\mathfrak b/2}\frac{\p{q^{\p{1+\mathfrak b}/2} y ; q }_\infty }{\p{q^{\p{1-\mathfrak b}/2} y ; q}_\infty}  =  \frac{ \p{-y}^{\mathfrak b/2} }{\p{q^{\p{1-\mathfrak b}/2} y ; q}_{\mathfrak b} }  =  \frac{ \p{-y^{-1} }^{\mathfrak b/2} }{\p{q^{\p{1-\mathfrak b}/2} y^{-1} ; q}_{\mathfrak b} } ~ .  	 
\end{align}
On the other hand, If $\mathfrak b \geq 1$, the situation is reversed as there are no regular fields in the kernel of $L_{ P}$ and    $\mathcal B_{ m_\sigmavar ,  m_\psi}$ do not contribute to  $Z^{\rm  CM}_{\text{1-L}}$; while   the eigenfunctions  $\Phi_{ n_\sigmavar ,  n_\psi}$ yield a factor of  
\begin{align}\label{eq: zphitwist}
	 Z^{\Phi}_{\text{1-L}} & = \prod_{n_\psi \in \mathbb Z} \prod_{n_\sigmavar =   -  \ff{ - \pS/ \nS } }^{  \ff{   \pN/ \nN } } \p{n_\psi + \amf n_\sigmavar - r \frac{\alpha_3}{2} -  q_G \varphi_G  }^{-1}  \nn\\
	 & = \prod_{n_\psi \in \mathbb Z} \prod_{ j =   0 } ^{ \mathfrak b - 1 } \comm{n_\psi + \amf \p{j + \frac{1-\mathfrak b + \mathfrak c}{2} }   + r \gamma_R +  q_G \gamma_G  }^{-1}  \nn\\	 
	 &  = \prod_{n_\psi \in \mathbb Z} \prod_{ j \in \mathbb N }  \frac{n_\psi + \amf \p{j + \frac{1+\mathfrak b + \mathfrak c}{2} }   + r \gamma_R +  q_G \gamma_G  }{j_\psi + \amf \p{j + \frac{1-\mathfrak b + \mathfrak c}{2} }   +r \gamma_R +  q_G \gamma_G  }   =  \frac{ \p{-y^{-1} }^{\mathfrak b/2} }{\p{q^{\p{1-\mathfrak b}/2} y^{-1} ; q}_{\mathfrak b} }  ~ ,
\end{align}
which is consistent with the result obtained by considering the eigenfunctions $\mathcal B_{ m_\sigmavar ,  m_\psi}$. We then   conclude that,  on topologically twisted $\spindle \times S^1$, the one-loop determinant for a chiral multiplet with weight $\rho$ in a  representation $\mathfrak R_G$ of a gauge group $G$ is
\begin{align}\label{eq: cm1loopspindles1twist}
	 Z^{\rm CM}_{\text{1-L}} &    = \prod_{\rho \in \mathfrak R_G}  \frac{ \p{-y^\rho }^{\mathfrak b/2} }{\p{q^{\p{1-\mathfrak b}/2} y^\rho ; q}_{\mathfrak b} }  = \prod_{\rho \in \mathfrak R_G}  \frac{ \p{-y^{-\rho} }^{\mathfrak b/2} }{\p{q^{\p{1-\mathfrak b}/2} y^{-\rho} ; q}_{\mathfrak b} } ~ ,  	 
\end{align}
as in \cite{Inglese:2023wky}. The property $\p{ \bullet ; q}_0 = 1$ enjoyed by the finite $q$-Pochhammer symbol  ensures that  $Z^{\rm CM}_{\text{1-L}}|_{\mathfrak b=0} =1$,  as anticipated. In the degenerate case  $\nN=\nS=1$ the spindle  becomes a smooth sphere and  the simplifications 
\begin{align}
	& \mathfrak b = 1 - q_G \mathfrak m - r  = 1 - q_G \flux_G - r \flux_R  ~ , & \mathfrak c = 0  ~ , \nn\\
	& 2 \rfug = - \alpha_3 \in \mathbb Z   ~ , & y = \p{-1}^{r \alpha_3} e^{2 \pi \im q_G \gfug} ~ ,
\end{align}
occurr. In turn,  (\ref{eq: cm1loopspindles1twist}) becomes identical to  the one-loop determinant for a chiral multiplet on topologically twisted $S^2\times S^1$ computed in \cite{Benini:2015noa}, where $\alpha_3=0$ was assumed. 

The one-loop determinant (\ref{eq: cm1loopspindles1twist}) can also be interpreted as the partition function of a weakly gauged chiral multiplet, namely a chiral multiplet coupled to a non-dynamical  background vector multiplet.  Such a partition function is not invariant under large gauge transformations shifting the effective  fugacity $\gfug$ by an integer $n$:
\begin{align}
	y \to e^{2 \pi \im \, n } y \qquad \implies  \qquad Z^{\rm CM}_{\text{1-L}} \to \p{-1}^{n \, \mathfrak b } Z^{\rm CM}_{\text{1-L}} ~ . 
\end{align}
This multi-valuedness of $Z^{\rm CM}_{\text{1-L}} $ is usually associated to a parity anomaly affecting the matter sector of the theory. Such an anomaly can be cured by  considering the classical contribution $Z^{\rm CS}_{\rm eff}  $ of an effective Chern-Simons term  with half-integer level, whose anomaly cancels   that of $Z^{\rm CM}_{\text{1-L}} $. In the case of $S^2 \times S^1$ the counterterm needed is $Z^{\rm CS}_{\rm eff}  = y^{\pm \flux_G /2} = y^{\pm \mathfrak m /2} $, while for $Z^{\rm CM}_{\text{1-L}} $ on $\spindle \times S^1$ it is
\begin{align}
	Z^{\rm CS}_{\rm eff} = \p{-y}^{\pm \mathfrak b/2} ~ .
\end{align}
We    discuss the differences between effective and canonical Chern-Simons  terms in Section \ref{eq: classcontr}, where we compute  the relevant classical contributions to the partition function of a gauge theory on $\spindle \times S^1$.


\paragraph{Anti-twist.}

In the case of the anti-twisted  $\spindle \times S^1$ the eigenfunction  $B_{ m_\sigmavar ,  m_\psi}\p{x}$   is non-singular near the north pole of the spindle  if 
\begin{align}
	 m_\sigmavar  \leq   \ff{ \frac{ q_G \mN+\p{r/2}-1}{\nN} }  =  \ff{ \frac{ \pN -1}{\nN} } = - 1 +   \cf{ \frac{ \pN }{\nN} }  = - 1 -   \ff{ -  \frac{ \pN }{\nN} } ~ ,
\end{align}
while it is non-singular near the south pole of the spindle   if  
\begin{align}
	 m_\sigmavar  \leq      \ff{ \frac{ q_G \mS+\p{r/2}-1}{\nS} }  =   \ff{ \frac{  \pS -1}{\nS} }   = - 1 +   \cf{ \frac{ \pS }{\nS} }  = - 1 -   \ff{ -  \frac{ \pS }{\nS } } ~ ,
\end{align}
with the two inequalities above being both true if
\begin{align}
	 m_\sigmavar  \leq      - 1 - \frac12 \p{ \ff{ -  \frac{ \pN }{\nN } } +   \ff{ -  \frac{ \pS }{\nS } } } -  \frac12\left|     \ff{ -  \frac{  \pN }{\nN} }    -  \ff{ -  \frac{  \pS }{\nS} }  \right|   ~ ,
\end{align}
where we employed  the identity 
\begin{align}
	\min\p{x_1 , x_2} = \frac{x_1+x_2}{2} - \left| \frac{x_1-x_2}{2} \right| ~ .
\end{align}
Analogously, the fields  $\phi_{ n_\sigmavar ,  n_\psi}\p{x}$  are non-singular near the north pole of the spindle   if 
\begin{align}
	 n_\sigmavar  \geq \cf{ \frac{r + q_G \mN }{\nN} } = \cf{ \frac{  \pN }{\nN} } = -  \ff{ -  \frac{  \pN }{\nN} }  ~ ,
\end{align}
while they are non-singular at the south pole of the spindle if   
\begin{align}
	 n_\sigmavar  \geq  \cf{ \frac{r + q_G \mS }{\nS} } = \cf{ \frac{  \pS }{\nS} } = -  \ff{ -  \frac{  \pS }{\nS} }   ~ ,
\end{align}
which are constraints  that are both satisfied if 
\begin{align}
	 n_\sigmavar  \geq    - \frac12 \p{  \ff{ -  \frac{  \pN }{\nN} }    +  \ff{ -  \frac{  \pS }{\nS} } } + \frac12\left|     \ff{ -  \frac{  \pN }{\nN} }    -  \ff{ -  \frac{  \pS }{\nS} }  \right|   ~ ,
\end{align}
where we used the formula
\begin{align}
	\max\p{x_1 , x_2} = \frac{x_1+x_2}{2} + \left| \frac{x_1-x_2}{2} \right| ~ .
\end{align}
Therefore,   the eigenfunctions $\mathcal B_{ m_\sigmavar ,  m_\psi}$ and   $\Phi_{ n_\sigmavar ,  n_\psi}$ both   contribute to the one-loop determinant of the chiral-multiplet partition function via the infinite product 
\begin{align}\label{eq: zphiBantitwist}
	 Z^{\mathcal B, \Phi}_{\text{1-L}} &  = \prod_{m_\psi \in \mathbb Z} \prod_{ j \in \mathbb N }  \frac{m_\psi + \amf \p{j + 1 + \frac{|1-\mathfrak b| - \mathfrak c}{2} }   - r \gamma_R -  q_G \gamma_G - \alpha_3 }{m_\psi + \amf \p{j + \frac{|1-\mathfrak b| + \mathfrak c}{2} }   +r \gamma_R +  q_G \gamma_G  }   \nn\\
	 &  = e^{2 \pi \im \Psi_{\rm AT}} \frac{\p{ q^{1 + \frac12 |1-\mathfrak b|} y^{-1} ; q}_\infty}{\p{ q^{\frac12|1-\mathfrak b|} y ; q}_\infty}   ~ ,
\end{align}
where the overall phase reads
\begin{align}
	\Psi_{\rm AT} = \frac{1}{4 \amf }\p{ 1 + \alpha_3 - \amf |1-\mathfrak b| }\comm{ \alpha_3 + 2 q_G \gamma_G + r \gamma_R + \amf\p{\mathfrak c - 1} } ~ . 
\end{align}
If we choose $\alpha_3 = - 1$, corresponding to anti-periodic Killing spinors on the spindle, then the phase factor simplifies  and the one-loop determinant becomes
\begin{align}
	 Z^{\mathcal B, \Phi}_{\text{1-L}} = \p{- y^{-1} \sqrt q }^{\frac12|1-\mathfrak b|}\frac{\p{ q^{1 + \frac12 |1-\mathfrak b|} y^{-1} ; q}_\infty}{\p{ q^{\frac12|1-\mathfrak b|} y ; q}_\infty}   ~ .
\end{align}
Furthermore,  if  $\nN=\nS=1$, then  
\begin{align}
	& \mathfrak b = 1 - q_G \mathfrak m   = 1 - q_G \flux_G   ~ , & \mathfrak c = 0  ~ , \nn\\
	& 2 \rfug = - \alpha_3 + \amf   ~ , & y = \p{-1}^{r \alpha_3}  q^{r/2} \, e^{2 \pi \im q_G \gfug} ~ ,
\end{align}
and up to overall factors we recover the generalized superconformal index for a chiral multiplet \cite{Kapustin:2011jm},
\begin{align}
	 Z^{\mathcal B, \Phi}_{\text{1-L}}|_{n_\pm = 1} 	 & = \p{ \p{-1}^{1-r }   q^{\frac{1-r}2} e^{-2 \pi \im q_G \gfug} }^{|q_G \mathfrak m|/2}\frac{\p{ \p{-1}^{r }   q^{ \frac12 |q_G \mathfrak m| + 1 - \frac{r}{2} }  e^{ - 2 \pi \im q_G \gfug}  ; q}_\infty}{\p{ \p{-1}^{r }  q^{\frac12|q_G \mathfrak m| + \frac{r}2 }    e^{2 \pi \im q_G \gfug} ; q}_\infty}   ~ .
\end{align}
In \cite{Dimofte:2011py} it was pointed out that the appearence of absolute values of fluxes into the index is spurious, meaning that the replacements   
\begin{align}\label{eq: removeabsvalflux}
& |q_G \mathfrak m | \to - q_G \mathfrak m ~ , & |1-\mathfrak b | \to \p{\mathfrak b - 1 } ~ ,
\end{align}
should be applied to $Z^{\rm CM}_{\text{1-L}}$. The described substitutions correspond to the absence of   regularity constraints on the eigenfunctions  $\mathcal{B}_{m\sigma, m_\psi}$ and $\Phi_{n_\sigma, n_\psi}$ at the north and south pole of the spindle. This approach becomes apparent when computing $Z^{\rm CM}_{\text{1-L}}$ through the index theorem, wherein calculations  are performed separately in patches on  $\spindle$. Notably, this method avoids generating absolute values in the one-loop determinants,  as shown  e.g. in \cite{Drukker:2012sr}.

In summary, the one-loop determinant for a chiral multiplet with weight $\rho$ in a  representation $\mathfrak R_G$ of a gauge group $G$ on  anti-twisted $\spindle \times S^1$, reads
\begin{align}\label{eq: cm1loopspindles1antitwist}
	 Z^{\rm CM}_{\text{1-L}} =  \prod_{\rho \in \mathfrak R_G}  \p{- y^{-\rho} \sqrt q }^{\frac12\p{\mathfrak b-1}}\frac{\p{ q^{ \frac12 \p{\mathfrak b+1}} y^{-\rho} ; q}_\infty}{\p{ q^{\frac12\p{\mathfrak b-1}} y^\rho ; q}_\infty}   ~ . 
\end{align}
Finally,   the identities enjoyed by $q$-Pochhammer symbols allows for the derivation of a unique  expression for $Z^{\rm CM}_{\text{1-L}}$ valid for  both topologically twisted and anti-twisted $\spindle\times S^1$:
\begin{align}\label{eq: cm1loopspindles1twistantitwist}
	 Z^{\rm CM}_{\text{1-L}} =  \prod_{\rho \in \mathfrak R_G}  \p{- y^{-\rho}  }^{\frac14\p{2\mathfrak b-1 + \st }} q ^{\frac18\p{1-\st}\p{\mathfrak b-1}}  \frac{\p{ q^{ \frac12 \p{\mathfrak b+1}} y^{-\rho} ; q}_\infty}{\p{ q^{\frac{\st}{2}\p{\mathfrak b-1}} y^{-\st \rho } ; q}_\infty}   ~ , 
\end{align}
which for $G=U(1)$ coincides with  the expression  reported in \cite{Inglese:2023wky}. The   one-loop determinant of a vector multiplet in a gauge group $G$ can be obtained from (\ref{eq: cm1loopspindles1twistantitwist}) after setting $r=2$ and $\mathfrak R_G={\rm adj}\, G$.


\subsubsection{Equivariant  index theorem on orbifolds}

One-loop determinants of supersymmetric partition functions can also be obtained from equivariant index theory \cite{Pestun:2007rz}. For instance, in the case of a chiral multiplet on a manifold $\mathscr M$ and of charge   $r$ with respect to the $U(1)_R$  $R$-symmetry  and  charge $q_G$ with respect to a gauge or flavour group $U(1)_G$,  the eigenvalues contributing to  $Z^{\rm CM}_{\text{1-L}} $ are encoded by the formula
\begin{align}\label{eq: equivariantindex}
	 {\rm ind}\p{ L_{\widetilde P} ; \widehat g} = \sum_{p \in \mathscr M^{\widehat g} } \frac{{\rm tr}_{\, \Gamma_0}\widehat g  - {\rm tr}_{ \, \Gamma_1 }\widehat g  }{\det\p{ 1 - J_p }} ~ ,
\end{align}
which is the index of the   operator $L_{\widetilde P}$ with respect to the  action of the group element $\widehat g  =  \exp\p{ - \im \eqp \delta^2 }$    induced by the square of the supersymmetry variation $\delta^2$ and tuned by the equivariant parameter $\eqp$.   The summation in (\ref{eq: equivariantindex}) spans across the fixed-point set $\mathscr M^{\widehat g}$ of the transformation $\widehat g$ on the manifold $\mathscr M $. At each fixed point $p\in \mathscr M$, the group element $\widehat g$ induces an action on the manifold $\mathscr M$, thereby effecting a mapping from the coordinate $z_p$ to $z'_p$. This transformation generates a non-trivial Jacobian denoted as $J_p = \partial z'_p/\partial z_p$,   appearing in the denominator of (\ref{eq: equivariantindex}). In addition, $\widehat g$ acts upon both $\Gamma_0$ and $\Gamma_1$, where $\Gamma_0$ represents the space of sections of  the $U(1)_R \times U(1)_G$-valued line bundle $\linbun$ over $\mathscr M$, while $\Gamma_1$ is the image of $\Gamma_0$ under   by $L_{\widetilde P}$. These combined contributions   yield the numerator of (\ref{eq: equivariantindex}). In the case of our interest  $\mathscr M = \spindle \times S^1$ and the neighborhoods $\mathcal U_p$ covering the fixed points $p\in \mathscr M^{\widehat g}$ are isomorphic to $\mathbb C/\mathbb Z_{n_p}$. In order for the computation to  accomodate such new structures, the orbifold version of  (\ref{eq: equivariantindex}) is required.  The fixed point formula in the framework of the equivariant orbifold index theorem is expressed as follows:
\begin{align}\label{eq: equivariantindexorb}
	 {\rm ind}_{\rm orb}\p{ L_{\widetilde P} ; \widehat g} = \sum_{p \in \mathscr M^{\widehat g} } \frac1{n_p } \sum_{ \widehat w \in \mathbb Z_{n_p} }   \frac{{\rm tr}_{\, \Gamma_0}\p{\widehat w \, \widehat g}  - {\rm tr}_{ \, \Gamma_1 }\p{\widehat w\, \widehat g}  }{\det\p{ 1 - W_p J_p }} ~ ,
\end{align}
where $W_p$ is the Jacobian resulting from the action of $\widehat w$ on the coordinates $z_p$ within $\mathcal U_p$. The elements $\widehat w \in \mathbb Z_p$ act on both sections and coordinates akin to the action of $\widehat g$, but their action is weighted by the roots of unity $w_p^\ell = e^{2 \pi \im \, \ell /n_p}$, with $\ell=0,\dots, \p{n_p - 1}$,  rather than by the equivariant parameter $\eqp$. Specifically,  on $\spindle \times S^1$ we have
\begin{align}
	 \delta^2 & = - 2 \im \p{ L_K + \mathcal G_{\Phi_G} } = - 2 \im \p{ \mathcal L_K - \im q_R \Phi_R - \im q_G \Phi_G }  ~ , \nn\\
	 \mathcal L_K & =  \knorm\p{ \amf \partial_\sigmavar + \partial_\psi }  ~ , 
\end{align}
where $\p{\Phi_R, \Phi_G}$ are to be evaluated on the BPS locus. The operator $\widehat g$ factorizes  as  $\widehat g   = \widehat g_\spindle \widehat g_{S^1}   $, with  
\begin{align}
	 \widehat g_\spindle &   =  \exp\comm{ - 2 t \p{ \knorm  \amf \partial_\sigmavar - \im q_R \Phi_R - \im q_G \Phi_G } }  ~ , \nn\\
	 \widehat g_{S^1} & = \exp\p{ - 2 t  \knorm   \partial_\psi }    ~ , 
\end{align}
being  group actions corresponding to the spindle $\spindle$ and to the circle $S^1$, respectively. In particular,   the  fixed-point set $\mathscr M^{\widehat g_{S^1}}$ is empty as $\widehat g_{S^1}$ acts freely on $S^1$.  Instead,   $\mathscr M^{\widehat g_\spindle}$ is non-trivial and it contains the poles of the spindle at $x=\pm1$. The actions of    $\widehat g_\spindle$ and $\widehat w$ upon the complex coordinates   defined in (\ref{eq: ccsnp}) and (\ref{eq: ccssp}) is  
\begin{align}
	& \widehat g_\spindle \zN  = \qN \zN ~ ,  & \qN = \exp\p{2 \im \eqp \knorm \amf/\nN} ~ ,  \nn\\
	  & \widehat g_\spindle \zS  = q_-^{-1} \zS  ~ ,  & \qS = \exp\p{2 \im \eqp \knorm \amf/\nS} ~ , \nn\\
	  & \widehat w_+ \zN  = \wN \zN ~ , & \wN = \exp\p{2 \pi \im /\nN} ~ ,  \nn\\
	  & \widehat w_- \zS  = w_-^{-1} \zS  ~ , &  \wS = \exp\p{2 \pi \im /\nS} ~ .
\end{align}
 The form of $L_{\widetilde P}$ as well as the action of $\widehat g_\spindle$ and $\widehat w$ on sections in $\Gamma_0$ and $\Gamma_1$ exhibit slight differences for twisted and anti-twisted spindles. Consequently, we analyze these two cases separately.
 

\paragraph{Topological twist.} 

In computing the contribution to the index formula (\ref{eq: equivariantindexorb})  coming from the north and south pole of the spindle we work     with   1-form fields whose components on $\spindle$ vanish at the origin of the   patches $\UN$ and $\US$. This makes the calculation simple as  $L_{\widetilde P}$ becomes a pure differential operator, for instance. Especially, the component $\sigmavar$ of the   $R$-symmetry field vanishes at the origin of $\UN$ and $\US$ if 
\begin{align}
	& \alpha_2|_{\UN}  = \frac1\nN ~ ,  &  \alpha_2|_{\US} = - \frac1\nS ~ .
\end{align}
Analogously, the  component $\ag_\sigmavar$ of the flavour field (\ref{eq: bpsgaugeconnectionshifted})    vanishes at $x=\pm1$ if
\begin{align}
	& \beta_2|_{\UN}  = - \frac\mN\nN ~ ,  &   \beta_2|_{\US} = - \frac\mS\nS ~ ,  
\end{align}
in the patches $\UN$ and $\US$ , respectively. In this setting,  
\begin{align}
	& L_{\widetilde P}|_{\UN} = - 2 \im \partial_+ ~ , & L_{\widetilde P}|_{\US} =  2 \im \partial_- ~ ,
\end{align}
where $\overline\partial_{\pm} $ is the complex conjugate of $\partial_{\pm} = \partial_{z_\pm}$.  In particular, the operators $L_{\widetilde P}|_{\UN}$ and $L_{\widetilde P}|_{\US}$ are transversally  elliptic with respect to the free action of $\widehat g_{S^1} \in U(1)$. Then, according to  \cite{Atiyah:1974obx},  the index of $L_{\widetilde P}|_{\UN} $ on $\spindle \times S^1$ is given by the  weighted   sum of the index on the quotient $\p{\spindle \times S^1}/S^1 \cong \spindle$  over all  irreducible representations of $U(1)$, where   the weights are the characters $\chi_{\mathfrak R_{U(1)}}\p{\widehat g_{S^1}} = \chi_{n}\p{\widehat g_{S^1}} = e^{- 2 \im \eqp \knorm n} $ with  $n$ being  the label of the representative $\phi_n \in \Gamma_0 $, where the latter is the space of sections of the line orbibundle 
\begin{align}
	\linbun=\mathcal O(-\mathfrak p)=\mathcal O(-\mathfrak p) = \mathcal O( - \nN \pS + \nS \pN  ) = \mathcal O\p{ - \p{r/2}\p{\nN + \nS} - q_G \mathfrak m } ~ .
\end{align}
The operator $\widehat g_{\spindle}$ acts  as
\begin{align}
	 \widehat g_\spindle|_{\UN} \phi_n & = \xi \,    \qN^{- q_G \mN  + \p{r/2} } \phi_n = \xi \,    \qN^{  -  \pN }  \phi_n     ~ , \nn\\
	 \widehat g_\spindle|_{\US} \phi_n & = \xi \,    \qS^{- q_G \mS -\p{r/2}  } \phi_n = \xi \,    \qS^{  -  \pS }  \phi_n   ~ , \qquad  \xi = e^{2 \im \eqp \knorm\comm{ \p{r/2}\alpha_3 + q_G \varphi_G } } ~ .
\end{align}
Moreover, the action of $\widehat g_{\spindle}$ on sections in the image $\Gamma_1$ reads
\begin{align}
	 \widehat g_\spindle|_{\UN} \partial_+\phi_n &   = \xi \,    \qN^{  -  \pN - 1 }  \partial_+  \phi_n     ~ , \nn\\
	 \widehat g_\spindle|_{\US} \partial_- \phi_n &  = \xi \,    \qS^{  -  \pS + 1 }  \partial_-  \phi_n   ~ .
\end{align}
Combining all these ingredients in the fixed point formula (\ref{eq: equivariantindexorb})  yields
\begin{align}
	 {\rm ind}_{\rm orb}\p{ L_{\widetilde P} ; \widehat g} & = \sum_{n\in\mathbb Z} {\rm ind}_{\rm orb}\p{ L_{\widetilde P} ; \widehat g_\spindle} \chi_n\p{\widehat g_{S^1}} \xi ~ , \nn\\
	 {\rm ind}_{\rm orb}\p{ L_{\widetilde P} ; \widehat g_\spindle} & = \frac1\nN \sum_{j=0}^{\nN-1} \frac{\wN^{-j\pN} \qN^{-\pN} }{1-\wN^j \qN} +  \frac1\nS \sum_{j=0}^{\nS-1} \frac{\wS^{-j\pS} \qS^{-\pS} }{1-\wS^{-j} \qS^{-1}} ~ ,
\end{align}
where we factored out the  flavour fugacity $\xi$. The identities
\begin{align}
	& \frac1{n_p}\sum_{j=0}^{n_p-1}\frac{w_p^{j \nu_p }}{1 - w_p^j \, q_p} = \frac1{n_p}\sum_{j=0}^{n_p-1}\frac{w_p^{-j \nu_p }}{1 - w_p^{-j} \, q_p} = \frac{q_p^{ \rem{ -  \nu_p}{n_p} }}{1 -   q_p^{n_p} }  ~ ,  & \frac{\bullet}{n_p} = \ff{\frac{\bullet}{n_p} } + \frac{\rem{\bullet}{n_p} }{n_p}   ~ , 
\end{align}
imply 
\begin{align}\label{eq: eqindorbspindles1twist}
	{\rm ind}_{\rm orb}\p{ L_{\widetilde P} ; \widehat g} & = \sum_{n\in\mathbb Z} {\rm ind}_{\rm orb}\p{ L_{\widetilde P} ; \widehat g_\spindle} \chi_n\p{\widehat g_{S^1}} \xi ~ , \nn\\
	  {\rm ind}_{\rm orb}\p{ L_{\widetilde P} ; \widehat g_\spindle}  & =  \frac{  \qe^{- \ff{\pN/\nN}} }{1-  \qe} +   \frac{  \qe^{\ff{-\pS/\nS} } }{1-  \qe^{-1}}   =  \frac{  \qe^{- \ff{\pN/\nN}}  -   \qe^{1+\ff{-\pS/\nS} } }{1-  \qe}  ~ ,
\end{align}
with $\qe = \qN^\nN  =\qS^\nS $. If
\begin{align} 
	- \ff{\pN/\nN} = 1+\ff{-\pS/\nS} \iff \mathfrak b = 0  ~ ,
\end{align}
then ${\rm ind}_{\rm orb}\p{ L_{\widetilde P} ; \widehat g_\spindle}   = 0$, where $\mathfrak b = \p{1 + \deg\linbun}$ defined in  (\ref{eq: bcpfug}) naturally appeared. Instead, if
\begin{align}
	- \ff{\pN/\nN} < 1+\ff{-\pS/\nS} \iff \mathfrak b \geq 1  ~ ,
\end{align}
then
\begin{align}
	  {\rm ind}_{\rm orb}\p{ L_{\widetilde P} ; \widehat g_\spindle}  = \qe^{- \ff{\pN/\nN}}   \frac{  1  -   \qe^{\mathfrak b } }{1-  \qe} = \qe^{- \ff{\pN/\nN}}   \p{ 1 + \qe + \dots + \qe^{\mathfrak b - 1} } ~ , 
\end{align}
where the overall factor $\qe^{- \ff{\pN/\nN}}  $ is the  \emph{lift of the equivariant action} \cite{Pestun:2016qko} at the north pole of the spindle. In fact, given that $L_{\widetilde P}$ equals   the Dolbeault operator $\partial$ in the vicinity of any fixed point, ${\rm ind}_{\rm orb}\p{ L_{\widetilde P} ; \widehat g_\spindle}$ represents the equivariant orbifold index of $\partial$ on $\mathbb{WCP}^1_{\comm{\nN, \nS}}$ in   presence of the line bundle $\linbun$. This index assigns a monomial $\qe^\ell$ to each element within the basis of  anti-holomorphic sections of $\linbun$ in the kernel  of $\partial$, where $\ell\in \mathbb N$,  $\qe^0=1$ corresponds to  the constant section and the polynomial's degree is $\p{\mathfrak b - 1} = \deg \linbun$. This   generalizes  what occurs in the case of the equivariant index of the Dolbeault operator on manifolds, where, modulo a sign, the degree of a line bundle equals the first Chern class of its representatives. Especially, in the non-equivariant limit $\qe\to1$ the index becomes
\begin{align}
{\rm ind}_{\rm orb}\p{ L_{\widetilde P} } = {\rm ind}_{\rm orb}\p{ \partial } = \mathfrak b = 1 + \deg \linbun ~ ,
\end{align}
reproducing  the Riemann-Roch-Kawasaki theorem for orbifolds of genus zero \cite{Closset:2018ghr}. The anti-holomorphic sections that contribute to a non-trivial ${\rm ind}_{\rm orb}\left( L_{\widetilde P} ; \widehat g_\spindle \right)$ for $\mathfrak{b} \geq 1$ exhibit the same  charges $\left(q_G, r\right)$ as those of  the eigenfunctions $\Phi_{n_\sigma, n_\psi} \in \ker L_{\widetilde P}$ reported in (\ref{eq: kerLPteigenfunctions}).  We can check that this is no coincidence by rewriting the index as
\begin{align}
	 {\rm ind}_{\rm orb}\p{ L_{\widetilde P} ; \widehat g}|_{\mathfrak b\geq1} & = \sum_{n\in\mathbb Z}   e^{- 2 \im \eqp \knorm n}    e^{2 \im \eqp \knorm\comm{ \p{r/2}\alpha_3 + q_G \varphi_G } } \sum_{\ell=- \ff{\pN/\nN}}^{\ff{-\pS/\nS}}   \qe^\ell  \nn\\
	 & = \sum_{n\in\mathbb Z} \sum_{\ell=- \ff{\pN/\nN}}^{\ff{-\pS/\nS}}   e^{2 \im \eqp \knorm\comm{ - n +  \p{r/2}\alpha_3 + q_G \varphi_G +  \amf \ell } }    ~ , 
\end{align}
and then exploiting the rule 
\begin{align}\label{eq: indextodeterminant}
	{\rm ind}\p{\mathscr D;   e^{ - \im \eqp \delta^2 } } = \sum_{j} d_j \, e^{ - \im \eqp  \lambda_j } \qquad \to \qquad  Z_{\rm ind} = \prod_j  \lambda_j^{ - d_j } ~ , 
\end{align}
relating the index of a differential operator $\mathscr D$ to the determinant $Z_{\rm ind}$, 
with $j$ being a multi-index,  $\lambda_j$ the $j$-th eigenvalue   of $\delta^2$ and $d_j$ the degeneracy of $\lambda_j$.  Applying the  prescription (\ref{eq: indextodeterminant})  to ${\rm ind}_{\rm orb}\p{ L_{\widetilde P} ; \widehat g}|_{\mathfrak b\geq1}$ provides the infinite product 
\begin{align}
	  Z_{\rm ind}|_{\mathfrak b\geq1}  & = \prod_{n\in\mathbb Z} \prod_{\ell = - \ff{\pN/\nN}}^{\ff{-\pS/\nS}}   \comm{  2  \knorm \p{  n -  \p{r/2}\alpha_3 - q_G \varphi_G -  \amf \ell } }^{-1}      \nn\\
	 & = \prod_{n\in\mathbb Z} \prod_{j = -\ff{-\pS/\nS} }^{  \ff{\pN/\nN} }   \comm{  2   \knorm \p{  n +  \amf j  -  \frac{r}{2}\alpha_3 - q_G \varphi_G } }^{-1}   ~ ,
\end{align}
exactly matching $Z^{\Phi}_{\text{1-L}}$ in (\ref{eq: zphitwist}) up to an  irrelevant normalization factor. Therefore, the  index  ${\rm ind}_{\rm orb}\left( L_{\widetilde P} ; \widehat g_\spindle \right)$ for $\mathfrak{b} \geq 1$ precisely encodes the eigenvalues associated with $\Phi_{n_\sigma, n_\psi}$.

Finally, if
\begin{align}
	- \ff{\pN/\nN} > 1+\ff{-\pS/\nS} \iff \mathfrak b \leq -1  ~ ,
\end{align}
then
\begin{align}
	  {\rm ind}_{\rm orb}\p{ L_{\widetilde P} ; \widehat g_\spindle} & = \qe^{\ff{-\pS/\nS}}   \frac{     \qe^{1-\mathfrak b } - \qe }{1-  \qe} = - \qe^{ 1 + \ff{-\pS/\nS}}   \p{ 1 + \qe^{2} + \dots + \qe^{ - \mathfrak b - 1 } }  \nn\\
	 & = - \qe^{ - \ff{\pS'/\nS}}   \p{ 1 + \qe^{2} + \dots + \qe^{ \deg \linbun' } } ~ , 
\end{align}
with
\begin{align}
	 \pN' & = \pN + 1 = q_G \mN - \frac{r-2}2~ ,  \nn\\
	   \pS' & = \pS -1 = q_G \mS + \frac{r-2}2 ~ , \nn\\
	 \linbun' & = \mathcal O\p{-\nN \pS' + \nS \pN'} = \mathcal O\comm{  q_G \mathfrak m  +   \frac{r-2}2\p{\nN + \nS} }  ~ , \nn\\
	 -\p{\mathfrak b + 1} & =    \ff{-\frac{\pN'}{\nN} } +   \ff{ \frac{\pS'}{\nS} } = \deg \linbun'   ~ .
\end{align}
The overall factor $\qe^{ \ff{\pS'/\nS}}  $ is  the   lift of the equivariant action  at the south pole of    $\spindle$. From $\linbun'$ we see that  the  sections counted by ${\rm ind}_{\rm orb}\left( L_{\widetilde P} ; \widehat g_\spindle \right)$ for $\mathfrak{b} \leq - 1$ have  charges $\p{q_G, r-2}$, which are  those of the Grassmann-odd (fermionic) eigenfunctions $\mathcal B_{m_\sigma, m_\psi} \in \ker L_{ P}$ written in (\ref{eq: kerLPeigenfunctions}), where $L_P\sim \overline \partial$ at the fixed points of $\widehat g_\spindle$. Indeed, by writing 
\begin{align}
	 {\rm ind}_{\rm orb}\p{ L_{\widetilde P} ; \widehat g}|_{\mathfrak b\leq-1} & = -  \sum_{n\in\mathbb Z}   e^{- 2 \im \eqp \knorm n}    e^{2 \im \eqp \knorm\comm{ \p{r/2}\alpha_3 + q_G \varphi_G } } \sum_{\ell=  1 + \ff{-\pS/\nS} }^{ - 1 - \ff{\pN/\nN}}   \qe^\ell  \nn\\
	 & = - \sum_{n\in\mathbb Z} \, \sum_{\ell=  1 + \ff{-\pS/\nS} }^{ - 1 - \ff{\pN/\nN}}    e^{2 \im \eqp \knorm\comm{ - n +  \p{r/2}\alpha_3 + q_G \varphi_G +  \amf \ell } }    ~ , 
\end{align}
and applying (\ref{eq: indextodeterminant}), we obtain 
\begin{align}
	 Z_{\rm ind}|_{\mathfrak b\leq-1} & =  \prod_{n\in\mathbb Z} \, \prod_{\ell=  1 + \ff{-\pS/\nS} }^{ - 1 - \ff{\pN/\nN}}    {2 \knorm\p{  n -  \frac{r}{2}\alpha_3 - q_G \varphi_G -  \amf \ell } }     \nn\\
	 & =  \prod_{n'\in\mathbb Z} \, \prod_{j=  1 + \ff{\pN/\nN} }^{ -   1 - \ff{-\pS/\nS} }    {2 \knorm\p{  n' -  \frac{r-2}{2}\alpha_3 - q_G \varphi_G +  \amf j } }     ~ , 
\end{align}
which is identical to $ Z^{\mathcal B}_{\text{1-L}}$ in (\ref{eq: zBtwist}), up to a  normalization. As the rule (\ref{eq: indextodeterminant}) maps ${\rm ind}_{\rm orb}\p{ L_{\widetilde P} ; \widehat g}|_{\mathfrak b=0}$   to $Z_{\rm ind}|_{\mathfrak b=0}=1$, we can conclude that the method of unpaired eigenvalues and the index theorem lead to the same result for the partition function of a chiral multiplet on topologically twisted $\spindle \times S^1$.


\paragraph{Anti-twist.}

In  the case of anti-twisted $\spindle \times S^1$ the   $\sigmavar$-component of the   $R$-symmetry background field vanishes at the poles of the spindle if 
\begin{align}
	& \alpha_2|_{\UN}  = - \frac1\nN ~ ,  &  \alpha_2|_{\US} = - \frac1\nS ~ ,
\end{align}
while   flavour field (\ref{eq: bpsgaugeconnectionshifted}) behaves as in the case of the topological twist. The pairing operator $L_{\widetilde P}$ is again transversally  elliptic with respect to the free action of $\widehat g_{S^1} \in U(1)$, and near the poles of $\spindle$ assumes the value  
\begin{align}
	& L_{\widetilde P}|_{\UN} = - 2 \im \overline  \partial_+ ~ , & L_{\widetilde P}|_{\US} =  2 \im \partial_- ~ .
\end{align}
 Repeating  the   methodology  employed in the case of  the topologically twisted $\spindle \times S^1$ yields 
\begin{align}\label{eq: eqindorbspindles1antitwist}
	 {\rm ind}_{\rm orb}\p{ L_{\widetilde P} ; \widehat g} & = \sum_{n\in\mathbb Z} {\rm ind}_{\rm orb}\p{ L_{\widetilde P} ; \widehat g_\spindle} \chi_n\p{\widehat g_{S^1}} \xi ~ , \nn\\
	 {\rm ind}_{\rm orb}\p{ L_{\widetilde P} ; \widehat g_\spindle} & =  \frac{  \qe^{ \ff{-\pN/\nN}} }{1-  \qe^{-1}} +   \frac{  \qe^{\ff{-\pS/\nS} } }{1-  \qe^{-1}}  =  \frac{  \qe^{ \ff{-\pN/\nN}} }{1-  \qe^{-1}} -   \frac{  \qe^{1+\ff{-\pS/\nS} } }{1-  \qe}    ~   .
\end{align}
Comparing (\ref{eq: eqindorbspindles1twist}) against (\ref{eq: eqindorbspindles1antitwist}) we see that the term in ${\rm ind}_{\rm orb}\p{ L_{\widetilde P} ; \widehat g_\spindle}$  involving $\pS$ is independent of the  $R$-symmetry twist present on $\spindle\times S^1$. The reason is that all these changes between twist and anti-twist case occurr  only in the northern patch $\UN$ of the spindle. Expanding ${\rm ind}_{\rm orb}\p{ L_{\widetilde P} ; \widehat g}$ in powers of $\qe$ gives
\begin{align}
	 {\rm ind}_{\rm orb}\p{ L_{\widetilde P} ; \widehat g} & =  \sum_{n\in\mathbb Z}      e^{2 \im \eqp \knorm\comm{ - n +  \p{r/2}\alpha_3 + q_G \varphi_G } }  \sum_{\ell\in \mathbb N}   \p{ \qe^{-\ell + \ff{-\pN/\nN} } - \qe^{\ell +1 + \ff{-\pS/\nS} } }     ~ , 
\end{align}
 which, according to the rule (\ref{eq: indextodeterminant}), corresponds to the infinite product
 \begin{align}
 	 Z_{\rm ind} & = \prod_{n\in \mathbb Z}\prod_{\ell \in \mathbb N}\frac{ - n +  \p{r-2}\p{\alpha_3/2} + q_G \varphi_G  + \amf\p{ \ell + 1 + \ff{- \pS/\nS } } 	}{  - n +  \p{r/2}\alpha_3 + q_G \varphi_G  + \amf\p{ - \ell + \ff{ - \pN/\nN } } }  \nn\\
 	 & = \prod_{n\in \mathbb Z}\prod_{\ell \in \mathbb N}\frac{ - n + \amf\p{ \ell + \frac{\mathfrak b + 1 - \mathfrak c }2 } - r\, \gamma_R - q_G \gamma_G - \alpha_3	}{  - n + \amf\p{ \ell + \frac{\mathfrak b - 1 + \mathfrak c }2 } + r \, \gamma_R + q_G \gamma_G   } ~   . 
	\end{align}
	The latter perfectly  reproduces the one-loop determinant of a chiral multiplet on   anti-twisted $\spindle\times S^1$   in (\ref{eq: zphiBantitwist}), already combined with the prescription   (\ref{eq: removeabsvalflux}), as anticipated.  
	
	
\subsubsection{One-loop determinants of 2d gauge theories on \texorpdfstring{$\spindle$}{Sigma}}
	
By index theorem we are also able to compute the one-loop determinants of  two-dimensional $\mathcal N=(2,2)$  chiral and vector multiplets on the spindle. Specifically, the index of the operator $L_{\widetilde P}$ acting on $\text{spindle}$ under any  $R$-symmetry twist yields
\begin{align} 
	I_\spindle = {\rm ind}_{\rm orb}\p{ L_{\widetilde P} ; \widehat g_\spindle} \xi_{\rm 2d}   = \p{ \frac{  \qe^{- \st  \ff{ \st \, \pN/\nN}} }{1-  \qe^{\st}} -   \frac{  \qe^{1+\ff{-\pS/\nS} } }{1-  \qe } }\xi_{\rm 2d}  ~ ,
\end{align}
with $\xi_{\rm 2d} = e^{2 \im \eqp \knorm q_G \varphi_G  }$.  Expanding the function $I_\spindle$ in a power series of $\qe$ gives
\begin{align} 
	I_\spindle & =   \sum_{\ell\in \mathbb N}  e^{2 \im \eqp \knorm q_G \varphi_G  } \p{   \qe^{\st \ell - \st  \ff{ \st \, \pN/\nN}}   -   \qe^{ \ell + 1+\ff{-\pS/\nS} }   }    \nn\\
	 & =   \sum_{\ell\in \mathbb N} \acomm{ e^{2 \im \eqp \knorm \comm{ q_G \varphi_G + \amf\p{\st \ell - \st  \ff{ \st \, \pN/\nN} } }  }         - e^{2 \im \eqp \knorm \comm{ q_G \varphi_G + \amf\p{  \ell + 1+\ff{-\pS/\nS} } }  }  }       ~ ,
\end{align}
which translates into the following infinite product:
\begin{align}\label{eq: zindspindleanytwist}
	Z_{\spindle}^{(\st)} = \prod_{\ell\in\mathbb N}\frac{  q_G \varphi_G + \amf\p{  \ell + 1+\ff{-\pS/\nS} }}{ \st q_G \varphi_G + \amf\p{  \ell -    \ff{ \st \, \pN/\nN} }} = \prod_{\ell\in\mathbb N}\frac{    \amf\p{\ell+\frac{1+\mathfrak b - \mathfrak c}{2}} - q_i \gamma^{\rm 2d}_i  }{     \amf\p{\ell+\st\frac{1-\mathfrak b - \mathfrak c}{2}} - \st \, q_i \gamma^{\rm 2d}_i   } ~ ,
\end{align}
where
\begin{align}\label{eq: rsymandgaugefug2d}
	& \gamma_R^{\rm 2d} = \amf \frac{\chi_{-\st} }{4} ~ , & \gamma_G^{\rm 2d} =  \frac{\amf}{2}\p{ \frac{\mS}{\nS} + \frac{\mN}{\nN}  } - \varphi_G ~ ,
\end{align}
	 are the  $R$-symmetry and gauge fugacity in two dimensions. Then, a $\mathcal N=(2,2)$   chiral multiplet  in the representation $\mathfrak{R}_G$ contributes to the partition function via the one-loop determinant
	 \begin{align}
	Z_{\spindle}^{{\rm CM}, (\st)}  & = \prod_{\rho\in\mathfrak R_G} \prod_{\ell\in\mathbb N}\frac{    \amf\comm{\ell+\frac{1+\mathfrak b\p{\rho\p{\mathfrak m}} - \mathfrak c\p{\rho\p{\mathfrak m}}}{2}} - r \gamma^{\rm 2d}_R - \rho\p{ \gamma^{\rm 2d}_G}   }{     \amf\comm{\ell+\st\frac{1-\mathfrak b\p{\rho\p{\mathfrak m}} - \mathfrak c\p{\rho\p{\mathfrak m}}}{2}} - \st \, r \gamma^{\rm 2d}_R   - \st \, \rho\p{ \gamma^{\rm 2d}_G }  }  \nn\\
	& = \prod_{\rho\in\mathfrak R_G} \frac{\Gamma\acomm{  \frac{\st}{2}\comm{ 1-\mathfrak b\p{\rho\p{\mathfrak m}} - \mathfrak c\p{\rho\p{\mathfrak m}} } - \st\, \amf^{-1}\comm{ r \gamma^{\rm 2d}_R + \rho\p{ \gamma^{\rm 2d}_G } } } }{ \Gamma\acomm{\frac{1}{2}\comm{ 1+\mathfrak b\p{\rho\p{\mathfrak m}} - \mathfrak c\p{\rho\p{\mathfrak m}} } - \amf^{-1}\comm{ r  \gamma^{\rm 2d}_R + \rho\p{ \gamma^{\rm 2d}_G}  } } } ~ ,
\end{align}
while the contribution of a $\mathcal N=(2,2)$ vector multiplet in the adjoint representation reads
 \begin{align}
	Z_{\spindle}^{{\rm VM}, (\st)}  & =  \prod_{\alpha\in {\rm adj}_G} \left. \frac{\Gamma\acomm{  \frac{\st}{2}\comm{ 1-\mathfrak b\p{\alpha\p{\mathfrak m}} - \mathfrak c\p{\alpha\p{\mathfrak m}} } - \st\, \amf^{-1}\comm{ 2 \gamma^{\rm 2d}_R + \alpha\p{ \gamma^{\rm 2d}_G } } } }{ \Gamma\acomm{\frac{1}{2}\comm{ 1+\mathfrak b\p{\alpha\p{\mathfrak m}} - \mathfrak c\p{\alpha\p{\mathfrak m}} } - \amf^{-1}\comm{ 2  \gamma^{\rm 2d}_R + \alpha\p{ \gamma^{\rm 2d}_G}  } } } \right|_{r=2} ~ .
\end{align}
 The quantities $ Z_{\spindle}^{{\rm CM}, (\st)} $, $\gamma_R^{\text{2d}} $ and $\gamma_G^{\text{2d}} $ represent the $S^1$ reduction of the corresponding three-dimensional objects $ Z_{\text{1-L}}^{\rm CM}$, $\gamma_R $ and $\gamma_G  $  on $\spindle\times S^1$ reported in  (\ref{eq: cm1loopspindles1twistantitwist}) and (\ref{eq: bcpfug}). In particular, if $\nN=\nS=1$ and $\st=-1$, we obtain
	 \begin{align}
	Z_{S^2}^{{\rm CM}, (-1)} 
	& = \prod_{\rho\in\mathfrak R_G} \frac{\Gamma\p{   \frac{r}{2} -  \frac{\rho\p{\mathfrak m}}{2}   + \amf^{-1}  \rho\p{ \gamma^{\rm 2d}_G } }  }{ \Gamma\p{  1 - \frac{r}{2} -  \frac{\rho\p{\mathfrak m}}{2}  - \amf^{-1} \rho\p{ \gamma^{\rm 2d}_G}  } } ~ ,
\end{align}
in agreement with \cite{Benini:2012ui}. Instead, if $\nN=\nS=1$ and $\st=+1$ we find
\begin{align}
	Z_{S^2}^{{\rm CM}, (+1)}  
	& = \prod_{\rho\in\mathfrak R_G} \amf^{\mathfrak r - 1}\p{ 1 - \frac{\mathfrak r}{2} - \amf^{-1} \rho\p{\gamma_G^{\rm 2d}} }_{\mathfrak r-1} ~ ,
\end{align}
in accordance with \cite{Closset:2015rna}, where $\mathfrak r = \rho\p{\mathfrak m} + r$ is the effective $R$-charge and 
\begin{align}
	& \p{z}_m = \prod_{\ell=0}^{m-1}\p{\ell + z } = \Gamma\p{z+m}/\Gamma\p{z} ~ ,  
\end{align}
is the Pochhammer symbol.


\subsection{Canonical and effective classical terms}\label{eq: classcontr}

 \subsubsection{Canonical terms}

General supersymmetric gauge theories include Chern-Simons and Fayet-Iliopoulos terms. The canonical Chern-Simons action at level $\kcs$  for a non-Abelian vector multiplet in the adjoint representation of a gauge group $G$ is
\begin{align}
	& S_{\rm CS} = \frac{\im \, \kcs }{4 \pi } \int \dd^3 x \, {\rm e} \, {\rm tr}\comm{ \epsilon^{\mu \nu \rho}\p{ \ag_\mu \partial_\nu \ag_\rho - \frac{2 \im}{3} \ag_\mu \ag_\nu \ag_\rho } + 2 \im D \sigma + 2 \widetilde \lambda \lambda} ~ , & {\rm e} =  \epsilon_{x\sigmavar\psi} ~ ,
\end{align}
 which,  once evaluated on the BPS locus,  contributes to the classical part of the partition function as follows:
\begin{align}
	Z^{\rm CS}_{\rm class}|_{\rm BPS} = e^{- S_{\rm CS} }|_{\rm BPS}  = e^{2 \pi \im \, \kcs  \gamma_G \mathfrak f_G} = e^{2 \pi \im \, \kcs \, \gamma_G  \, \mathfrak m/\p{\nN \nS}}  ~ .
\end{align}
The latter can be generalized to a mixed Chern-Simons term at level $\kcs_{ij}$ that combines  two vector multiplets in the adjoint of  different gauge groups $G_{(i)}$ and $G_{(j)}$ with $i\neq j$:
\begin{align}
	& S_{\rm MCS} = \frac{\im \, \kcs_{ij} }{2 \pi } \int \dd^3 x \, e \,  \p{ \epsilon^{\mu \nu \rho} \ag_\mu^{(i)} \partial_\nu \ag_\rho^{(j)}  + \im D^{(i)}  \sigma^{(j)}  + \im D^{(j)}  \sigma^{(i)}   +  \widetilde \lambda^{(i)}  \lambda^{(j)}  +  \widetilde \lambda^{(j)}  \lambda^{(i)} } ~ ,  
\end{align}
yielding 
  \begin{align}
	Z^{\rm MCS}_{\rm class}|_{\rm BPS} =  e^{2 \pi \im \, \kcs_{ij} \p{  \gamma_{G_i}  \mathfrak f_{G_j}  +   \gamma_{G_j}  \mathfrak f_{G_i} }}  ~ .
\end{align}
The gauge symmetry can also be mixed with the  $R$-symmetry via an  $R$-symmetry-gauge Chern-Simons term, whose action reads \cite{Closset:2012ru}
  \begin{align}
	& S_{\rm RCS} = \frac{\im \, \kcs_R }{2 \pi } \int \dd^3 x \, e  \,\comm{ \epsilon^{\mu \nu \lambda} \ag_\mu \partial_\nu\p{ A_\lambda  -\frac12 V_\lambda } +  \im D \sigma +  \frac{\im \,\sigma}{4}\p{R + 2 V_\mu V^\mu + 2 H^2}} ~ ,
\end{align}
giving 
\begin{align}
	Z^{\rm RCS}_{\rm class}|_{\rm BPS} =  e^{2 \pi \im \, \kcs_R \p{  \gamma_{R} \mathfrak f_G  +  \gamma_{G} \mathfrak f_{R} }}  ~ ,
\end{align}
which depends on the twist  through the  $R$-symmetry fugacity $\gamma_R$ defined in (\ref{eq: bcpfug}) and the flux $\mathfrak f_R = \chi_\pm/2$. Finally, for any Abelian factor in a gauge group $G$,  a classical contribution to the partition function can also descend  from  Fayet-Iliopoulos terms with parameter $\zeta_{\rm FI}$ and action
  \begin{align}
	S_{\rm FI} & = \frac{  \zeta_{\rm FI} }{2 \pi } \int \dd^3 x \, e \p{  D - \ag_\mu V^\mu - \sigma H } ~ . 
\end{align}
In fact, thanks to the relation \cite{Closset:2012ru}  
\begin{align}
	& V^\mu = - \im \epsilon^{\mu \nu \lambda}\partial_\nu \mathcal C_\lambda   ~ ,
\end{align}
 the Fayet-Iliopoulos term can be interpreted as a mixed Chern-Simons term coupling  fields in an Abelian vector multiplet  to the background scalar $H$ and    the graviphoton field   $\mathcal C_\mu$. Hence,
\begin{align}
	Z^{\rm FI}_{\rm class}|_{\rm BPS} = e^{2 \pi \im \, \zeta_{\rm FI}\p{  \gamma_{\mathcal C} \mathfrak f_G  +  \gamma_{G} \mathfrak f_{\mathcal C} }}    ~ , 
\end{align}
where $\gamma_{\mathcal C}$ and $\mathfrak f_{\mathcal C}$ are the fugacity and the flux associated to $\mathcal C_\mu$, respectively.

 \subsubsection{Effective terms}
  
  In general,  the classical contributions obtained from evaluating on the BPS locus the canonical actions above are not invariant under large gauge transformations along $S^1$. For example, under a large gauge transformation acting as $\gamma_G \to \gamma_G + \ell $, with $\ell\in\mathbb N$,   the  classical part of the canonical Chern-Simons partition function term behaves as follows: 
  \begin{align}
	Z^{\rm CS}_{\rm class}|_{\rm BPS}  = e^{2 \pi \im \, \kcs \, \gamma_G  \, \mathfrak m/\p{\nN \nS}}  \qquad \to \qquad \widetilde Z^{\rm CS}_{\rm class}|_{\rm BPS}  = e^{2 \pi \im \, \kcs  \ell  \mathfrak m/\p{\nN \nS}} Z^{\rm CS}_{\rm class}|_{\rm BPS}   ~ .
\end{align}
In the case in which the spindle is a sphere,  then  $\nN=\nS=1$ and  the partition function exhibits invariance under large gauge transformations as $\widetilde Z^{\rm CS}_{\rm class}|_{\rm BPS}=Z^{\rm CS}_{\rm class}|_{\rm BPS}$. Conversely, for a general spindle $\spindle=\mathbb{WCP}^1_{\comm{\nN, \nS}}$,   the parameters $\p{\nN,\nS}$ are positive  coprime integers and $\widetilde Z^{\rm CS}_{\rm class}|_{\rm BPS}\neq Z^{\rm CS}_{\rm class}|_{\rm BPS}$ unless either the Chern-Simons level $\kcs$ or $\mathfrak m$ is a multiple of $\p{\nN \nS}$.

A similar circumstance arises in supersymmetric theories on the manifold with   boundary $D^2 \times S^1$, where $D^2$ represents a two-dimensional disk or hemisphere. The canonical Chern-Simons term lacks supersymmetry and gauge invariance on $D^2 \times S^1$. Restoring these properties requires taking into  account  degrees of freedom localized at the boundary $\partial\p{D^2 \times S^1}=T^2$, with $T^2=S^1 \times S^1$ being  a two-dimensional torus. A convenient method  for promptly deriving the correct expression for the effective Chern-Simons term on $D^2 \times S^1$ was elucidated in \cite{Beem:2012mb}, where it was demonstrated that  the partition function for a Chern-Simons term at level $\kcs=\pm1$ equals the collective partition function of two anomaly-free chiral multiplets $\p{\phi_{(1)}, \phi_{(2)}}$ coupled by a superpotential $W\p{\phi_{(1)}, \phi_{(2)}} = m_0 \phi_{(1)} \phi_{(2)}$, where $m_0$ encodes a mass. Such a procedure allowed to show that the Chern-Simons partition function on $\partial\p{D^2 \times S^1}=T^2$  is  given by the theta function $\theta\p{{\bf x}; {\bf q}} = \p{ - \sqrt {\bf q}/{\bf x} ; {\bf q} }_\infty  \p{ - \sqrt {\bf q}\,{\bf x} ; {\bf q} }_\infty$ depending on gauge and flavour fugacities ${\bf x}$, as well as on an equivariant parameter $\epsilon$  appearing in  the fugacity ${\bf q} = e^{  \epsilon}$ for the  angular momentum on $D^2$. In particular, gauge invariance of such an effective Chern-Simons term is ensured by the analytic properties of the Theta function.

We can apply the same procedure to the case of $\spindle \times S^1$. Two chiral multiplets   $\p{\phi_{(1)}, \phi_{(2)}}$ coupled by the aforementioned  superpotential $W\p{\phi_{(1)}, \phi_{(2)}} = m_0 \phi_{(1)} \phi_{(2)}$ have charges satisfying
\begin{align}
	& r^{(1)}+ r^{(2)} = 2 ~ ,  &  q_G^{(1)}+ q_G^{(2)} = 0  ~ ,   
\end{align}
which, in the case of the twisted spindle\footnote{Similar relations hold in the case of the anti-twisted spindle $\spindle$.}, imply 
\begin{align}
	& \mathfrak b^{(1)}+ \mathfrak b^{(2)} = 0 ~ ,    \qquad \mathfrak c^{(1)}+ \mathfrak c^{(2)} = - \chi_- ~ ,   & y_{(1)} y_{(2)} = 1 ~ ,
\end{align}
where  $\p{\mathfrak b^{(i)} , \mathfrak c^{(i)} ,  y_{(i)}  }$ are  $\p{\mathfrak b  , \mathfrak c  ,  y   }$ evaluated at $r = r^{(i)}$ and $q_G = q_G^{(i)}$ with $i=1,2$. In particular,  $y_{(1)} y_{(2)}=1$ follows from the constraint $ \gamma_R - \p{\amf \chi_-/4} \in \mathbb Z$. As in \cite{Beem:2012mb}, we consider anomaly-free chiral multiplets, whose partition function is obtained from $Z^{\rm CM}_{\text{1-L}}$ by removing the phase factors and considering the $q$-Pochhammer part only. For instance, in the case of  the topologically twisted $\spindle \times S^1$, we have
\begin{align}
	Z^{(2)} & = \frac1{\p{ q^{\p{1-\mathfrak b_2}/2} y_{(2)}^{-1} ; q}_{\mathfrak b_2} } = \frac1{\p{ q^{\p{1+\mathfrak b_1}/2} y_{(1)} ; q}_{-\mathfrak b_1} }  \nn\\
	& = \p{-y_{(1)}}^{\mathfrak b_1} \p{ q^{\p{1-\mathfrak b_1}/2} y_{(1)}^{-1} ; q}_{\mathfrak b_1} =  \p{-y_{(1)}}^{\mathfrak b_1} / {Z^{(1)}}   ~ ,
\end{align}
yielding
\begin{align}
	Z^{(1)} Z^{(2)}  =  \p{-y_{(1)}}^{\mathfrak b_1}    ~ ,
\end{align}
which,  according to \cite{Beem:2012mb}, is  the   partition function of the effective Chern-Simons term at level $\kcs = \pm1$. At  arbitrary level $\kcs$ we have
 \begin{align}\label{eq: zcseff}
	Z^{\rm CS}_{\rm eff}  =  \p{-y}^{\kcs \, \mathfrak b}    ~ .
\end{align}
Since we have fixed only the sum of the $R$-charges $\p{q_G^{(i)} , r^{(i)} }$ but not their specific values, $Z^{\rm CS}_{\rm eff}$   contains information on all global and gauge symmetries affecting the chirals  $\p{\phi_{(1)}, \phi_{(2)}}$. For instance, if $\nN=\nS=1$, the spindle $\spindle$ boils down to a two-sphere and 
\begin{align}
	& Z^{\rm CS}_{\rm eff}|_{\nN = \nS = 1}  = \p{ -  e^{ - \im \pi r \alpha_3 }e^{2 \pi \im q_G \gamma_G } }^{ \kcs\p{ 1 - r - q_G \mathfrak m} }    ~ .
\end{align}
In particular, the latter manifestly depends on the spin structure on $\spindle$ via $\alpha_3 \in \mathbb Z$. Indeed, the periodic spin structure corresponds to even $\alpha_3$, giving
\begin{align}
	 Z^{\rm CS (+)}_{\rm eff}|_{\nN = \nS = 1}  = \p{ -   e^{2 \pi \im q_G \gamma_G } }^{ \kcs\p{ 1 - r - q_G \mathfrak m} }     ~ ;
\end{align}
whereas the anti-periodic spin structure requires  odd $\alpha_3$, yielding 
\begin{align}
	& Z^{\rm CS (-)}_{\rm eff}|_{\nN = \nS = 1}  = \p{  \p{-}^{1-r} e^{2 \pi \im q_G \gamma_G } }^{ \kcs\p{ 1 - r - q_G \mathfrak m} }    ~ .
\end{align}
The effective Chern-Simons terms $Z^{\rm CS (\pm)}_{\rm eff}|_{\nN = \nS = 1}$ perfectly match those computed in \cite{Closset:2017zgf} for $r=1$ and $q_G=1$. The dependence of $y$ on the spin structure via $\alpha_3$ in $\gamma_R$ is present also in the one-loop determinants as $y$ appears in the arguments of the $q$-Pochhammer symbols. 

More generally, an effective   Chern-Simons term mixing two gauge groups $\p{G_i, G_j}$ reads
\begin{align}\label{eq: zmcseff}
	Z^{\rm MCS}_{\rm eff}  =  \p{-y_{(i)}}^{\kcs_{ij} \, \mathfrak b_j} \p{-y_{(j)}}^{\kcs_{ij} \, \mathfrak b_i}    ~ .
\end{align}
 The partition function (\ref{eq: zcseff}) is manifestly invariant under large gauge transformations:
\begin{align}
	Z^{\rm CS}_{\rm eff}|_{\rm BPS}   \qquad \to \qquad \widetilde Z^{\rm CS}_{\rm eff}|_{\rm BPS}  = e^{2 \pi \im \, \kcs  \ell  \mathfrak b} Z^{\rm CS}_{\rm eff}|_{\rm BPS} =   Z^{\rm CS}_{\rm eff}|_{\rm BPS}   ~ .
\end{align}
In presence of matter, the Chern-Simons level $\kcs$ is affected by non-trivial corrections: for instance, a chiral multiplet causes an half-integer shift $\kcs \to \kcs' = \kcs + \p{1/2}$, as we see from (\ref{eq: cm1loopspindles1twistantitwist}). 

Drawing an analogy from the case of $D^2 \times S^1$, we conjecture  that the modifications to the canonical Chern-Simons theory's   stem from   degrees of freedom localized  at the conical singularities of the spindle $\spindle=\mathbb{WCP}^1_{[\nN, \nS]}$. We leave  the proof of this hypothesis  for future investigation; however, in Section \ref{sec: susydualities} we  assess the validity of (\ref{eq: zcseff}) through the application of supersymmetric dualities. We use partition functions related to the topological twist, as the anti-twist case is similar.


\subsection{Examples and dualities}\label{sec: susydualities}

\subsubsection{Effective \texorpdfstring{$U(1)$}{U(1)} Chern-Simons theory at level \texorpdfstring{$\kcs$}{k}}

As a first example let us calculate the   partition function of an effective   Chern-Simons theory with gauge group $U(1)$   at level $\kcs$.  The corresponding contour integral reads
\begin{align}
	 Z = -  \sum_{\mathfrak m \in \mathbb Z} \int_{\mathscr C} \frac{\dd y}{2 \pi \im y} \p{-y}^{\kcs \mathfrak b} \p{-y}^{ \mathfrak n} z^{\mathfrak b} ~ ,
\end{align}
where $y$ is the gauge fugacity, $\mathfrak m$ the label of the gauge flux $\mathfrak f_G = \mathfrak m/\p{\nN \nS}$, while $z$ and $\mathfrak n$   are background flux and fugacity, respectively. The integer $\mathfrak b$ is given by
\begin{align}
	& \mathfrak b = \mathfrak b\p{\mathfrak m} = 1 + \ff{\frac{\mN}{\nN}} +  \ff{-\frac{\mS}{\nS}} = 1 + \ff{\frac{\mathfrak m \, \aN}{\nN}} +  \ff{-\frac{\mathfrak m \, \aS}{\nS}}   ~ , & \nN \aS - \nS \aN = 1 ~ .
\end{align}
We write $\mathfrak m\in\mathbb Z$ as
\begin{align}
	& \mathfrak m = \nN \nS \mathfrak m' + \rf  ~ , & \rf = 0, \dots, \p{\nN \nS - 1} ~ ,
\end{align}
with $\mathfrak m'\in\mathbb Z$. The integral then becomes
\begin{align}
	 Z & = -  \sum_{\mathfrak m \in \mathbb Z} \int_{\mathscr C} \frac{\dd y}{2 \pi \im y} y^{\kcs \mathfrak b + \mathfrak n} z^{\mathfrak b}  \nn\\
	 & = - \sum_{\rf = 0}^{\nN \nS - 1}  \sum_{\mathfrak m' \in \mathbb Z} \int_{\mathscr C} \frac{\dd y}{2 \pi \im y} y^{- \kcs \mathfrak m' + \kcs \mathfrak b\p{\rf} + \mathfrak n} z^{-  \mathfrak m' +  \mathfrak b\p{\rf}}  \nn\\
	 & = - \sum_{\rf = 0}^{\nN \nS - 1}  \sum_{\mathfrak m'' \in \mathbb Z} \int_{\mathscr C} \frac{\dd y}{2 \pi \im y} y^{ \kcs \mathfrak m''   + \mathfrak n} z^{ \mathfrak m''  } ~ .
\end{align}
The latter is non-vanishing only if the contour $\mathscr C$ surrounds   $y=0$ and if
\begin{align}
	 \kcs \mathfrak m''   + \mathfrak n = 0 ~ ,
\end{align}
which is  a Diophantine equation that can  be solved by conveniently splitting $\mathfrak n$ as
\begin{align}
	   \mathfrak n = \kcs \mathfrak n' + \rem{\mathfrak n}{\kcs} ~ .
\end{align}
Indeed,   plugging the expression above into the Diophantine equation gives
\begin{align}
	   & \mathfrak m'' = - \mathfrak n' ~ , &  \rem{\mathfrak n}{\kcs} = 0 ~ ,
\end{align} 
which   yields the final result
\begin{align}
	 Z & =  \nN \nS \,  z^{ -  \mathfrak n'  } \, \delta_{\rem{\mathfrak n}{\kcs} , 0 } =  \nN \nS \,  z^{ - \ff{  \mathfrak n /\kcs } }  \, \delta_{\rem{\mathfrak n}{\kcs} , 0 } ~ , 
\end{align}
meaning that $Z\neq0$ only if  $\rem{\mathfrak n}{\kcs} = 0 $. Especially, if $\mathfrak n=0$, we find $Z = \p{\nN \nS}$, suggesting  that the effective Chern-Simons theory on $\spindle=\mathbb{WCP}^1_{\comm{\nN, \nS}}$ has $\p{\nN \nS}$ vacua. Technically, this degeneracy stems  from the fact that for any $\overline{\mathfrak b}\in \mathbb Z$ there are $\p{\nN \nS}$ values of  $\mathfrak m \in \mathbb Z$ such that $\mathfrak b\p{\mathfrak m} = \overline{\mathfrak b}$.


\subsubsection{Effective \texorpdfstring{$U(1)$}{U(1)} Chern-Simons theory coupled to a chiral multiplet}

As another example we compute the partition function of a chiral multiplet coupled to an effective Chern-Simons term with gauge group $U(1)$ at level $\kcs = 1$. We also add a mixed Chern-Simons term with flavour fugacity $\p{-z}$ and effective flux $\widetilde{\mathfrak b}$. We again consider a countour ${\mathscr C} $ such that all contributions to the integral comes from the pole at  $y=0$. The collective partition function is
\begin{align}
	Z = \sum_{\mathfrak m \in \mathbb Z}\int_{\mathscr C} \frac{\dd y}{2 \pi \im \, y} \frac{\p{-z}^{\mathfrak b} y^{\widetilde{\mathfrak b}}y^{\mathfrak b - 1}}{\p{y \, q^{\p{1-\mathfrak b}/2} ; q }_{\mathfrak b} } ~ ,
\end{align}
where the finite $q$-Pochhammer symbol corresponds to a chiral multiplet on topologically twisted $\spindle \times S^1$, while $y^{\mathfrak b - 1}$ is the spindle counterpart of $e^{2 \pi \im \gamma_G \mathfrak m}$, which is the Chern-Simons term on $S^2 \times S^1$. In analogy with  \cite{Benini:2015noa}, we inserted    in the $Z$ above   sign factors such as  $\p{-1}^{\mathfrak b}$ and $\p{-1}^{\widetilde{\mathfrak b}}$ for convenience. 

It is convenient to expand the integrand as 
\begin{align}
	Z & = \sum_{\mathfrak m \in \mathbb Z}\int \frac{\dd y}{2 \pi \im y} \frac{\p{-z}^{\mathfrak b} y^{\widetilde{\mathfrak b} + \mathfrak b - 1}}{\p{y \, q^{\p{1-\mathfrak b}/2} ; q }_{\mathfrak b} }   = \sum_{\mathfrak m \in \mathbb Z}\sum_{\ell \in \mathbb N} \int \frac{\dd y}{2 \pi \im y} \p{-z}^{\mathfrak b} y^{\widetilde{\mathfrak b} + \mathfrak b - 1 + \ell }\frac{\p{ q^{\mathfrak b} ; q}_\ell}{\p{q ; q }_\ell } q^{\ell \p{1-\mathfrak b}/2} ~ ,  
\end{align}
where we employed the $q$-binomial theorem
\begin{align}
	 \frac{1}{\p{t ; q}_{\mathfrak b} } = \frac{\p{ t \, q^{\mathfrak b}; q}_\infty}{\p{t ; q}_\infty  } =  \sum_{\ell\in \mathbb N} \frac{\p{q^{\mathfrak b} ; q }_\ell}{\p{q; q}_\ell} t^\ell  ~ ,
\end{align}
valid for both finite and infinite $q$-Pochhammer symbols. The contour integral providing $Z$ is non-vanishing if
\begin{align}
	\widetilde{\mathfrak b} + \mathfrak b - 1 + \ell = 0~ ,
\end{align}
which is an integer-valued equation that can be solved by writing  $\mathfrak m$ as
\begin{align}
	  & \mathfrak m   =  \nN \nS \mathfrak m' + \rf ~ , &  \mathfrak m'  \in \mathbb Z \qquad \rf =0, \dots, \p{\nN \nS - 1} ~ ,
\end{align}
yielding
\begin{align}
	\widetilde{\mathfrak b} - \mathfrak m'  + \mathfrak b\p{\rf} - 1 + \ell = 0 ~ .
\end{align}
Inserting the  latter in the expression for $Z$ gives 
\begin{align}
	Z &  = - \sum_{\rf=0}^{\nN \nS - 1}  \sum_{\ell \in \mathbb N}  \p{-z}^{ 1 - \widetilde{\mathfrak b} - \ell }  \frac{\p{ q^{1 - \widetilde{\mathfrak b} - \ell } ; q}_\ell}{\p{q ; q }_\ell } q^{\ell \p{\mathfrak n + \ell}/2}  \nn\\
	&  = - \nN \nS \sum_{\ell \in \mathbb N}  \p{-z}^{ 1 - \widetilde{\mathfrak b} - \ell }  \frac{\p{ q^{ \widetilde{\mathfrak b}  } ; q}_\ell}{\p{q ; q }_\ell } q^{\ell \p{ \widetilde{\mathfrak b} + \ell}/2}\p{- q^{1 - \widetilde{\mathfrak b}  - \ell }}^{\ell}q^{\ell\p{\ell-1}/2}  \nn\\
	&  = - \nN \nS \sum_{\ell \in \mathbb N}  \p{-z}^{ 1 - \widetilde{\mathfrak b}  }  \frac{\p{ q^{ \widetilde{\mathfrak b}  } ; q}_\ell}{\p{q ; q }_\ell }  \p{ q^{ \p{1 - \widetilde{\mathfrak b} }/2 }  z^{-1}}^{ \ell }  \nn\\
	&  = - \nN \nS     \frac{ \p{-z}^{ 1 -  \widetilde{\mathfrak b} } }{\p{  q^{ \p{1 - \widetilde{\mathfrak b} }/2 }  z^{-1}  ; q }_{\widetilde{\mathfrak b}} }  \nn\\
	&  =  \nN \nS     \frac{ z   }{\p{ z \,  q^{ \p{1 - \widetilde{\mathfrak b} }/2 }    ; q }_{\widetilde{\mathfrak b}}   } ~ , 
\end{align}
where in the second, fourth and fifth equalities  we exploited  the identity
\begin{align}
	\p{ t ; q}_\ell = \p{t^{-1} \, q^{1 - \ell} ; q}_\ell \p{- t \,  q^{\p{\ell-1}/2} }^\ell~ .
\end{align}
Therefore, we find that on $\spindle \times S^1$ the partition function of a chiral multiplet coupled to an effective Chern-Simons theory is dual to a free chiral multiplet,  interpreted as describing a monopole operator. The overall factor  $\p{\nN \nS}$ originates from a sum over $\rf$ in the integer range going from $0$ to $\p{\nN \nS - 1}$, spanning degenerate sectors of the original theory.


\section{Localization on the branched-squashed lens space}\label{sec: localizationbranchedlens}
 
 In this section we perform supersymmetric localization  on the branched-squashed lens space $\blens$, yielding the partition functions of three-dimensional gauge theories compactified on this orbifold. Such a calculation extends previous localization computations  carried out for theories defined on both round lens spaces   \cite{Alday:2012au} and squashed lens spaces   \cite{Imamura:2012rq}, as well as those on branched coverings of the three-sphere  \cite{Nishioka:2013haa}.

\subsection{BPS locus}

We parametrize  Abelian gauge or flavour  fields on  $\blens$ as follows:
\begin{align}\label{eq: bpsgaugeconnectionblens}
& \ag = \ag_\sigmavar\p{x} \dd \sigmavar +  \frac{h}{n} \dd \psi  ~ , & \flux_G = \frac{1}{2\pi} \int_\spindle \dd \ag = \frac{\mm}{\nN \nS} ~ , \qquad \exp\p{\im \oint_{S^1}  \ag} = e^{2 \pi \im h/n} ~ ,
\end{align}
with  $\mm\in \mathbb Z$ labelling the flux through the spindle base and    $h=0, \dots, \p{n-1}$  the $\mathbb Z_n$ holonomy along the Hopf fiber. As we do not impose reality conditions on the BPS locus, supersymmetry allows for $\mathfrak f_G \neq 0$ and $\mathfrak m\neq0$, in principle. Then, the BPS locus (\ref{eq: bpsgaugeconnectionblens})   includes and generalizes  those employed e.g. in \cite{Alday:2012au,Imamura:2012rq}, which are recovered in the subcase $\mathfrak m=0$. In fact, we will explicitly check that the integer deformation tuned by $\mathfrak m$ is perfectly under control and has the effect of shifting the value of the holonomy $h$ in the final result of the partition function. Similar shifts   were observed in  \cite{Drukker:2012sr}, where vortex configurations on the Hopf base of  $S^3$ were considered. Notice that we  again gauged away the component $\ag_x\p{x}$. As in the case of $\spindle \times S^1$, the  behaviour of $\ag_\sigmavar$ is
\begin{align}
  & \ag_\sigmavar\p{0} = \frac{\mN}{\nN}  ~ , \qquad \ag_\sigmavar\p{\frac\pi2} = \frac{\mS}{\nS}  ~ ,  & \nN \mS - \nS \mN = \mm ~ ,
\end{align}
with   $\mathfrak m_\pm = \mathfrak m \, \mathfrak a_\pm$  and $\nN \aS - \nS \aN = 1$. In this context, the configurations characterized by $\p{\mN, \mS}$ assume a role akin to that of $H$   in \cite{Drukker:2012sr}.

 If we  do not impose reality conditions on fields, then the  locus (\ref{eq: bpsgaugeconnectionblens}) is supersymmetric with respect to  the  vector-multiplet BPS equations $\delta \lambda = \delta \widetilde \lambda=0$, which  require that on the supersymmetric locus the dynamical scalar $\sigma$ and  the auxiliary scalar $D$ are
\begin{align}
	& \sigma|_{\rm BPS} = \sigma_c + \im\p{\nN b_1^{-1} + \nS b_2^{-1}  }\ag_\sigmavar\p{x}    ~ ,  \nn\\
	& D|_{\rm BPS} = \frac{1}{b_1 b_2 f}\comm{ \im \, b_1 b_2 \, \sigma|_{\rm BPS}  + \p{ \nN b_2 \cot x - \nS b_1 \tan x}\ag_\sigmavar'\p{x}     }  ~ ,
\end{align}
where $\sigma_c$ is a constant real mass in the cause of a weakly gauged theory with a non-dynamical background vector multiplet. Instead, $\sigma_c$ is integrated over in the path integral in the case of a dynamical vector multiplet. 

The chiral multiplet BPS-equations  $\delta \psi =  \delta \widetilde \psi = 0$, for generic values of flavour and  $R$-symmetry fugacities require a trivial supersymmetric locus for matter fields: 
\begin{align}
	\phi|_{\rm BPS} = \widetilde \phi|_{\rm BPS} = F|_{\rm BPS} = \widetilde F|_{\rm BPS} = 0 ~ .
\end{align}


\subsection{One-loop determinants}

We calculate the one-loop determinant for a chiral multiplet on $\blens$ via the same technique adopted in the case of $\spindle \times S^1$. Especially, the eigenfunction  $B_{ m_\sigmavar ,  m_\psi}\p{x}$   is non-singular at $x=0, \pi/2$  if the integers $\p{m_\sigmavar, m_\psi}$ do not violate the following bounds:
\begin{align}
	 & -\nN m_\sigmavar + n \, \tN m_\psi - q_G \tN h + q_G \mN \geq 0    ~ ,  & \samf \p{ \nS m_\sigmavar - n \,  \tS m_\psi + q_G  \tS h - q_G \mS } \geq 0     ~ .
\end{align}
The latter are satisfied if 
 \begin{align}
	 m_\sigmavar  & =  \tS \p{ q_G  \mN - n \ell_\sigmavar - j_\sigmavar } - \tN \p{ \samf  n \ell_\psi + \samf  j_\psi + q_G  \mS  } ~ , \nn\\
	 m_\psi  &= - \samf \nN \ell_\psi - \nS \ell_\sigmavar + \ff{\frac{ - \samf \nN j_\psi + q_G\p{h-\mathfrak m} }{n}} - \ff{\frac{ \nS j_\sigmavar }{n}}    ~ , \nn\\
	 \rem{\nS j_\sigmavar}{n} &= \rem{- \samf \nN j_\psi + q_G\p{h-\mathfrak m} }{n}   ~ , \nn\\
	  \ell_\psi, \ell_\sigmavar & \in \mathbb N ~ , \qquad j_\sigmavar, j_\psi = 0, \dots, \p{n-1} ~ .
\end{align}
Correspondingly, the eigenfunction  $\phi_{ n_\sigmavar ,  n_\psi}\p{x}$   is non-singular at $x=0, \pi/2$  if  
 \begin{align}
	n_\sigmavar  & =  \tS \p{ n \ell_\sigmavar +  k_\sigmavar  + q_G  \mN } + \tN \p{ \samf n \ell_\psi + \samf  k_\psi  - q_G  \mS } ~ , \nn\\
	 n_\psi  & = \nS \ell_\sigmavar + \samf \nN \ell_\psi + \ff{\frac{ \nS k_\sigmavar + q_G\p{h-\mathfrak m} }{n}} - \ff{\frac{ - \samf  \nN k_\psi }{n}}    ~ , \nn\\
	 \rem{ - \samf \nN k_\psi}{n} & = \rem{\nS k_\sigmavar + q_G\p{h-\mathfrak m} }{n}   ~ , \nn\\
	    k_\sigmavar, k_\psi & = 0, \dots, \p{n-1} ~ .
\end{align}
Altogether, the one-loop determinant generated by  $\mathcal B_{ m_\sigmavar ,  m_\psi}$ and   $\Phi_{ n_\sigmavar ,  n_\psi}$ is 
\begin{align}
	 Z^{\mathcal B, \Phi}_{\text{1-L}} &  =  \prod_{j_1=0}^{n-1}\prod_{ k, \ell \in \mathbb N}\frac{  \prod_{j_2\in \mathbb J_-\p{j_1, \mathfrak h }} \comm{ \frac1{b}\p{k + \frac{j_2}{n} } + b \p{\ell+\frac{j_1}{n} } + \frac{Q}{2 n}+ \frac{\im u}{n} 
} }{  \prod_{j_3\in \mathbb J_+\p{j_1, \mathfrak h }} \comm{ \frac1{b}\p{k + \frac{j_1}{n} } + b \p{\ell+\frac{j_3}{n} } + \frac{Q}{2 n} - \frac{\im u}{n} }  } ~ , \nn\\
	 b   & = \sqrt{\frac{ b_1}{ b_2} } ~ , \qquad Q  =  \frac1b + b    ~ , \qquad \mathfrak h =  h - \mathfrak m  ~ , \nn\\
	    u &  = q_G   \comm{ \im \p{ \frac{\mN}{b} - \samf \mS  b   }  - \samf  \sqrt{b_1 b_2} \sigma_c   } + \im   \p{r-1} \frac{Q}{2}   ~ , \nn\\
	 \mathbb J_-\p{j_1,  \mathfrak h}   & =\acomm{  j_2 =0, \dots, \p{n-1}  : \rem{ \nS  j_2}{n} = \rem{ - \samf  \nN  j_1 + q_G \mathfrak h }{n} }   ~ , \nn\\
	 \mathbb J_+\p{j_1,  \mathfrak h}   & =\acomm{  j_3 =0, \dots, \p{n-1}  : \rem{ - \samf  \nN  j_3}{n} = \rem{ \nS  j_1 + q_G \mathfrak h }{n} }   ~ .
\end{align}
Hence, on $\blens$ the one-loop determinant for a chiral and a vector multiplet with weights $\rho\in \mathfrak R_G$ and $\alpha \in {\rm adj}\, G $ for a gauge group $G$ read
\begin{align}
	 Z^{\rm CM}_{\text{1-L}}  & =   \prod_{\rho \in \mathfrak R_G} Z^{\mathcal B, \Phi}_{\text{1-L}}\p{ \alpha\p{u}, \alpha\p{\mathfrak h}  }  ~ , \nn\\
	 Z^{\rm VM}_{\text{1-L}}  & =    \prod_{\alpha \in {\rm adj}\, G} Z^{\mathcal B, \Phi}_{\text{1-L}}\p{ \alpha\p{u}, \alpha\p{\mathfrak h}  }|_{r=2}  ~ .
\end{align}
These match the one-loop determinants on the squashed lens space computed in \cite{Imamura:2012rq} upon setting $\samf=-1$,  $n_\pm = 1$, $\mathfrak h  = h_{\rm there}$ and employing the function
\begin{align}
	 \langle x \rangle_n  = \frac1n\p{\rem{x}{n} + \frac12} - \frac12  ~ .
\end{align}
Further setting $b = 1$ makes  our computation agree with the analysis conducted in \cite{Alday:2012au} for the unsquashed lens space. Eventually, if $  n = 1$, the one-loop determinant above reduces to that of the squashed three-sphere. In this limit the one-loop determinant above also matches the determinant of fluctuations for matter fields computed in  \cite{Nishioka:2013haa} on the resolved space of the branched sphere, once the identifications $b_1 = q_{\rm there}$ and $b_2 = p_{\rm there}$ are made.


\section{Spindles from the squashed three-sphere}\label{sec: spindlefromthreesphere}

In this section we  derive  one-loop determinants for matter and gauge fields on the pure two-dimensional spindle starting  from three dimensions. Indeed, by exploiting the fact that  the spindle non-trivially embeds into  a squashed three-sphere $S^3_{b_s, b_c}$,  we reduce the one-loop determinant associated with an $\mathcal N=2$ chiral multiplet  on  $S^3_{b_s, b_c}$ to the one-loop determinant  of  an $\mathcal N=(2,2)$ chiral multiplet  on $\spindle$. We first focus on  chiral multiplets coupled to $U(1)$ vector multiplets.   Eventually, we shall consider one-loop determinants of   vector multiplets, corresponding to chiral multiplets in the representation $\mathfrak R_G={\rm adj}\, G$ with $R$-charge $r=2$. The method that we employ below is different from analogous reduction techniques discussed in \cite{Benini:2012ui} and extended in \cite{Lundin:2021zeb,Lundin:2023tzw}.

\subsection{General construction}

Let 
\begin{align}\label{eq: lineelementss}
    \dd s^2 = f^2 \dd x^2 + b_s^2  \sin^2 x \, \dd \varphi_1^2 +  b_c^2  \cos^2 x \, \dd \varphi_2^2  ~ ,
\end{align}
be the line element on the squashed three-sphere $S^3_{b_s, b_c}$, with $f$ being the squashing function 
\begin{align}
	f = - \sqrt{ b_s^2 \cos^2 x + b_c^2 \sin^2 x } ~ ,
\end{align}
while the ranges of the  coordinates are $x\in\comm{0,\pi/2}$ and $ \varphi_{1},\varphi_{2}\in \comm{0,2\pi }$. We can embed  $S^3_{b_s, b_c}$ in $\mathbb C^2$ by means of the   complex coordinates  
\begin{align}\label{eq: ccs3inc2}
& z_1 = b_s e^{\im \varphi_1 }\sin x ~ , \qquad z_2 = b_c e^{\im \varphi_2 }\cos x ~ , &  \frac{|z_1|^2}{b_s^2} + \frac{|z_2|^2}{b_c^2} = 1 ~ .
\end{align}
 We then define the following map from the coordinates $\p{x, \varphi_1, \varphi_2}$  to   $\vartheta\in\comm{0,\pi}$ and $ \varphi, \psi\in \comm{0,2\pi }$:
\begin{align}
    x & = \p{\pi - \vartheta}/2  ~ , \nn\\
    \varphi_1 & =     n_{1} \psi + t_{1} \varphi   ~ , \nn\\
    \varphi_2 & =    n_{2} \psi + t_{2} \varphi    ~ , 
\end{align}
where $\p{n_1, n_2,t_1, t_2}\in \mathbb Z$, which satisfy $n_1 t_2 - n_2 t_1 = 1$,  can be interpreted as entries of a matrix in   $ SL(2,\mathbb Z)$. The inverse change of coordinates reads
\begin{align}
    \vartheta & = \pi - 2 \, x  ~ , \nn\\
    \varphi  & =  n_{1} \varphi_2 -  n_{2} \varphi_1   ~ , \nn\\
    \psi & =   - t_{1} \varphi_2 + t_{2} \varphi_1  ~ .
\end{align}
In $\p{\vartheta, \varphi, \psi}$ coordinates the line element of $S^3_{b_s, b_c}$ reads
\begin{align}
    \dd s'^2 & =   \p{f^2/4} \dd \vartheta^2 + {\rm h}_{11}\dd \varphi^2 + {\rm h}_{22}\p{\dd \psi + A^{(1)}}^2 ~ , \nn\\
    {\rm h}_{11} & = \frac{ b_s^2 b_c^2 \, \sin^2\vartheta }{ 4 \p{ b_s^2 \, n_{1}^2 \cos^2\frac{\vartheta}{2} + b_c^2 \, n_{2}^2 \sin^2\frac{\vartheta}{2} } } ~ , \nn\\
    {\rm h}_{22} & =   b_s^2 \, n_{1}^2 \cos^2\frac{\vartheta}{2} + b_c^2 \, n_{2}^2 \sin^2\frac{\vartheta}{2}  ~ , \nn\\
    A^{(1)} & =  \frac{ b_s^2 \, n_{1} t_{1} \cos^2\frac{\vartheta}{2} + b_c^2 \, n_{2}t_{2} \sin^2\frac{\vartheta}{2} }{   b_s^2 \, n_{1}^2 \cos^2\frac{\vartheta}{2} + b_c^2 \, n_{2}^2 \sin^2\frac{\vartheta}{2}  } \dd \varphi    ~ .
\end{align}
Near   $\vartheta=0$ the metric becomes
\begin{align}
    \dd s^2|_{\mathcal U_+} & =  \frac{  b_c^2}{4} \p{ \dd \vartheta^2  + \frac{\vartheta^2}{n_{1}^2}  \dd \varphi^2 } +   b_s^2 \, n_{1}^2\p{\dd \psi + \frac{t_{1}}{n_{1}} \dd \varphi}^2 ~ ,
\end{align}
while   at $\vartheta=\pi$ the metric reads
\begin{align}
    \dd s^2|_{\mathcal U_-} & =  \frac{  b_s^2}{4} \p{ \dd \vartheta^2  + \frac{\p{\pi-\vartheta}^2}{n_{2}^2}  \dd \varphi^2 } +   b_c^2 \, n_{2}^2\p{\dd \psi + \frac{t_{2}}{n_{2}} \dd \varphi}^2 ~ ,
\end{align}
corresponding to a $\mathcal O\p{-1}$ circle  bundle whose base  has  the topology of a  spindle $\spindle = \mathbb{WCP}^1_{\comm{n_1 , n_2}}$: indeed, the orbifold Euler characteristic of the base is  
\begin{align}
    \frac{1}{4\pi}\int_{\spindle}\dd \vartheta \dd \varphi \sqrt{g_\spindle} R_\spindle   =  \frac{1}{n_2} + \frac{1}{n_1}    ~ ,
\end{align}
while the flux of $A^{(1)}$ reads
\begin{align}
     &  \frac{1}{2 \pi} \int_\spindle \dd A^{(1)} =  \frac{t_2}{n_2} -  \frac{t_1}{n_1} = \frac{1}{n_1 n_2}  ~ .
\end{align}
In terms of $\p{\varphi, \psi}$ the complex coordinates (\ref{eq: ccs3inc2}) read
\begin{align}
& z_1 = b_s e^{\im \p{  n_{1} \psi + t_{1} \varphi } }\sin x ~ , \qquad z_2 = b_c e^{\im \p{  n_{2} \psi + t_{2} \varphi }}\cos x ~ ,  
\end{align}
which, under a  shift along the fiber  $\psi \to \psi + \Delta \psi$ with $\Delta\psi\in \mathbb R$ and $\rem{\Delta\psi}{\p{2 \pi}}\neq0$, behave as
\begin{align}
& z_1  \to \lambda_\psi^{n_1} z_1 ~ , \qquad z_2  \to \lambda_\psi^{n_2} z_2 ~ ,  & \lambda_\psi = e^{\im \,  \Delta\psi} \in U(1) \subset \mathbb C^* ~ .
\end{align}
The  base $\spindle$ is given by  the quotient of $S^3_{b_s, b_c}\subset \mathbb C^2$ with respect to the $\mathbb C^*$ action generated by $\lambda_\psi$, where  the coordinates (\ref{eq: ccs3inc2}) satisfy the identification
\begin{align}
& \p{z_1 , z_2} \sim \p{\lambda_\psi^{n_1} z_1 , \lambda_\psi^{n_2} z_2 } ~ , 
\end{align}
 again confirming that  $\spindle = \mathbb{WCP}^1_{\comm{n_1, n_2}}$. The spinors 
 \begin{align}
 	& \zeta_\alpha = e^{\frac{\im}{2} \alpha_2 \varphi_1 + \frac{\im}{2} \alpha_3 \varphi_2} \begin{pmatrix} \sqrt{1+\sin x} \\ \frac{\cos x}{\sqrt{1+\sin x}} \end{pmatrix}_\alpha ~ ,
 	 & \widetilde \zeta_\alpha = \samf e^{-\frac{\im}{2} \alpha_2 \varphi_1 - \frac{\im}{2} \alpha_3 \varphi_2} \begin{pmatrix} \frac{\cos x}{\sqrt{1+\sin x}} \\ - \sqrt{1+\sin x}  \end{pmatrix}_\alpha ~ ,
 \end{align}
 with $\samf=\pm1$, fulfil the Killing spinor equation on $S^3_{b_s, b_c}$ with background fields
 \begin{align}
 	A^C & = \frac12\p{\frac{b_s}{ f} + \alpha_2}\dd\varphi_1 + \frac12\p{\frac{\samf b_c}{ f} + \alpha_3}\dd\varphi_2 ~ ,\nn\\
 	V & = - \frac{  b_s}{f}\p{ 1  +  \im \,\samf f H}\sin^2 x \, \dd\varphi_1 - \frac{ b_c}{f}\p{\samf + \im f H}\cos^2 x \, \dd\varphi_2  ~ , 
	\end{align}
and $A = A^C + (3 \, V/2)$. Killing spinors and background fields are smooth on the whole $S^3_{b_s, b_c}$ if $\alpha_2 = 1$ and $\alpha_3=\samf$. In   $\p{\varphi, \psi}$ coordinates the phase of  the Killing spinors reads
\begin{align}
      \exp\p{\frac{\im}{2}  \alpha_2 \varphi_1 + \frac{\im}{2}  \alpha_3 \varphi_2}  & = \exp \comm{ \frac{\im}{2}   \p{ t_{1} \alpha_2 + t_{2}\alpha_3 } \varphi    + \frac{\im}{2}   \p{ n_{1} \alpha_2 + n_{2}\alpha_3 } \psi  } ~ .
\end{align}
If we denote by $c_R$ and $\lambda_R$ the coefficient of $\varphi$ and $\psi$ in the phase above, we have   
\begin{align}\label{eq: 2dRsymfluxes3dphases}
     c_R & = \frac12\p{t_{1} \alpha_2 + t_{2}\alpha_3 } = \frac12\p{ t_1 + \samf t_2 }     ~ , \nn\\
    \lambda_R  & = \frac12\p{n_{1} \alpha_2 + n_{2}\alpha_3 } = \frac12\p{ n_1 + \samf n_2 }   ~ ,
\end{align}
with inverse relations
\begin{align}\label{eq: 3dphasesfrom2dRsymfluxes}
      - 2 n_2 c_R + 2 t_2  \lambda_R  & = \alpha_2  = 1    ~ , \nn\\
           2 n_1 c_R - 2 t_1   \lambda_R & = \alpha_3  = \samf     ~ .
\end{align}
In particular, $\lambda_R$ is an   $R$-symmetry  flat connection along the $S^1$ fiber parameterized. In \cite{Benini:2012ui} flat connections along the Hopf fiber were shown to be related to fluxes on the base. In fact,   $\lambda_R$ precisely coincides with $\p{n_1 n_2}$ times the   $R$-symmetry flux affecting the Killing spinors on the spindle base, where $\samf=\pm1$ correspond  to twist and anti-twist, respectively.

The eigenfunctions of $\delta^2 = K^\mu D_\mu$ contributing  to the one-loop determinant of a chiral multiplet of $R$-charge $r$ on $S^3_{b_s, b_c}$ have the form 
\begin{align}
	\mathcal B_{m_1^{(1)} , m_2^{(1)}}\p{x, \varphi_1, \varphi_2} & = e^{ \im m_1^{(1)} \varphi_1 + \im m_2^{(1)} \varphi_2}  B_{m_1^{(1)} , m_2^{(1)}}\p{x}    ~ , \nn\\
	\Phi_{m_1^{(2)} , m_2^{(2)}}\p{x, \varphi_1, \varphi_2} & = e^{ \im m_1^{(2)} \varphi_1 + \im m_2^{(2)} \varphi_2}  \phi_{m_1^{(2)} , m_2^{(2)}}\p{x}      ~ ,
\end{align}
where the modes $\mathcal B_{m_1^{(1)} , m_2^{(1)}} $ have $R$-charge $\p{r-2}$ while $\Phi_{m_1^{(2)} , m_2^{(2)}} $ have $R$-charge $r$. The conditions $\mathcal B_{m_1^{(1)} , m_2^{(1)}} \in {\rm Ker} \, L_P $ and $\Phi_{m_1^{(2)} , m_2^{(2)}} \in {\rm Ker} \, L_{\widetilde P} $ require 
\begin{align}
	& m_1^{(1)} =  k_1   ~ ,  & m_2^{(1)} =  \samf k_2   ~ , \nn\\
	& m_1^{(2)} = - \ell_1   ~ ,  & m_2^{(2)} = - \samf \ell_2   ~ , 
\end{align}
 with  $k_1, k_2, \ell_1, \ell_2\in \mathbb N$. In $\p{\varphi, \psi}$ coordinates the phases of the Fourier modes of the cohomological fields $\mathcal B_{m_1^{(1)} , m_2^{(1)}} $ and $\Phi_{m_1^{(2)} , m_2^{(2)}} $  read
\begin{align}
      \exp\p{ \im \, m_1^{(i)} \varphi_1 + \im \, m_2^{(i)} \varphi_2}  = \exp \p{ \im \,  j^{(i)} \varphi    + \im \,   \mathfrak f^{(i)} \psi  } ~ ,
\end{align}
with $i=1,2$ and
\begin{align}\label{eq: 2dmodesandfluxesfrom3dmodes}
      & j^{(i)}  = t_1 m_1^{(i)} + t_2 m_2^{(i)}   ~ ,  & \mathfrak f^{(i)}  =  n_1 m_1^{(i)} +  n_2 m_2^{(i)}      ~ ,  
\end{align}
along with the inverse relations
\begin{align}\label{eq: 3dmodesfrom2dmodesandfluxes}
     & m_1^{(i)}  =  - n_2 j^{(i)}   +  t_2 \mathfrak f^{(i)}   ~ ,  & m_2^{(i)}  =  n_1 j^{(i)} - t_1 \mathfrak f^{(i)}      ~ .
\end{align}
In analogy with $\lambda_R$, we expect  that $\mathfrak f^{(1)}$ and $\mathfrak f^{(2)}$  corresponds to $\p{n_1 n_2}$ times the   flux affecting  $\mathcal B_{m_1^{(1)} , m_2^{(1)}} $ and $\Phi_{m_1^{(2)} , m_2^{(2)}} $, respectively. Similarly,  the Killing vector $K = ( \zeta \sigma^\mu \widetilde \zeta ) \partial_\mu$ written in $\p{\varphi, \psi}$ coordinates is
\begin{align}
    K & = \knorm b_c \p{ \frac{\samf}{b_s}\partial_{\varphi_1} + \frac{1}{b_c}\partial_{\varphi_2} }   =  \knorm'  \p{ \amf_{\rm 2d}\partial_\varphi    + \partial_\psi     }    ~ , 
\end{align}
where we introduced 
\begin{align}
     & \knorm' =  \frac{\knorm\p{ t_2 \samf b_c - t_1 b_s }}{b_s}    ~ , & \amf_{\rm 2d} =  \frac{ n_1 b_s -\samf n_2 b_c }{\samf t_2 b_c - t_1 b_s}     ~ .
\end{align}
The one-loop determinant of a chiral multiplet coupled to an Abelian vector multiplet on $S^3_{b_s,b_c}$ is
\begin{align}
    Z_{\text{1-L}}^{\rm CM} & = \frac{  \prod_{k_1, k_2 \in \mathbb N} \comm{ b_c k_1 + b_s k_2 +\p{1-\frac{r}{2}}\p{b_c+b_s}  - \im \samf b_c b_s   \sigma_c}  }{  \prod_{\ell_1, \ell_2 \in \mathbb N} \comm{ b_c \ell_1 + b_s \ell_2 + \frac{r}{2}\p{b_c+b_s}  + \im \samf b_c b_s  \sigma_c} }  \nn\\
    & = \frac{  \prod_{m^{(1)}_1, m^{(1)}_2  } \comm{ b_c m^{(1)}_1 + \samf b_s m^{(1)}_2 +\p{1-\frac{r}{2}}\p{b_c \alpha_2+ \samf b_s \alpha_3}  - \im \samf b_c b_s   \sigma_c}  }{  \prod_{m^{(2)}_1, m^{(2)}_2  } \comm{ -   b_c m^{(2)}_1 - \samf b_s m^{(2)}_2 + \frac{r}{2}\p{b_c \alpha_2+ \samf b_s \alpha_3}  + \im \samf b_c b_s   \sigma_c } } ~ , 
\end{align}
which, by means of (\ref{eq: 3dphasesfrom2dRsymfluxes}) and (\ref{eq: 3dmodesfrom2dmodesandfluxes}),  can be reformulated exclusively in terms of two-dimensional quantities: 
\begin{align}\label{eq: spindle1loopfromsl2zreduction}
 Z_{\rm 2d}^{\rm CM} & = \frac{  \prod_{j^{(1)}} \acomm{ \amf_{\rm 2d} \comm{j^{(1)} + \p{2-r} c_R } + \mathfrak f^{(1)} +\p{2-r} \lambda_R + \im \, \sigma_{\rm 2d } } }{  \prod_{j^{(2)}} \comm{  \amf_{\rm 2d} \p{j^{(2)} - r \, c_R } + \mathfrak f^{(2)} - r \lambda_R + \im \, \sigma_{\rm 2d } } } ~ , 
\end{align}
with
\begin{align}
	\sigma_{\rm 2d} = \frac{ b_c b_s \sigma_c }{t_1 b_s - \samf t_2 b_c } ~ ,
\end{align}
while the ranges of $j^{(1)}$ and $j^{(2)}$ are determined by  
\begin{align}
    &   - n_2 j^{(1)}   +  t_2 \mathfrak f^{(1)} =  k_1    ~ ,   &       n_1 j^{(1)} - t_1 \mathfrak f^{(1)}    = \samf k_2  ~ , \nn\\
     &  - n_2 j^{(2)}   +  t_2 \mathfrak f^{(2)}  =  - \ell_1   ~ ,   &    - n_1 j^{(2)} + t_1 \mathfrak f^{(2)}  =  \samf \ell_2    ~ ,  
\end{align}
taking into account that $k_1, k_2, \ell_1$ and $\ell_2$ are all integers. We recall that the numerator of $Z_{\text{1-L}}^{\rm CM}$ arises from the fermionic modes associated to  the chiral-multiplet  cohomological field $B$ defined in Appendix \ref{sexcohomo}, while the denominator corresponds to the contribution of bosonic degrees of freedom encoded by the chiral-multiplet scalar $\phi$. This is consistent  with relating the integers $\p{j^{(1)}, \mathfrak f^{(1)} }$ and $\p{j^{(2)}, \mathfrak f^{(2)} }$ with $B$ and $\phi$, respectively.  We now proceed to apply the formula (\ref{eq: spindle1loopfromsl2zreduction})  to the cases  where the bases of the fibrations are either  topologically twisted or  anti-twisted spindles.


\subsection{Topologically twisted spindle}

If $\samf=+1$, then
\begin{align}
    & c_R = \frac12\p{t_1 + t_2} ~ , & \lambda_R = \frac12\p{n_1 + n_2} ~ ,   
\end{align}
and
\begin{align}
    &   - n_2 j^{(1)}   +  t_2 \mathfrak f^{(1)}  \geq 0   ~ ,   &       n_1 j^{(1)} - t_1 \mathfrak f^{(1)}    \geq 0  ~ , \nn\\
     &  - n_2 j^{(2)}   +  t_2 \mathfrak f^{(2)}  \leq 0   ~ ,   &    - n_1 j^{(2)} + t_1 \mathfrak f^{(2)}    \geq 0   ~ ,  
\end{align}
implying
\begin{align}
    &  \cf{ t_1 \mathfrak f^{(1)} /n_1}  \leq   j^{(1)}    \leq \ff{  t_2 \mathfrak f^{(1)} / n_2 } ~ ,      &   \cf{ t_2 \mathfrak f^{(2)} /n_2} \leq j^{(2)}  \leq  \ff{ t_1 \mathfrak f^{(2)} /n_1}   ~ . 
\end{align}
As anticipated, we interpret $\mathfrak f^{(1)}$ and $\mathfrak f^{(2)}$   as $\p{n_1 n_2}$ times the   total fluxes attached to  $B $ and $\phi $: 
\begin{align}
	& \mathfrak f^{(1)}  = \p{r-2}\lambda_R + \mathfrak m ~ ,  & \mathfrak f^{(2)}  = r \, \lambda_R + \mathfrak m~ , 
\end{align}
with $\mathfrak m\in \mathbb Z$. Thanks to the identities 
\begin{align}
	 \cf{ \frac{t_1 \mathfrak f^{(1)} }{n_1} } & = 1 + \ff{ \frac{  \mathfrak p_1 }{n_1} } +  \p{r-2} c_R   ~ , \nn\\
	 \ff{ \frac{t_2 \mathfrak f^{(1)} }{n_2} } & = -1 - \ff{ - \frac{  \mathfrak p_2 }{n_2} } +  \p{r-2} c_R   ~ ,  \nn\\
	 \cf{ \frac{t_2 \mathfrak f^{(2)} }{n_2} } & =   \cf{ \frac{  \mathfrak p_2 }{n_2} } +   r \, c_R   ~ , \nn\\
	 \ff{ \frac{t_1 \mathfrak f^{(2)} }{n_1} } & =   \ff{  \frac{  \mathfrak p_1 }{n_1} } +   r \, c_R   ~ , 
\end{align}
we obtain 
\begin{align}\label{eq: sl2zreductionttranges}
        &  1 + \ff{ \frac{  \mathfrak p_1 }{n_1} }     \leq   \ell^{(1)}    \leq  -1 - \ff{ - \frac{  \mathfrak p_2 }{n_2} }  ~ ,  
        & - \ff{ - \frac{  \mathfrak p_2 }{n_2} }   \leq \ell^{(2)}  \leq  \ff{  \frac{  \mathfrak p_1 }{n_1} }   ~ , 
\end{align}
 where we introduced
 \begin{align}
 	& \mathfrak p_1 = t_1 \mathfrak m - \p{r/2}   ~ ,   & \mathfrak p_2 = t_2 \mathfrak m + \p{r/2}     ~ ,  \nn\\
     & j^{(1)}  = \ell^{(1)} + \p{r-2} c_R    ~ ,   &  j^{(2)}  = \ell^{(2)}   +   r \, c_R  ~ .  
\end{align}
The intervals (\ref{eq: sl2zreductionttranges}) are mutually exclusive: if 
\begin{align}
	- \ff{ - \frac{  \mathfrak p_2 }{n_2} }  \leq  \ff{  \frac{  \mathfrak p_1 }{n_1} } ~ ,
\end{align}
then only the modes $\Phi_{m_1^{(2)} , m_2^{(2)}} $   contribute, giving
\begin{align} 
 Z_{\rm 2d}^{\Phi}  & =   \prod_{\ell^{(2)}= - \ff{ -   \mathfrak p_2 /n_2 }}^{\ff{   \mathfrak p_1 /n_1 }} \p{  \amf_{\rm 2d} \, \ell^{(2)} + \mathfrak m + \im \, \sigma_{\rm 2d } }^{-1}  ~ . 
\end{align} 
On the other hand, if
\begin{align}
	1 + \ff{ \frac{  \mathfrak p_1 }{n_1} }     \leq    -1 - \ff{ - \frac{  \mathfrak p_2 }{n_2} }  ~ , 
\end{align}
then only the eigenfunctions $\mathcal B_{m_1^{(1)} , m_2^{(1)}} $      contribute, providing
\begin{align}
 Z_{\rm 2d}^{\mathcal B}  & =   \prod_{\ell^{(1)}=1 + \ff{ {  \mathfrak p_1 }/{n_1} }    }^{ -1 - \ff{ - {  \mathfrak p_2 }/{n_2}} } \p{ \amf_{\rm 2d} \, \ell^{(1)} + \mathfrak m + \im \, \sigma_{\rm 2d } }  ~ . 
\end{align}
Both $ Z_{\rm 2d}^{\Phi}$ and $ Z_{\rm 2d}^{\mathcal B}$ are encoded by the formula
\begin{align}\label{eq: Z2dcmttfromsl2zreduction}
	Z_{\rm 2d}^{\rm CM} = \prod_{\ell\in \mathbb N}\frac{\amf_{\rm 2d} \p{\ell + 1 + \ff{-\mathfrak p_2/n_2} } - \mathfrak m - \im \sigma_{\rm 2d} } {\amf_{\rm 2d} \p{\ell - \ff{\mathfrak p_1/n_1} } - \mathfrak m - \im \sigma_{\rm 2d} }  ~ ,
\end{align}
which, after  relabelling $ \p{\mathfrak m + \im \sigma_{\rm 2d}} \to \p{ - \varphi_G } $ and
\begin{align}\label{eq: sl2zreductiondictionary}
	 & \p{t_1, n_1} \to \p{\tN, \nN} ~ , & \p{t_2, n_2} \to \p{\tS, \nS} ~ , \nn\\
	 & \p{ t_1 \mathfrak m, \mathfrak p_1} \to \p{\mN, \pN} ~ , & \p{ t_2 \mathfrak m, \mathfrak p_2} \to \p{\mS, \pS} ~ ,
\end{align}
precisely corresponds to \eqref{eq: zindspindleanytwist} for $\st=+1$, which represents  the one-loop determinant of a chiral multiplet coupled to an Abelian vector multiplet on a topologically twisted spindle. Furthermore,    setting $\nN=\nS=+1$ and following the methodology detailed in Section \ref{sec: localizationspindle} gives the chiral-multiplet one-loop determinant on an A-twisted $S^2$, as calculated in \cite{Closset:2015rna}.


\subsection{Anti-twisted spindle}

If $\samf=-1$, then
\begin{align}
    & c_R = \frac12\p{t_1 - t_2} ~ , & \lambda_R = \frac12\p{n_1 - n_2} ~ ,   
\end{align}
and
\begin{align}
    &   - n_2 j^{(1)}   +  t_2 \mathfrak f^{(1)}  \geq 0   ~ ,   &       n_1 j^{(1)} - t_1 \mathfrak f^{(1)}    \leq 0  ~ , \nn\\
     &  - n_2 j^{(2)}   +  t_2 \mathfrak f^{(2)}  \leq 0   ~ ,   &     n_1 j^{(2)} - t_1 \mathfrak f^{(2)}    \geq 0   ~ ,  
\end{align}
meaning
\begin{align}
	&  j^{(1)} \leq \ff{t_2 \mathfrak f^{(1)}/n_2 }  ~ ,   &    j^{(1)} \leq \ff{t_1 \mathfrak f^{(1)}/n_1 }  ~ , \nn\\
	&  j^{(2)} \geq \cf{t_2 \mathfrak f^{(2)}/n_2 }  ~ ,   &    j^{(2)} \geq \cf{t_1 \mathfrak f^{(2)}/n_1 }  ~ .
\end{align}
The inequalities   above would lead to the inclusion of absolute values of fluxes in the    one-loop determinant. This can be circumvented by selecting ranges from dual poles, as prescribed by the index theorem. Indeed, if
\begin{align}
	j^{(1)} & \leq \ff{t_2 \mathfrak f^{(1)}/n_2 } =  -1 - \ff{ -   \mathfrak p_2 / n_2 } +  \p{r-2} c_R   ~ , \nn\\
	   j^{(2)} & \geq \cf{t_1 \mathfrak f^{(2)}/n_1 }   =   \cf{    \mathfrak p_1 / n_1 } +   r \, c_R  ~ , 
\end{align}
we have
\begin{align}
	j^{(1)}  & = - \ell  -1 - \ff{ -   \mathfrak p_2 / n_2 } +  \p{r-2} c_R   ~ , \nn\\
	   j^{(2)}  & = \ell +  \cf{    \mathfrak p_1 / n_1 } +   r \, c_R = \ell - \ff{ -    \mathfrak p_1 / n_1 } +   r \, c_R  ~ . 
\end{align}
Plugging such $j^{(1)}$ and $j^{(2)}$ into (\ref{eq: spindle1loopfromsl2zreduction})  yields
\begin{align}\label{eq: Z2dcmatfromsl2zreduction}
 Z_{\rm 2d}^{\rm CM} & = \prod_{ \ell \in \mathbb N } \frac{  \amf_{\rm 2d} \p{  \ell  + 1 + \ff{ -   \mathfrak p_2 / n_2 }   } - \mathfrak  m - \im \, \sigma_{\rm 2d }  }{  \amf_{\rm 2d} \p{  \ell -   \ff{ -    \mathfrak p_1 / n_1 }   } + \mathfrak m + \im \, \sigma_{\rm 2d }  } ~ , 
\end{align}
exactly matching  \eqref{eq: zindspindleanytwist} for $\st=-1$,  after applying the dictionary (\ref{eq: sl2zreductiondictionary}) and substituting $ \p{\mathfrak m + \im \sigma_{\rm 2d}}$ with $ \p{-\varphi_G}$. Consequently,  (\ref{eq: Z2dcmatfromsl2zreduction}) perfectly reproduces the one-loop determinant of a chiral multiplet coupled to an Abelian vector multiplet on an anti-twisted spindle. If  $\nN=\nS=+1$,   (\ref{eq: Z2dcmatfromsl2zreduction})  reduces to the one-loop determinant of a chiral multiplet on a two-sphere without  $R$-symmetry twist, as computed in \cite{Benini:2012ui}.


\section{Conclusions}\label{sec: conclusions}

In this paper we considered three-dimensional ${\cal N} = 2$ supersymmetric gauge theories compactified  on a large class of three-dimensional orbifolds.  Our results include the calculation of the partition function of gauge theories on $\spindle \times S^1$ in the presence of both anti-twist and twist for the  $R$-symmetry connection, which  were 
 anticipated in \cite{Inglese:2023wky}.  We computed such one-loop determinants via the eigenvalue-pairing method  and  by  the equivariant orbifold index theorem. We also calculated  both the  canonical expressions  and the effective corrections to   Chern-Simons theories and Fayet-Iliopoulos terms, which were crucial for testing our partition functions via non-perturbative dualities. Furthermore, we derived by  supersymmetric localization the refined partition function on the branched-squashed  lens space, which is a generalization of the usual lens space thanks to the inclusion of conical singularities, thus 
 extending the partition function on branched coverings of the  three-sphere.

Moreover, we computed  the one-loop determinants for two-dimensional quantum field theories on the spindle, again with both twist and anti-twist.   The method employed  in Section \ref{sec: spindlefromthreesphere}  to derive this result exploited the interpretation of the squashed three-sphere as a smooth non-trivial circle orbi-bundle over a spindle. This approach naturally extends to the computation of   one-loop determinants of four-dimensional supersymmetric theories on orbifolds by deriving them from  one-loop determinants of a parent five-dimensional theory. Similarly, it holds promise for calculating the one-loop determinants of gauge theories defined on six-dimensional orbifolds through reduction from seven-dimensional counterparts \cite{Polydorou:2017jha}.

The results of this paper grant several avenues for future exploration. A significant next step is to extract the microscopic entropy of accelerating black holes in four-dimensional anti-de Sitter spacetime \cite{Ferrero:2020twa,Cassani:2021dwa} from the large-$N$ limit of the spindle index of ${\cal N}=2$ theories with AdS$_4$ duals. This result is expected to match the large-$N$ predictions recently derived from supergravity considerations in \cite{Boido:2023ojv}, as has been explicitly demonstrated in \cite{Colombo:2024mts}.  Another interesting direction  is probing the large-$N$ limit of partition functions  of SCFTs  defined on the branched-lens space $\blens$, along with  the investigation of their possible holographic counterparts   as asympotically locally AdS$_4$   supergravity solutions. To this aim, a possibility is to  extend the supergravity solutions constructed in \cite{Martelli:2011fu,Martelli:2011fw,Martelli:2012sz,Martelli:2013aqa,Toldo:2017qsh}  to encompass geometries whose   conformal boundaries are  
non-trivial circle fibrations over a spindle. It would also be interesting to study the possible implications of our findings on the supersymmetric Renyi entropy
\cite{Nishioka:2013haa} and its gravity dual \cite{Nishioka:2014mwa}.

On the   field theory side, further investigation of supersymmetric dualities is interesting to gain a deeper comprehension of the quantum moduli space pertaining to SQFTs   on orbifolds. For instance, in the study of gauge theories on   manifolds, the phenomenon of partition functions factorizing into holomorphic blocks is well-known \cite{Pasquetti:2011fj,Beem:2012mb,Nieri:2015yia,Longhi:2019hdh}. Extending such results to gauge theories on orbifolds characterized by conical singularities, as those explored in this article, is  an intriguing line of research. Another interesting outlook lies in generalizing the findings established for twisted gauge theories on non-compact manifolds \cite{Pittelli:2018rpl,Iannotti:2023jji} to cover the case of non-compact orbifolds. Moreover, gauge theory partition functions on the two-sphere are known to provide the instanton corrections to the K\"{a}hler moduli space of   Calabi-Yau manifolds, which directly correlate with Gromov-Witten invariants, as shown in \cite{Gomis:2012wy}. Extending this understanding to incorporate two-dimensional gauge theories defined on spindles presents an intriguing area for exploration.  Relatedly, it would be interesting to explore possible applications of our spindle index  in the context of enumerative geometry,  possibly generalizing the results of  \cite{Crew:2023tky,Dedushenko:2023qjq}.

It would be important to generalize our results to higher-dimensional  models, including either $\mathcal N=1$ or $\mathcal N=2$ four-dimensional gauge theories on the direct product of a spindle  and a Riemann surface,  on $\spindle \times \spindle$, $\mathbb{WCP}^2_{\comm{n_1, n_2, n_3}}$ (about which some results were anticipated in \cite{Martelli:2023oqk}), or more general toric four-dimensional orbifolds. This could be done by exploiting  the cohomological formalism developed e.g. in \cite{Festuccia:2020yff}.  Orbifold partition functions of five-dimensional supersymmetric theories would allow to make contact with the supergravity solutions describing D4 branes wrapped on various orbifolds constructed in 
\cite{Faedo:2021nub,Giri:2021xta,Suh:2022olh,Couzens:2022lvg,Faedo:2022rqx,Colombo:2023fhu,Faedo:2024upq} and should also be relevant for six-dimensional SCFTs \cite{Hosseini:2021mnn} compactified on orbifolds. 
Finallly, it is natural to wonder whether enlarging to the realm of orbifolds the  equivariant localization program pursued in 
\cite{DelZotto:2021gzy,Nekrasov:2021ked,Cassia:2022lfj}
would allow to uncover interesting aspects of  (K-theoretic) Donaldson-Thomas and Gromov-Witten invariants associated to toric geometries.

 
 \paragraph{Conflict of interest and data availability.} 
 
The corresponding author assures that, on behalf of all authors, there are no conflicts of interest to disclose. Furthermore, no additional data beyond what is already reported in this article  are required  to validate the  research findings presented in this paper.


\section*{Acknowledgements}

The authors acknowledge partial support by the INFN. We thank E. Colombo, L. Cassia, P. Ferrero  and F. Nieri for useful conversations.  We also thank A. Zaffaroni for comments on a draft of this paper. 


\appendix
\setcounter{equation}{0}
\renewcommand{\theequation}{\thesection.\arabic{equation}}

\section{Notation, conventions and identities}

\subsection{Spinors and Clifford algebra}

We employ the spinor conventions of  \cite{Dedushenko:2016jxl}: especially, indices of two-component spinors in the {\bf 2} representation  of $SU(2)$ are raised and lowered by acting on
the left with the anti-symmetric tensors $\lc^{\alpha\beta}$ and $\lc_{\alpha\beta}$,  respectively, where $\lc^{12}  =\lc_{21} = +1$. Given two spinors $\chi$ and $\eta$, their $SU(2)$-invariant contraction  is defined as follows:
\begin{align}
\p{\chi \eta}  =\chi^\alpha \eta_\alpha = - \chi_\alpha \eta^\alpha= \lc^{\alpha \beta} \chi_\beta \eta_\alpha = \lc_{\alpha \beta} \chi^\alpha \eta^\beta ~ . 
\end{align}
Let $\chi_i$, with $i=1,2,3$, be three commuting or anti-commuting  spinors with components $\chi_{i,\alpha}$. Then, the Fierz identity
\begin{align}
\chi_{1,\alpha} (\chi_2 \chi_3) + (\chi_1 \chi_2) \chi_{3,\alpha} + \chi_{1,\beta}\chi_{2,\alpha}\chi_3^\beta = 0 ~ . 
\end{align}
holds for any $\alpha = 1,2$. In the case of four commuting spinors, namely $\chi_j$ with $j=1,\dots ,4$,  the identity above implies 
\begin{align}
(\chi_1 \chi_2)(\chi_3 \chi_4) = (\chi_1 \chi_3)(\chi_2 \chi_4) + (\chi_1 \chi_4)(\chi_3 \chi_2) ~ .
\end{align}
The Euclidean gamma-matrices  are proportional to the Pauli matrices $\sigma^i$ with $i=1 , 2 , 3 $:
\begin{align}
& \gamma_1 = \sigma^2 ~ , \qquad \gamma_2 = \sigma^3 ~ , \qquad \gamma_3 = \sigma^1 ~ , & \gamma^i = \delta^{ij} \gamma_j ~ .
\end{align}
We sometimes used the skew-symmetric product of two gamma matrices, 
\begin{align}
 & \gamma_{ij} = \f12 \p{ \gamma_i \gamma_j - \gamma_j \gamma_i } = \im \, \lc_{ijk} \gamma^k ~ , 
\end{align}
where the three-dimensional Levi-Civita tensor $\lc_{ijk}$ satisfies
 \begin{align}
& \lc^{ijm} \lc_{k \ell m} = {\delta^i}_k {\delta^j}_\ell - {\delta^i}_\ell {\delta^j}_k ~ , \qquad \lc^{i \ell m} \lc_{k \ell m} = 2 {\delta^i}_k  ~ , & \lc^{123} = \lc_{123} = +1 ~ .
\end{align}
The following identities hold:
\begin{align}
 \gamma_i \gamma_j  & = \delta_{ij} + \im \, \lc_{ijk} \gamma^k  ~ , \nn\\
 \gamma_k \gamma^{k i}  & =  \gamma^{i k} \gamma_k  = 2 \, \gamma^i  ~ , \nn\\
 { \gamma^i}_k \gamma^{kj} & = 2 \, \delta^{ij}   +  \gamma^{ij} ~ ,  
\end{align}
as well as the completeness relation
\begin{align}
{\p{ \gamma^j}_\alpha}^\beta {\p{ \gamma_j}_\gamma}^\delta = 2 \, {\delta_\alpha}^\delta {\delta_\gamma}^\beta - {\delta_\alpha}^\beta {\delta_\gamma}^\delta ~ ,
\end{align}
implying  
\begin{align}
( \chi_1  \gamma^j  \chi_2 ) (\chi_3 \gamma_j  \chi_4 ) = 2 (\chi_1 \chi_4) (\chi_3 \chi_2) - (\chi_1 \chi_2) (\chi_3 \chi_4) ~ ,
\end{align}
with $\chi_{1,2,3,4}$ being four commuting spinors.


\subsection{Differential geometry}

The spin connection $\omega_{\mu \, ab}$ used in the main text   is defined by the relations  
\begin{align}
& \omega_{\mu \, ab} = {e_a}^\nu  \nabla_\mu e_{b \nu } = - e_{b \nu }\nabla_\mu {e_a}^\nu ~ , & \dd e_a + {\omega_a}^b \wedge e_b = 0 ~ .
\end{align}
The spin connection allows for defining the  covariant derivative of a spinor $\Psi$,
\begin{align}
\nabla_\mu \Psi = \partial_\mu \Psi + \f14 \omega_{\mu \, ab} \gamma^{ab} \Psi = \partial_\mu \Psi + \f\im4 \omega_{\mu \, ab} \lc^{abc} \gamma_c\Psi ~ .
\end{align}
The Riemann tensor $R_{\mu \nu a b}$, the Ricci tensor $R_{\mu\nu}$ and the Ricci scalar $R$ read 
\begin{align}
R_{\mu \nu a b }  = \partial_\mu \omega_{\nu \, a b } - \partial_\nu \omega_{\mu \, a b } + {\omega_{\mu \, a } }^c \omega_{\nu \, c b} - {\omega_{\nu \, a } }^c \omega_{\mu \, c b}  ~ , \qquad R_{\mu \nu} = {R^\rho}_{\mu \rho \nu} ~ , \qquad R = {R^\mu}_\mu ~ .
\end{align}
The Lie derivatives along a vector $X^\mu$ of a scalar $\Phi$ and of a 1-form $A_\mu$ respectively are
\begin{align}
& \mathcal L_X \Phi = X^\nu \partial_\nu \Phi  ~ , & \p{\mathcal L_X A}_\mu  = X^\nu \partial_\nu A_\mu + A_\nu \partial_\mu X^\nu   ~ ,
\end{align}
while the Lie derivative along a vector $X^\mu$ of a spinor $\Psi$ reads
\begin{align}
\mathcal L_X \Psi = X^\mu \nabla_\mu \Psi + \f14 \p{\nabla_\mu X_\nu} \gamma^{\mu\nu} \Psi = X^\mu \nabla_\mu \Psi + \f\im4 \lc^{\mu \nu \lambda} \p{\nabla_\mu X_\nu} \gamma_{\lambda} \Psi  ~ .
\end{align}


\section{Special functions}\label{sec: specialfunctions}

\subsection{Zeta-functions and gamma-functions with multiple arguments}

The multiple zeta-function is defined as
 \begin{align}
 & \zeta_N( s , w | \vec a_N) = \sum_{ \vec m_N \in \mathbb N^N} \f1{\p{w + \vec a_N \cdot \vec m_N}^s} ~ , & {\rm Re}(s) > N ~ ,
 \end{align}
 which is related to the multiple gamma-function by
 \begin{align}
 \Gamma_N(w | \vec a_N) = \exp{\comm{\partial_s \zeta_N( s , w | \vec a_N) |_{s=0} }} = \prod_{ \vec m_N \in \mathbb N^N} \f1{w + \vec a_N \cdot \vec m_N} ~ .
 \end{align}
 If ${\rm Im}(\alpha_i) > 0$ with $i=1, \dots  , N$, the identity
 \begin{align}
 \Gamma_{N+1}(w | 1 ,  \vec a_N) \Gamma_{N+1}(1-w | 1 , - \vec a_N) = e^{-\im\pi\zeta_{N+1}(0 , w | 1 , \vec \alpha_N)} \prod_{\vec m_N \in \mathbb N^N}\f1{1-e^{2\pi \im(w + \vec a_N \cdot \vec m_N)}} ~ ,
 \end{align}
 holds. A special value of the multiple zeta-function employed in the main text is
 \begin{align}
 \zeta_2(0, w | a_1 , a_2 ) = \f1{12 \, a_1 a_2}\comm{6 w^2 - 6(a_1 + a_2)w + a_1^2 + a_2^2 + 3 a_1 a_2} ~ .
 \end{align}


\section{Conformal Killing spinor equation}

 Let us begin by recalling that a  Killing spinor $\z$ is also a conformal Killing-spinor: if $\z$ fulfils the KSE with background fields $(A_\mu, V_\mu, H)$, then it exists a background field $A^{C}_\mu$ such that  $\z$ satisfies  the following CKSE:
\begin{align}
& \p{\nabla_\mu - \im  A^{C}_\mu }\z =  \f13 \g_\mu  \g^\nu \p{ \nabla_\nu - \im  A^{C}_\nu } \z   ~ , 
\end{align}
where 
\begin{align}
A^C_\mu = A_\mu - \f32 V_\mu ~ . 
\end{align}
Indeed, by using
\begin{align}
& \gamma_\mu \gamma_\nu = g_{\mu\nu} + \im \lc_{\mu\nu\rho}\gamma^\rho ~ , &      \g^\nu \p{ \nabla_\nu - \im A_\nu } \z =  - \f{3H}2   \z   ~ ,
\end{align}
we can rewrite the  KSE for $\z$ as follows:
\begin{align}
\p{\nabla_\mu - \im A_\mu}\z = & - \f H2 \g_\mu \z - \im V_\mu \z - \f12 \lc_{\mu\nu\rho}V^\nu \g^\rho \z  \nn \\
= &  \f13 \g_\mu  \g^\nu \p{ \nabla_\nu - \im A_\nu } \z  - \im V_\mu \z + \f\im2 V^\nu \g_\mu \g_\nu \z - \f\im2 V_\mu \z ~ , 
\end{align}
namely
\begin{align}
& \p{\nabla_\mu - \im  A^{C}_\mu }\z =  \f13 \g_\mu  \g^\nu \p{ \nabla_\nu - \im  A^{C}_\nu } \z   ~ ,  & A^{C}_\mu = A_\mu - \f32 V_\mu ~ .
\end{align}
The corresponding equation for $\zt$ is readily found by applying the maps $A_\mu \to - A_\mu$ and $V_\mu \to - V_\mu$, yielding $A^C_\mu \to - A^C_\mu$ and
\begin{align}
& \p{\nabla_\mu + \im  A^{C}_\mu }\zt =  \f13 \g_\mu  \g^\nu \p{ \nabla_\nu + \im  A^{C}_\nu } \zt    ~ .
\end{align}
Now, let $\Dc_\mu$ be the $R$-symmetry covariant derivative at $V_\mu = 0$, namely 
\begin{align}
\Dc_\mu = \nabla_\mu - \im q_R A_\mu ~ .
\end{align}
Then, the Killing spinors $(\z , \zt)$ of $R$-charges $(+1, -1)$ satisfy the following Killing-spinor equations (KSEs):
\begin{align}
& \Dc_\mu \z = (\nabla_\mu - \im A_\mu)\z = - \f H2 \g_\mu \z - \im V_\mu \z - \f12 \lc_{\mu\nu\rho} V^\nu \g^\rho \z ~ , \nn \\
& \Dc_\mu \zt = (\nabla_\mu + \im A_\mu)\zt = - \f H2 \g_\mu \zt + \im V_\mu \zt + \f12 \lc_{\mu\nu\rho} V^\nu \g^\rho \zt ~ . \nn 
\end{align}
The latter imply the relations
\begin{align}
& \g^\mu \Dc_\mu \z = \g^\mu \p{ \nabla_\mu - \im A_\mu } \z = - \f{3 H}2 \z ~ , & \g^\mu \Dc_\mu \zt =  \g^\mu \p{ \nabla_\mu + \im A_\mu }  \zt = - \f{3 H}2 \zt ~ ,
\end{align}
which are also useful to connect the KSE to the corresponding conformal Killing-spinor equations (CKSEs).


\section{Alternative representation of the supersymmetric background}
   
   The supersymmetry conditions  can also be rewritten in terms of two  $R$-symmetry neutral objects $(\xi , \Phi)$ defined as follows:
\begin{align}
 \xi_\mu = \zzt^{-1} K_\mu  ~ , \qquad \Phi_{\mu \nu} = \lc_{\mu \nu \rho} \xi^\rho  = \im P_{[\mu} \widetilde P_{\nu]}  ~ , 
\end{align}
satisfying
\begin{align}
 & \iota_\xi \xi = 1 ~ ,  & g_{\mu\nu} = \xi_\mu \xi_\nu - {\Phi_\mu}^\rho \Phi_{\rho \nu} = \xi_\mu \xi_\nu - P_{(\mu} \widetilde P_{\nu)} ~ , 
\end{align}
as well as
\begin{align}
  \Phi_{\mu \nu}\xi^\nu = 0~,
  \qquad  \Phi_{\mu \nu} P^\nu = - \im P^\mu ~ , \qquad  \Phi_{\mu \nu}  \widetilde P^\nu = + \im \widetilde P^\mu ~ .
\end{align}
When $\zt=\zeta^c$ coincides with the charge conjugate of the spinor $\z$, the 
one-form $\xi$ and the two-form $\Phi$ are both real and define an almost contact metric structure \cite{Closset:2012ru}. However, one can use these variables to write the background fields even if the background is not ``real'',  as in our set-up. The derivatives of the bilinears are
\begin{align}
 & \nabla_\mu \log \zzt  = \Phi_{\mu \rho} V^\rho ~ ,  & \nabla_\mu \xi_\nu = \im H \Phi_{\mu\nu} - \lc_{\mu\nu\lambda} V^\lambda - \Phi_{\mu\rho} V^\rho \xi_\nu  ~ ,  
\end{align}
implying 
\begin{align}
 V^\lambda & = \nabla_\mu \Phi^{\mu\lambda} + \p{ \im H - \f12 \xi_\rho \nabla_\mu \Phi^{\mu\rho} }\xi^\lambda = \Phi^{\lambda \mu} \nabla_\mu \log \zzt + \p{ \im H + \f12 \xi_\rho \nabla_\mu \Phi^{\mu\rho} }\xi^\lambda  ~ ,  
\end{align}
and 
\begin{align}
A_\mu & = \f\im4 \widetilde P^\nu \nabla_\mu P_\nu + \f{\im H}2 \xi_\mu + V_\mu  ~ .  
\end{align}
The background field $A^C$ appearing in the conformal Killing spinor equation is independent of $H$ and reads
\begin{align}
A^C_\mu & = A_\mu - \f32 V_\mu    \nn\\  
& = \f\im4 \widetilde P^\nu \nabla_\mu P_\nu + \f{\im H}2 \xi_\mu - \f12 V_\mu   \nn\\  
& = \f\im4 \widetilde P^\nu \nabla_\mu P_\nu  - \f12 \nabla_\rho {\Phi^\rho}_{\mu} + \f14 \p{  \xi_\rho \nabla_\lambda \Phi^{\lambda\rho} }\xi_\mu   \nn\\  
& = \f\im4 \widetilde P^\nu \nabla_\mu P_\nu - \f12 {\Phi_\mu}^\lambda \nabla_\lambda \log \zzt - \f14 \p{ \xi_\rho \nabla_\lambda \Phi^{\lambda \rho} } \xi_\mu ~ .  
\end{align}


\section{Supersymmetry  and cohomology}
\label{sexcohomo}

\subsection{Vector multiplet}

\paragraph{Supersymmetry transformations.}

Given a gauge group $G$, a non-Abelian $\mathcal N=2$ vector multiplet in three dimensions contains a gauge field $\Ac_\mu$, a scalar field $\sigma$, the two-component  spinors $\la_\alpha$, $\lat_\alpha$ and an auxiliary field $D$, all transforming in the adjoint representation of $G$. The fields $(\Ac_\mu , \sigma, \la , \lat , D)$ respectively have $R$-charges $(0, 0 , 1 , -1 , 0)$. The corresponding supersymmetry transformations are 
\begin{align}\label{eq: vm_susy}
\delta \Ac_\mu & = - \im \z \g_\mu \lat - \im \zt \g_\mu \la ~ , \nn \\
\delta \sigma & = -  \z  \lat + \zt  \la ~ , \nn \\
\delta \la & = - \f\im2 \lc^{\mu\nu\rho}\g_\rho \z \Fc_{\mu\nu} + \im \z(D+\sigma H) - \g^\mu \z(\im \Dr_\mu\sigma - V_\mu \sigma) ~ , \nn \\
\delta \lat & = - \f\im2 \lc^{\mu\nu\rho}\g_\rho \zt \Fc_{\mu\nu} - \im \zt(D+\sigma H) + \g^\mu \zt(\im \Dr_\mu\sigma + V_\mu \sigma) ~ , \nn \\
\delta D & = \Dr_\mu\p{\z\g^\mu\lat - \zt\g^\mu\la} - \im V_\mu\p{\z\g^\mu\lat + \zt \g^\mu\la} - \comm{\sigma, \z\lat + \zt\la} - H\p{\z \lat - \zt \la} ~ ,
\end{align}
where  $\Fc_{\mu\nu}  =\partial_\mu \Ac_\nu - \partial_\nu \Ac_\mu - \im \comm{\Ac_\mu , \Ac_\nu}$, while the gauge-covariant derivative $\Dr_\mu$ acts on a field $\Psi$ of $R$-charge $q_R$, central charge $q_Z$ and in the representation $\Rc$ of the gauge group $G$ as follows:
\begin{align}
& \Dr_\mu \Psi = \comm{ \nabla_\mu  - \im q_R\p{A_\mu - \f12 V_\mu}  - \im q_Z \Cc_\mu  - \im \Ac_\mu \circ_\Rc } \Psi  ~ ,  
\end{align}
with, for instance, 
\begin{align}
 \Ac \circ_{\rm adjoint} \Psi  =\comm{\Ac , \Psi} ~ , \qquad \Ac \circ_{\rm fundamental} \Psi  = \Ac  \Psi ~ , \qquad \Ac \circ_{\rm anti-fundamental} \Psi  = - \Psi \Ac   ~ .
\end{align}
The fields belonging to the vector multiplet have vanishing central charge, hence $\Cc_\mu$  appears in (\ref{eq: vm_susy}) only implicitly, via $V_\mu$. 


\paragraph{Cohomological complex.}

The supersymmetry variation $\delta = \delta_\z + \delta_\zt$ is an equivariant differential. Indeed,  the supersymmetry transformations for the vector multiplet can be rewritten as a cohomological complex containing the gauge field $\Ac$, the scalar $\sigma$, the Grassmann-odd 0-form $\chi = \im(\zt\la + \z\lat)/(2 \zzt)$ and their $\delta$-differentials: 
\begin{align}\label{eq: vm_cc}
 \delta \Ac & = \Lambda ~ ,   & \delta \Lambda  & = - 2 \im \p{   L_K   + \Gc_{\Phi_G}  } \Ac ~ , \nn \\
\delta \sigma  & = - (\im/\zzt) \iota_K \Lambda ~ , &     \delta {\Phi_G}  & = 0 ~ , \nn \\
 \delta \chi  & = \Delta  ~ ,  & \delta \Delta  & = - 2 \im \p{  L_K   + \Gc_{\Phi_G}   } \chi ~ ,
\end{align}
with $L_K$ being the covariant Lie derivative
\begin{align}
L_K  & =   \Lc_K  - \im q_R\p{ \iota_K A - \f12 \iota_K V } - \im q_Z ( \iota_K \Cc )    + \zzt (  q_Z - q_R H  )  \nn\\
& =   \Lc_K  - \im q_R \Phi_R - \im q_Z ( \iota_K \Cc + \im \zzt)   ~ . 
\end{align}
In fact,  $L_K  =   \Lc_K $ in (\ref{eq: vm_cc}) as all fields in (\ref{eq: vm_cc}) have vanishing   central charge $q_Z$ and $R$-charge $q_R$. The cohomological complex of the vector multiplet is independent of the $U(1)_R$ bundle. In (\ref{eq: vm_cc}) also appears  $\Gc_{\Phi_G} $, which is a gauge transformation with parameter ${\Phi_G}$,  respectively acting on $\Ac$ and on a field $X\neq\Ac$ in the representation $\Rc$ of the gauge group as
\begin{align}
& \Gc_{\Phi_G}  \Ac = - \dd_A {\Phi_G} = -  \dd {\Phi_G} + \im \comm{\Ac, {\Phi_G}}   ~ , & \Gc_{\Phi_G} X = - \im  {\Phi_G} \circ_\Rc X  ~ .
\end{align}
Explicitly, the Grassmann-odd 1-form $\Lambda_\mu$, the Grassmann-odd 0-form $\chi$ and the Grassmann-even 0-forms ${\Phi_G}, \Delta$ are defined by
\begin{align}
\Lambda_\mu & = - \im \p{ \z \g_\mu \lat + \zt \g_\mu \la } ~ , \nn \\
\chi &  =  \f{\im}{2 \zzt} \p{ \zt \la + \z \lat}~ ,  \nn \\
{\Phi_G} & = \iota_K \Ac - \im \zzt \sigma ~ ,  \nn \\
\Delta  & = D + \f1\zzt\iota_K \p{\star \Fc} + \f\im\zzt  \sigma \comm{    \iota_K V - \im \zzt H } = D + \f\im 2 P^\mu \widetilde P^\nu \Fc_{\mu\nu} +\f\im\zzt \sigma \comm{   \iota_K V - \im \zzt H }  ~ ,  
\end{align}
with the  map from  $(\Lambda, \chi)$ to  $(\la, \lat)$  being
\begin{align}
& \la  = \im \p{ \f{1}{2 \zzt^2} \iota_K \Lambda +  \chi }\z - \f{\im}{2 \zzt}\p{\iota_P\Lambda} \zt ~ ,  & \lat  = \im \p{ \f{1}{2 \zzt^2} \iota_K \Lambda -   \chi }\zt - \f{\im}{2 \zzt}\p{\iota_\Pt\Lambda} \z ~ .  
\end{align}


\subsection{Chiral multiplet}

\paragraph{Supersymmetry transformations.}

A three-dimensional $\mathcal N=2$ chiral multiplet of $R$-charge $r$, central charge $z$ and in a representation $\Rc$ of a  gauge group $G$  contains a complex scalar field $\phi$,  a two-component  spinor $\psi_\alpha$  and an auxiliary field $F$. The fields $(\phi, \psi, F)$ respectively have $R$-charges $(r, r-1, r-2)$ and central charge $z$. The corresponding supersymmetry transformations are
\begin{align}
\delta \phi& = \sqrt 2 \z\psi  ~ , \nn \\
\delta \psi & =  \sqrt 2 \z F + \im \sqrt 2 \p{\sigma + r H - z }\zt \phi - \im \sqrt 2 \g^\mu \zt \Dr_\mu \phi  ~ , \nn \\
\delta F & = \im \sqrt 2 \comm{ z - \sigma - (r-2)H }\zt \psi - \im \sqrt2 \Dr_\mu\p{\zt \g^\mu \psi} + 2 \im (\zt \lat)\phi   ~ . 
\end{align}
Analogously, a three-dimensional $\mathcal N=2$ anti-chiral multiplet of $R$-charge $-r$, central charge $-z$ and in the conjugate representation $\overline\Rc$ of a  gauge group $G$  contains a complex scalar field $\phit$,  a two-component  spinor $\psit_\alpha$  and an auxiliary field $\Ft$. The fields $(\phit, \psit, \Ft)$ respectively have $R$-charges $(-r, -r+1, -r+2)$ and central charge $-z$. In this case, the supersymmetry transformations read
\begin{align}
\delta \phit & = - \sqrt 2 \zt\psit  ~ , \nn \\
\delta \psit & =  \sqrt 2 \zt \Ft - \im \sqrt 2 \phit  \p{ \sigma + r H - z   }\z + \im \sqrt 2 \g^\mu \z \Dr_\mu \phit    ~ , \nn \\
\delta \Ft & = \im \sqrt 2 (\z\psit) \comm{ z - \sigma - (r-2)H }  - \im \sqrt2 \Dr_\mu\p{\z \g^\mu \psit }   +  2 \im \phit  (\z \la) ~ .
\end{align}


\paragraph{Cohomological complex.}

Similarly to the case of the vector multiplet, the supersymmetry variation $\delta = \delta_\z + \delta_\zt$ acts on the fields of the chiral multiplet as an equivariant differential, which induces  the following cohomological complex:
\begin{align}
& \delta \phi =    C ~ , & \delta C = & - 2 \im  \p{ L_K  + \Gc_{\Phi_G}   } \phi ~ , \nn \\
& \delta B  =    \Theta ~ ,  & \delta \Theta  = & -2 \im \p{L_K + \Gc_{\Phi_G} } B  ~ , 
\end{align}
where the Grassmann-odd 0-forms $B, C$ and the  Grassmann-even 0-form $\Theta$ are defined by
\begin{align}
B & = - \f{\zt \psi}{\sqrt2 \, \zzt} ~ , \nn \\
C & = \sqrt2 (\z \psi) ~ , \nn \\
\Theta & =  F + \im  \comm{ \Lc_\Pt  - \im r \p{\iota_\Pt A - \f12 \iota_\Pt V}  - \im z (\iota_\Pt \Cc)  - \im (\iota_\Pt \Ac)  } \phi ~ . 
\end{align}
The inverse map from the Grassmann-odd scalars $(B,C)$ to the spinor $\psi$ is
\begin{align}
& \psi = \sqrt 2 B \z + \f{C}{\sqrt 2 \, \zzt} \zt ~ .
\end{align}
For completeness, we report the intermediate form of the supersymmetry transformations, the one  obtained before introducing $\Theta$:
\begin{align}
\delta \phi = &  C ~ , \nn \\
\delta C = & - 2 \im  \comm{   \Lc_K  - \im  r\p{ \iota_K A - \f12 \iota_K V }  + \zzt(  z - r H  )  - \im z ( \iota_K \Cc )   - \im {\Phi_G}    } \phi ~ , \nn \\
\delta B  = &  F + \im  \comm{ \Lc_\Pt  - \im r \p{\iota_\Pt A - \f12 \iota_\Pt V}  - \im z (\iota_\Pt \Cc)  - \im (\iota_\Pt \Ac)  } \phi ~ ,    \nn \\
\delta F  = & -2 \im \comm{ \Lc_K - \im (r-2)\p{\iota_K A - \f12 \iota_K V} - \im z \iota_K \Cc + \zzt \comm{z-\p{r-2}H} - \im {\Phi_G} } B \nn \\
& - \im \comm{ \Lc_\Pt - \im r \p{ \iota_\Pt A - \f12 \iota_\Pt V } - \im z ( \iota_\Pt \Cc ) - \im ( \iota_\Pt \Ac) } C - (\iota_\Pt \Lambda) \phi ~ .
\end{align}
Analogously, the cohomological complex of the anti-chiral multiplet reads 
\begin{align}
& \delta \phit =  \Ct ~ , & \delta \Ct = & - 2 \im  \p{ L_K  + \Gc_{\Phi_G}   } \phit ~ , \nn \\
& \delta \Bt  =    \Tht ~ ,  & \delta \Tht  = & -2 \im \p{L_K + \Gc_{\Phi_G} } \Bt  ~ , 
\end{align}
where the Grassmann-odd 0-forms $\Bt, \Ct$ and the  Grassmann-even 0-form $\Tht$ are defined by
\begin{align}
\Bt & =  \f{\z \psit}{\sqrt2 \, \zzt} ~ , \nn \\
\Ct & = - \sqrt2 (\zt \psit) ~ , \nn \\
\Tht & =  \Ft + \im  \comm{ \Lc_P + \im r \p{\iota_P A - \f12 \iota_P V}  + \im z (\iota_P \Cc)   } \phit - \phit  (\iota_P \Ac)     ~ ,
\end{align}
with the inverse map from the Grassmann-odd scalars $(\Bt,\Ct)$ to the spinor $\psit$ being
\begin{align}
& \psit = \sqrt 2 \Bt \zt + \f{\Ct}{\sqrt 2 \, \zzt} \z ~ .
\end{align}
The intermediate step in the derivation of the cohomological complex of the anti-chiral multiplet is given by the transformations
\begin{align}
\delta \phit = &  \Ct ~ , \nn \\
\delta \Ct = & - 2 \im  \comm{   \Lc_K  + \im r\p{ \iota_K A - \f12 \iota_K V }  + \zzt ( -  z + r H  )  + \im z ( \iota_K \Cc )     } \phit  + 2  \phit  {\Phi_G}     ~ , \nn \\
\delta \Bt  = &  \Ft + \im  \comm{ \Lc_P  + \im r \p{\iota_P A - \f12 \iota_P V}  + \im z (\iota_P \Cc)    } \phit -   \phit ( \iota_P \Ac)    ~ ,    \nn \\
\delta \Ft   = & -2 \im \comm{ \Lc_K + \im (r-2)\p{\iota_K A - \f12 \iota_K V} + \im z ( \iota_K \Cc ) + \zzt \comm{- z - \p{- r + 2}H} } \Bt +2 \Bt {\Phi_G}  \nn \\
& - \im \comm{ \Lc_P + \im r \p{ \iota_P A - \f12 \iota_P V } + \im z ( \iota_P \Cc ) } \Ct + \Ct ( \iota_P \Ac) + \phit (\iota_P \Lambda)  ~ .
\end{align}


\end{document}